\documentclass[a4paper,fleqn,usenatbib]{mnras}

\usepackage{newtxtext,newtxmath}
\usepackage[T1]{fontenc}
\usepackage{ae,aecompl}

\usepackage{graphicx}	% Including figure files
\usepackage{amsmath}	% Advanced maths commands
\usepackage{amssymb}	% Extra maths symbols
\usepackage{adjustbox} % Adjust tables
\usepackage{lipsum}
\usepackage{caption} % so caption works in tabe with notes
\usepackage[para,online,flushleft]{threeparttable}

\newcommand{\Me}{$M_\oplus$}

\newcommand{\Msun}{$M_{\sun}$}

\newcommand{\Mp}{M_\mathrm{p}}

\newcommand{\rmax}{r_{\max}}

\newcommand{\rc}{r_\mathrm{c}}

\newcommand{\Sigmag}{\Sigma_{\rm G}}
\newcommand{\Sigmadot}{\dot{\Sigma}^{+}_\mathrm{coll}}

\newcommand{\rb}{r_\mathrm{belt}}
\newcommand{\rbb}{r_\mathrm{BB}}
\newcommand{\tc}{t_\mathrm{c}}
\newcommand{\Qd}{{Q_\mathrm{D}^{\star}}}
\newcommand{\dotCO}{\dot{M}_\mathrm{CO}}
\newcommand{\fir}{f_\mathrm{IR}}
\newcommand{\Dc}{D_\mathrm{c}}
\newcommand{\Dcc}{D_\mathrm{cc}}
\newcommand{\Xc}{X_\mathrm{c}}

\title[Exocometary gas evolution around A stars]{Population synthesis
  of exocometary gas around A stars}
\author[S. Marino et al.]{S. Marino$^{1}$\thanks{E-mail:
    sebastian.marino.estay@gmail.com}, M. Flock$^{1}$, Th. Henning$^{1}$, Q. Kral$^{2}$, L. Matr{\`a}$^{3}$ and M. C. Wyatt$^{4}$\\ %\newauthor{and M. C. Wyatt$^{1}$} \\
  $^{1}$Max Planck Institute for Astronomy, K\"onigstuhl 17, 69117 Heidelberg, Germany\\
  $^{2}$LESIA, Observatoire de Paris, Universit{\'e} PSL, CNRS, Sorbonne Universit{\'e}, Univ. Paris Diderot, Sorbonne Paris Cit{\'e}, 5 place Jules Janssen, 92195 Meudon, France\\
  $^{3}$Harvard-Smithsonian Center for Astrophysics, 60 Garden Street, Cambridge, MA 02138, USA\\
  $^{4}$Institute of Astronomy, University of Cambridge, Madingley Road, Cambridge CB3 0HA, UK\\
}

\date{Accepted XXX. Received YYY; in original form ZZZ}

\pubyear{2019}

\begin{document}
\label{firstpage}
\pagerange{\pageref{firstpage}--\pageref{lastpage}}
\maketitle

\begin{abstract}

The presence of CO gas around 10-50~Myr old A stars with debris discs
has sparked debate on whether the gas is primordial or
secondary. Since secondary gas released from planetesimals is poor in
H$_2$, it was thought that CO would quickly photodissociate never
reaching the high levels observed around the majority of A stars with
bright debris discs. \cite{Kral2019} showed that neutral carbon
produced by CO photodissociation can effectively shield CO and
potentially explain the high CO masses around 9 A stars with bright
debris discs. Here we present a new model that simulates the gas
viscous evolution, accounting for carbon shielding and how the gas
release rate decreases with time as the planetesimal disc loses
mass. We find that the present gas mass in a system is highly
dependant on its evolutionary path. Since gas is lost on long
timescales, it can retain a memory of the initial disc mass. Moreover,
we find that gas levels can be out of equilibrium and quickly evolving
from a shielded onto an unshielded state. With this model, we build
the first population synthesis of gas around A stars, which we use to
constrain the disc viscosity. We find a good match with a high
viscosity ($\alpha\sim0.1$), indicating that gas is lost on timescales
$\sim1-10$~Myr. Moreover, our model also shows that high CO masses are
not expected around FGK stars since their planetesimal discs are born
with lower masses, explaining why shielded discs are only found around
A stars. Finally, we hypothesise that the observed carbon cavities
could be due to radiation pressure or accreting planets.

\end{abstract}

  %% The planet mass should be even lower if the
  %% gap was primordial, i.e. already present in the dust distribution
  %% that formed the planetesimal disc


%%%%%%%%%%%%%%%%%%%%%%%%%%%%%%%%%%%%%%%%%%%%%%%%%%

%%%%%%%%%%%%%%%%% BODY OF PAPER %%%%%%%%%%%%%%%%%%

\section{Introduction}
\label{sec:intro}

% gas knowledge (absorption and emission)

The discovery of circumstellar gas around young main-sequence stars
with debris discs dates back more than two decades, with a few systems
showing molecular emission at millimetre wavelengths
\citep[e.g][]{Zuckerman1995} and others atomic absorption in the UV
\citep{Slettebak1975}, some of which varies with time
\citep{Ferlet1987}. Thanks to Herschel, it also became possible to
study the gas component in the far-IR through atomic emission from
ionised carbon and neutral oxygen
\citep[e.g.][]{Riviere-Marichalar2012, Roberge2013, Cataldi2014},
adding a third tool for the observational study of gas in systems with
debris discs. Today we know of about 20 systems with debris discs and
gas detected in emission, and another $\sim11$ systems (with and
without infrared excess) with circumstellar gas detected in absorption
only \citep[e.g.][]{Montgomery2012, Welsh2018, Rebollido2018,
  Iglesias2018}.

% origin

While there has been some consensus on the origin of the absorption
lines as arising from \textit{falling evaporating bodies}
\citep[FEBs,][]{Ferlet1987, Beust1990, Beust1996, Kiefer2014b}, no
model has been completely successful at explaining the colder emitting
gas at tens of au that is now found around $\sim20$ systems
\citep[e.g.][]{Moor2011a, Dent2014, Moor2015gas, Marino2016,
  Greaves2016, Lieman-Sifry2016, Marino2017etacorvi,
  Matra2017fomalhaut, Moor2017, Matra2019twa7}, mostly through CO
detections. Since CO molecules exposed to stellar and interstellar UV
radiation will photodissociate in short-timescales
\citep[$\lesssim120$~yr,][]{Hudson1971, vanDishoeck1988, Visser2009},
the CO detected in some of these systems suggests that we are
observing these systems at a very particular moment (e.g. at the later
stages of protoplanetary disc dispersal, hereafter primordial origin)
or CO and other gas species are being replenished in these systems
(hereafter secondary origin). \cite{Zuckerman2012} proposed that gas
could be released in the same collisional cascade that replenishes the
dust levels in debris discs. The amount of gas and dust, could then be
used to infer the volatile composition of planetesimals.  This
scenario would be consistent with systems with the low CO levels
observed, for example, around $\beta$~Pic, HD~181327 and
Fomalhaut. However, a significant fraction of the systems with
detected CO would require planetesimals much richer in CO and/or
CO$_{2}$ compared to Solar System comets, or gas release rates that
are decoupled from the inferred dust production rates
\citep{Kral2017CO}. Furthermore, the systems with the highest CO
content have ages between 10-50~Myr and their content is not
necessarily correlated with their debris disc-like dust
\citep{Moor2017}. Therefore, these have been tagged as \textit{hybrid
  discs} meaning they have primordial gas leftovers and secondary dust
\citep{Kospal2013}, although it is not clear how primordial gas has
survived for so long in those systems.

A potential pathway to alleviate these tensions in the secondary
origin scenario, is that the CO becomes shielded from UV radiation in
systems with high gas content and its lifetime is much longer than in
the unshielded case. While debris discs are optically thin hence dust
cannot effectively shield the gas, CO can become self-shielded or
shielded by molecular hydrogen \citep{vanDishoeck1988,
  Visser2009}. This is difficult since H$_{2}$ is unlikely to be
present at high densities in this secondary origin scenario, and CO
would need to be released at a very high rate to reach the necessary
column densities to become self-shielded \citep{Kral2017CO}. However,
carbon that is produced through CO photodissociation can also shield
CO from UV photons, becoming ionised \citep{Rollins2012}. Carbon and
oxygen are expected to viscously evolve and form an atomic accretion
disc \citep{Kral2016}, even if the stellar luminosity is high enough
to blow out carbon since it can remain bound due to self-shielding and
interactions with more bound species such as oxygen
\citep{Fernandez2006, Kral2017CO}. \cite{Kral2019} found that if
viscosities were low or the CO input rate high, enough carbon could
accumulate to shield CO and explain hybrid discs as shielded discs of
secondary origin.

So far, none of the above studies has considered the viscous evolution
of an exocometary gas disc coupled with the time dependent gas release
rate. If the release of gas is regulated by the mass loss rate in the
planetesimal disc, then we expect this to decrease steeply with time
relaxing into a collisional equilibrium \citep[e.g.][]{Dominik2003,
  Krivov2006, Wyatt2007hotdust}. This means that the total gas mass
input into a system depends strongly on its previous evolution if the
gas lifetime is similar or longer than the age of the system (or
collisional timescales). If viscosities are low, this could well be
the case for shielded discs, thus collisional evolution cannot be
neglected a priori. In this paper we consider the effect of an
evolving mass input rate on the viscous evolution of the gas in the
system, particularly we focus on CO and carbon. This paper is
structured as follows. We first present our model and numerical
simulations in \S\ref{sec:model}. Then in \S\ref{sec:population} we
use this model to produce a population synthesis of A stars with
debris discs that release gas, which we compare with observations. In
\S\ref{sec:fgk} we show that the same model applied to FGK stars is
also consistent with observations. In \S\ref{sec:discussion} we
discuss our results and some of our assumptions, and finally in
\S\ref{sec:conclusions} we report our conclusions.

%% Whether there is enough carbon to shield CO in
%% systems with high CO content depends on the rate at which carbon is
%% produced (i.e. rate at which CO is released and photodissociated) and
%% dispersed through viscous evolution. \cite{Kral2016} found that 

%% into carbon and oxygen, they will disperse and
%% potentially form a viscous accretion disc \citep{Kral2016}. Even if
%% the stellar luminosities are high enough to blow out carbon, we still
%% expect it to remain bound to the system due to self-shielding and
%% interactions with more bound species such as oxigen
%% \citep{Fernandez2006, Kral2017CO}.....

%% until recently only CO self-shielding was
%% considered was rollins2012
%% , and thus the rate at which gas is released
%% from collisions is lower and thus consistent with it can survive
%% icnrease CO lifetime... self-shielding... carbon..

% shielded discs

% absence of coll evolution modelling

% Sections in the paper

\section{Model}
\label{sec:model}
The gas evolution is treated with a simple viscous evolution 1D model,
focusing only on the radial dimension and assuming axisymmetry
\citep[similar to][]{Moor2019}. In this model, gas is input into the
system at the radial locations of the planetesimal belt, from where it
is allowed to viscously evolve. The gas surface density evolution,
$\Sigmag$, is then set by the usual viscous evolution equation
\citep{Lynden-Bell1974} with the addition of an input source that
arises from a collisional cascade of planetesimals, $\Sigmadot(r,t)$,
\begin{equation}
  \frac{\partial\Sigmag}{\partial t} = \frac{3}{r} \frac{\partial}{\partial r} \left[r^{1/2} \frac{\partial}{\partial r} (\nu \Sigmag r^{1/2})  \right] + \Sigmadot(r,t), \label{eq:stot}
\end{equation}
where $\nu$ is the kinematic viscosity. In Equation \ref{eq:stot}, we
have also made the usual approximation that the orbital velocity does
not significantly differ from circular Keplerian rotation.

Rather than studying $\Sigmag$, in this paper we are mainly interested
in following the evolution of CO, C and O, which can have
significantly different distributions \citep{Kral2016}. Therefore,
instead of solving Equation \ref{eq:stot}, we solve for the evolution
of the surface density of each gas species $\Sigma_i$ by advecting its
mass at the viscous radial velocity, $\varv_r$, hence
\begin{eqnarray}
  \frac{\partial\Sigma_i}{\partial t} &=& - \frac{1}{r}\frac{\partial}{\partial r} (r \varv_r \Sigma_i) + \dot\Sigma_i^{\pm}(r,t), \label{eq:si} \\
  \Sigmag \varv_r &=& - \frac{3}{\sqrt{r}} \frac{\partial}{\partial r} (\nu \Sigmag \sqrt{r}), \label{eq:vr}\\
  \Sigmag &=& \sum_{i} \Sigma_i
\end{eqnarray}
where $\dot\Sigma_i^{\pm}(r,t)$ represents the additional production
and destruction processes (e.g. CO photodissociation, see below) and
$\varv_r$ the radial velocity of the gas. Equation \ref{eq:si} is a
continuity equation with a source term, while Equation~\ref{eq:vr} is
the conservation of angular momentum which is applicable to the total
mass. We solve Equations \ref{eq:si} and \ref{eq:vr} using first-order
explicit finite-volume update following \citet{Bath1981} and
\citet{Booth2017gas}. We also include the diffusion of the gas
species, that is given by
\begin{equation}
  \frac{\partial f_i}{\partial t} = \frac{1}{r\Sigmag} \left( r \nu \Sigmag \frac{\partial f_i}{\partial r}\right), \label{eq:diff}
\end{equation}
where $f_i=\Sigma_i/\Sigmag$ and we have taken a Schmidt number of
unity \citep{Stevenson1990, Turner2006}. The addition of
Equation~\ref{eq:diff} accounts for the diffusion due to turbulent
mixing rather than an accretion flow. It does not affect the total
surface density, but the relative abundances of CO, carbon and oxygen.

%% viscosity
For the viscosity we make the standard assumption of an alpha disc
model, with $\nu=\alpha c_s^2/\Omega_\mathrm{K}$, where $c_s$ is the
sound speed and $\Omega_\mathrm{K}$ the Keplerian speed. The sound
speed is calculated from the gas temperature, which is fixed and taken
equal to the blackbody temperature for simplicity ($T=278
L_\star^{1/4} r^{-1/2}$) and a mean-molecular weight that can vary
between 28 and 14 depending on if the gas is dominated by CO or by
atomic oxygen and carbon. This choice of parametrization of $\nu$ is
arbitrary, but it offers a form simple enough to understand the gas
evolution. Note that we expect $\alpha$ to be smaller than unity,
although its value has only been constrained between $10^{-4}-0.5$
\citep{Kral2016, Kral2016b, Kral2019, Moor2019}.

%% input of CO

The input of gas in the system happens through the release of volatile
species that escape after the break up of volatile rich bodies in a
collisional cascade. This release is spatially confined to a
planetesimal belt that extends in radius and is radially resolved by
our simulations. Furthermore, we assume that the released gas is
completely dominated by CO. In principle, other molecular species such
as CO$_2$, H$_2$O, HCN, etc. could also be released, however, current
observations are roughly consistent with CO dominated gas release. The
addition of other gas species is discussed in \S\ref{dis:othergas}.

%% photodissociation

Once CO is released and exposed to interstellar UV radiation it will
photodissociate into carbon and oxygen on a short timescale of 120~yr
\citep{Visser2009}, or even shorter depending on the strength of the
stellar UV flux \citep{Matra2017betapic}. For example, a stellar
luminosity higher than $20~L_\odot$ will shorten the CO
photodissociation timescale to less than 10 year within $\sim100$~au
if unshielded \citep[see Figure 1 in][]{Kral2017CO}. This timescale,
however, can be much longer if the CO column densities are high enough
to shield itself or if neutral carbon is present. The latter is of
special interest since depending on how fast CO is released and carbon
viscously spreads, it could explain the existence of large amounts of
CO in 10--50 Myr old systems \citep{Kral2019}.

In this work we will focus solely on the effect of interstellar UV
radiation and will neglect the stellar contribution. This is justified
by the two following reasons. First, most of the discs that we want to
explain have stellar luminosities below $20\ L_\odot$ and debris disc
sizes of $\sim100$~au, which means that if CO was completely
unshielded its lifetime would be of 120~yr and dominated by
ISRF. Second, the column densities of CO and carbon along the radial
direction will be 1--5 orders of magnitude larger than in the vertical
direction because the radial distribution of gas is typically much
broader than the disc scale height. This means that CO will be highly
shielded in the radial direction (see estimates in
\S~\ref{dis:photodissociation}). Therefore, UV interstellar radiation
from the top and bottom of the disc will be the predominant factor
that sets the CO lifetime.

Then, $\dot\Sigma_i^{+,-}(r,t)$ for CO and C/O becomes \citep{Kral2019}
\begin{eqnarray}
  \dot\Sigma_\mathrm{CO}^{\pm}(r,t)&=& \dot{\Sigma}^{+}_\mathrm{CO}  - \frac{\Sigma_\mathrm{CO}}{T_\mathrm{ph}(\Sigma_\mathrm{CO}, \Sigma_\mathrm{C})}, \\
  \dot\Sigma_\mathrm{C}^{\pm}(r,t)&=& + \frac{3}{7}\frac{\Sigma_\mathrm{CO}}{T_\mathrm{ph}(\Sigma_\mathrm{CO}, \Sigma_\mathrm{C})}, \\
  T_\mathrm{ph}(\Sigma_\mathrm{CO}, \Sigma_\mathrm{C})&=& \mathrm{120\ yr} \frac{\exp[\Sigma_\mathrm{C}/ \Sigma^\mathrm{c}_\mathrm{C}]}{\Theta(\Sigma_\mathrm{CO})}, 
\end{eqnarray}
where $\dot{\Sigma}^{+}_\mathrm{CO}$ is the input rate of CO and
$T_\mathrm{ph}$ is the photodissociation timescale of CO which depends
both on the column or surface density of CO (self-shielding) and
neutral carbon C. The self-shielding function $\Theta$ is calculated
by interpolating tabulated values in \citet[][Table
  6]{Visser2009}. This has a power-law-like dependence, with the CO
lifetime increased by a factor of 2.7 when the average column density
is $10^{15}$~cm$^{-2}$ or a surface density of
$4\times10^{-9}$~\Me~au$^{-2}$ ($1.1\times10^{-7}$~g~cm$^{-2}$). On
the other hand, shielding by neutral carbon has an exponential
dependence \citep{Rollins2012}, with a critical surface density
$\Sigma^\mathrm{c}_\mathrm{C}=10^{-7}$~\Me~au$^{-2}$
($2.7\times10^{-6}$~g~cm$^{-2}$). A caveat in the calculation of
$T_\mathrm{ph}$ is that in reality it will depend on the amount of
material along every line of sight since the 120~yr reference lifetime
assumes radiation is coming from all directions, therefore the
\textit{average} CO lifetime is expected to be slightly longer
depending on the disc scale height.

Our model has two further simplifications. First, we consider that the
ionisation fraction of carbon is negligible, or at least that the
fraction of neutral carbon is not much lower than unity for discs in
which neutral carbon plays an important role (this assumption is
discussed in \S\ref{dis:ionisation}). Second, we neglect the stellar
radiation pressure acting on carbon. In principle, neutral carbon
could be blown out by radiation pressure for stellar luminosities
greater than $\sim20\ L_\odot$ \citep[][note that the specific
  threshold depends on the model spectrum used]{Fernandez2006}. This,
however, probably does not happen in the discs we typically observe
since even with a thousandth of the CO input rate estimated in
$\beta$~Pic, carbon is continuously produced and can reach densities
that are high enough to become self-shielded from stellar UV and stay
bound to the system \citep{Kral2017CO}. The effect of radiation
pressure on carbon is further discussed in \S\ref{dis:carbon} and
\S\ref{dis:removal}.

Finally, we consider the following boundary conditions. We use an
inner radius of 1~au and we set
$\nu_\mathrm{in}\Sigma_\mathrm{G,in}=\mathrm{const}$, such that the
accretion rate is constant between the inner two cells and
$\Sigma_\mathrm{G}(r)$ tends to the expected solution that is
inversely proportional to $\nu(r)$. For the surface density at the
outer edge of our simulations ($r_\mathrm{out}$), we take whichever is
smaller between a power-law extrapolation or
$\nu_\mathrm{out}\Sigma_\mathrm{G,out}=\mathrm{const}$. We find that
when the CO mass input rate is constant, $r_\mathrm{out}=30 \rb$ is
large enough and solutions converge. However, for CO mass input rates
varying in time (\S\ref{sec:collevol}) we find that convergence
between solutions is achieved when $r_\mathrm{out}$ is a few times
larger than the viscous characteristic radius ($\rb(1+
t/t_\mathrm{\nu})$), thus we set $r_\mathrm{out}=3 \rb(1+
t_\mathrm{f}/t_\mathrm{\nu})]$, where $t_\mathrm{f}$ and
  $t_\mathrm{\nu}$ are the length of the simulation and viscous
  timescale at $\rb$ ($t_\nu=\rb^2/\nu(\rb)$). This ensures that the
  result from our simulations are not sensitive to the outer boundary
  conditions.

\subsection{Constant CO input rate}

 \begin{figure*}
  \centering
  \includegraphics[trim=0.4cm 0.5cm 0.5cm 0.4cm, clip=true, width=0.99\textwidth]{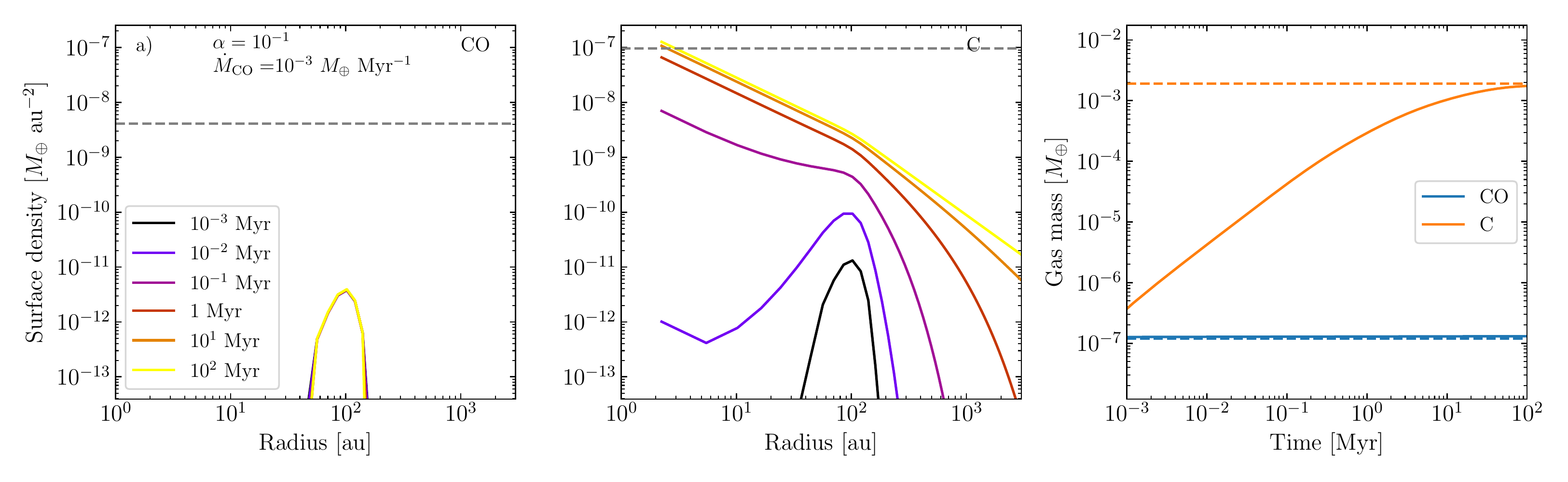}
  \includegraphics[trim=0.4cm 0.5cm 0.5cm 0.4cm, clip=true, width=0.99\textwidth]{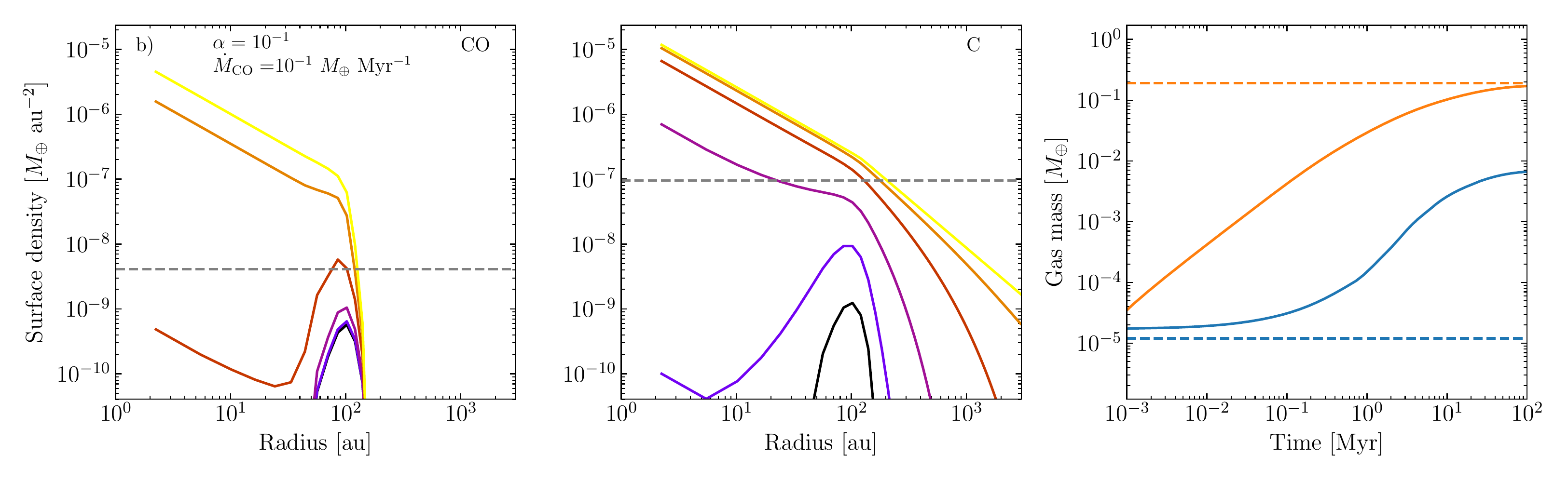}
  \includegraphics[trim=0.4cm 0.5cm 0.5cm 0.5cm, clip=true, width=0.99\textwidth]{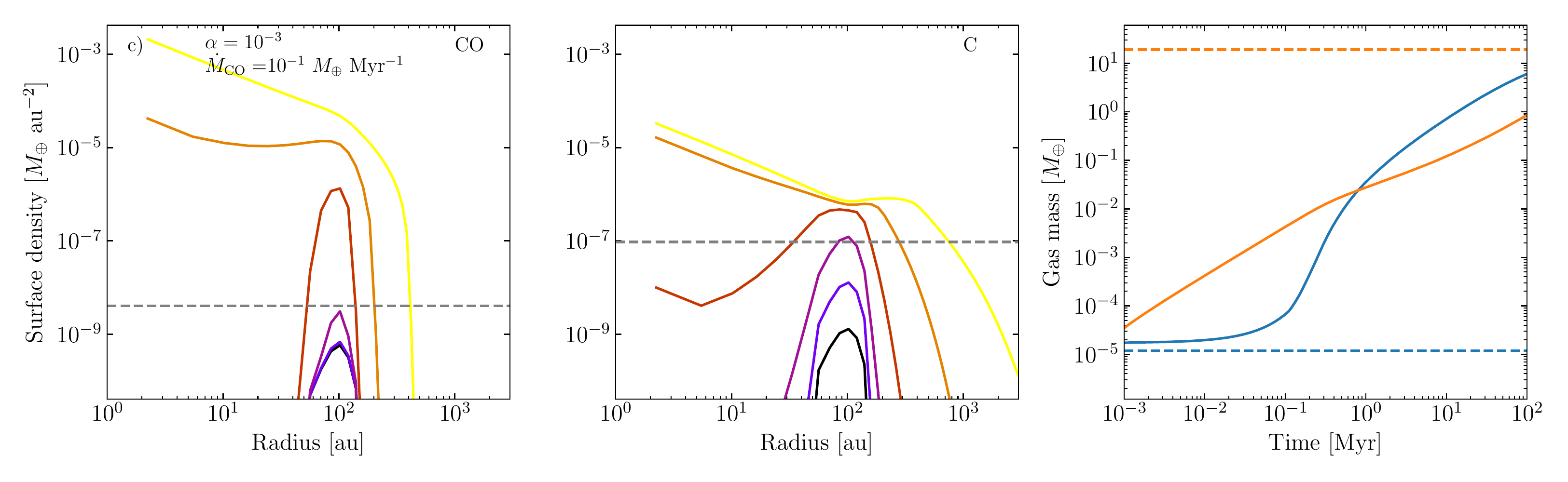}
 \caption{Viscous evolution of a gaseous disc over 100 Myr, with a
   constant input rate of CO over time and released roughly between
   75--125 au, for three different cases. \textbf{\textit{Top:}}high
   viscosity and low CO input rate. \textbf{\textit{Middle:}} high
   viscosity and high CO input rate. \textbf{\textit{Bottom:}} low
   viscosity and high CO input rate. The left and middle columns show
   the evolution of the CO and carbon surface density,
   respectively. The right column shows the evolution of the total gas
   mass of CO (blue continuous line) and carbon (orange continuous
   line). The dashed blue and orange lines represent the CO and carbon
   mass expected in steady state when CO is unshielded and the mass
   input rate of CO is constant over time. }
 \label{fig:constMdot}
\end{figure*}

Consider a disc of planetesimals around a 2~$M_\odot$ (16~$L_\odot$)
star, sustaining a collisional cascade that releases both dust and CO
gas. Let's assume that CO is released at a constant rate
$\dot{\Sigma}^{+}_\mathrm{CO}$, parametrized as a Gaussian centred on
$\rb=100$~au (belt mid radius) and with FWHM $\Delta \rb=50$~au. These
are arbitrary choices, but overall consistent with the observed
distribution of planetesimal disc radii and width around high
luminosity stars \citep{Matra2018mmlaw}. In Figure~\ref{fig:constMdot}
the resulting evolution of the CO (left panels) and carbon (middle
panels) surface densities and total masses (right panels) are shown
for different CO mass input rates and values of $\alpha$: 0.1 and
0.001. Hereafter we will refer to these $\alpha$'s and resulting
viscosities as high and low, respectively.

The top panel shows a simple case that would be applicable to systems
that are observed to have a low CO gas mass \citep[e.g. $\beta$~Pic,
  HD~181327 and Fomalhaut,][]{Dent2014, Marino2016,
  Matra2017fomalhaut}. CO is input in the system and in a few hundred
years it reaches steady state with
$\Sigma_\mathrm{CO}=\dot{\Sigma}^{+}_\mathrm{CO} \times
120\ \mathrm{yrs}$ and thus is confined in space and co-located with
planetesimals and large dust grains (top left). Carbon instead
viscously spreads forming an accretion disc which reaches a steady
state in a few times the viscous timescale at the belt location
($t_\nu=\rb^2/\nu=0.6$~Myr, top middle). Neither the surface density
of CO nor C reach the necessary levels to shield CO (grey dashed
lines), except for carbon within a few AU where no CO is present. The
final CO and C mass (top right) are consistent with analytic
predictions where $M_\mathrm{CO}=\dot{M}_\mathrm{CO}\times
120\ \mathrm{yrs}$ and $M_\mathrm{C}$ calculated by integrating
equation B17 in \citet{Metzger2012} assuming that the viscosity scales
linearly with radius (true for our case if $\mu$ is constant),
\begin{equation}
M_\mathrm{C}= \frac{2 \dot{M}_\mathrm{CO} \rb^2 }{7 \nu_\mathrm{belt}} \left[ 2 \left( \frac{r_{\max}}{\rb} \right)^{1/2}-1 \right] \label{eq:mc}
\end{equation}
where $r_{\max}$ is the maximum radius in the simulation. Note that
$M_\mathrm{C}\propto r_{\max}^{1/2}$, thus the total mass reported
here is dominated by low density gas at large radii to which
observations can be less sensitive. Therefore the carbon masses
reported in this section cannot be compared directly with
observations. Later in \S\ref{sec:population} we report instead the CO
and carbon masses within 500~au which is a more sensible estimate to
compare with typical observations. Note that if the viscosity did not
scale linearly with radius, then the bulk of the mass could reside at
small radii.

%% lecan be orders of magnitude higher than what could be
%% detected in is sensitive to the choice of considered here to calculate
%% the total mass ($r$) and $\nu_\mathrm{belt}$ is the viscosity at
%% $\rb$.

%%  This
%% particular choice of radius is arbitrary (and affects the carbon mass
%% estimates since carbon can vicously spread beyond that radius), but is
%% motivated by the maximum radius at which, for example, ALMA
%% observations would be sensitive enough to detect cold gas emission in
%% nearby systems.

In the middle panels, the CO mass input rate is now 100 times
higher. While the CO is slightly self-shielded within the first 0.1
Myr of evolution (note that the blue line in the right panel is above
the dashed line), by 1~Myr the surface density of carbon reaches the
critical level to shield CO at 100~au. The subsequent evolution has CO
viscously spreading inwards forming an accretion disc and slightly
diffusing outwards, although it is truncated at $\sim170$~au where the
CO lifetime becomes much shorter than the viscous timescale. Over time
the carbon mass increases similarly to the upper panel, although the
carbon mass saturates at a slightly lower level than the theoretical
maximum because some of the CO is never photodissociated into carbon
and oxygen ($\sim15\%$).

In the lower panels we show a case with a low viscosity and high CO
mass input rate. Due to the lower viscosity, carbon is locally piling
up and so reaching a higher surface density much faster than it can
spread out. By 0.1~Myr CO becomes shielded by carbon and starts to
build up and viscously spread. With CO's significantly longer
lifetime, the production rate of carbon is much lower, and the mass of
CO overcomes the carbon mass after 1 Myr. Nevertheless, the carbon
surface density is still comparable to the high viscosity case near
the belt location, but orders of magnitude larger beyond 100~au. This
means that CO can also significantly spread outwards since it is
shielded out to 1000~au, although the viscous timescale to reach those
regions can be shorter than its lifetime. Given the longer viscous
timescale at $\rb$ (60 Myr), only by the end of the simulation the
surface density of CO and carbon are close to steady state at the belt
location. Something worth noticing in this last scenario is that the
surface density of carbon after 10 Myr has a flat slope in between
from $\rb$ out to where the CO surface density drops exponentially. In
the two cases where CO becomes shielded, we find that CO spreads
inwards reaching our simulation boundary at 1~au, and spreads outwards
out to a radius where its lifetime against photodissociation
(determined by both carbon and CO surface densities) is too short for
CO to viscously spread or diffuse from where it is produced.

The CO and carbon evolution presented here is overall consistent with
the 0D model presented in \citet[][see their Figure
  18]{Kral2019}. The main difference is that the predicted masses of
carbon and CO (when completely shielded) are overall larger in this
study for the same $\alpha$ since in \citet{Kral2019} we used the
\textit{local} viscous timescale which is a twelfth of the viscous
timescale. Moreover, here we are considering the total gas mass in the
system, which can be orders of magnitude higher than the gas mass
around the belt location. Therefore, the disc mass saturates at a
lower level in our previous simulations and in a shorter timescale.

\subsection{Collisional evolution}
\label{sec:collevol}
An important ingredient that so far has not been considered in
previous gas evolution studies \citep[e.g.][]{Kral2016, Kral2017CO,
  Kral2019} is the disc collisional evolution. These have assumed
that CO gas is released at a constant rate equal to the product
between the CO mass fraction in planetesimals, $f_\mathrm{CO}$, and
the mass loss rate due to collisions $\dot{M}$. While the first might
stay constant overtime if planetesimals do not devolatise, the second
typically decreases with time \citep[e.g.][]{Dominik2003, Krivov2006,
  Wyatt2007hotdust, Thebault2007, Lohne2008, Kral2013} as
planetesimals are ground down and mass is removed from the system
through the blow out of small grains \citep{Burns1979}. Since the rate
at which planetesimals suffer disruptive collisions is proportional to
the total disc mass, $M$, it is easy to show that when assuming a
single slope size distribution the disc mass should decrease with time
as $1/t$, with
\begin{eqnarray}
  M(t)&=&\frac{M_0}{t/\tc+1}, \label{eq:Mt}\\
  \dot{M}(t)&=& - \frac{M(t)^2}{M_0\tc}, \label{eq:Mdot}
\end{eqnarray}
where $\tc$ is the collisional lifetime of the largest planetesimals
at $t=0$ and $M_0$ is the initial disc mass. Note that $t$ here
represents the time since planetesimals were stirred enough to cause
catastrophic collisions. Here we assume debris discs are pre-stirred
and collisions start immediately after the protoplanetary disc is
dispersed.

These equations have two strong implications for the evolution of the
disc mass. First, for $t\gg\tc$ the disc mass will be simply
proportional to $1/t$ and independent of $M_0$. Thus two systems could
have similar masses today although one started its evolution with a
much higher disc mass than the other. Therefore the total gas mass
released into the system over time can be orders of magnitude
different depending on the initial conditions. Second, the mass loss
rate decreases as $t^{-2}$, which is much faster than the $t^{-1/2}$
and $t^{-3/2}$ expected for the total mass and surface density of a
gaseous disc that evolves viscously according to Equation
\ref{eq:stot} (without input sources). This means that the amount of
gas in a debris disc can be out of equilibrium given its current gas
release rate which has been decreasing with time.

The collisional lifetime is a function of the initial disc mass and
other parameters such as the planetesimal strength, $\Qd$, maximum
planetesimal size, $\Dc$, mean eccentricity, $e$, mean inclination,
$I$, disc mean radius, $r$, and width, $dr$. More generally, $\tc$ is
given by \citep{Wyatt2007Astars}\footnote{Note that there is a typo in
  \cite{Wyatt2007Astars}'s equation as it should be in years rather
  than Myr}
\begin{equation}
\tc=\frac{3.8\rho r^{7/2} (dr/r) \Dc}{M_\star^{1/2} M}\frac{8}{9 G(\Xc, q)}, \label{eq:tc}
\end{equation}
in years, where $\rho$ is the internal density of solids in
kg~m$^{-3}$, $r$ and $dr$ are units of au, $\Dc$ is in units of km,
$M_\star$ is in units of $M_\odot$. $G(X_\mathrm{c}, q)$ is a factor
defined by Equation 10 in \cite{Wyatt2007Astars} that is a function of
$q$ (mass distribution exponent, $N(m)\propto m^{-q}$) and
$\Xc=\Dcc/\Dc$, with $\Dcc$ as the size of the smallest planetesimal
that has sufficient energy to destroy a planetesimal of size
$\Dc$. Its value depends on the relative velocities and $\Qd$, and can
be written as \citep{Wyatt2002}
\begin{equation}
  \Xc=1.3\times10^{-3} \left(\frac{\Qd r M_\star^{-1}}{1.25e^2+I^2} \right)^{1/3},
\end{equation}
with $\Qd$ in units of J~kg$^{-1}$. Assuming $\rho=2700$~kg~m$^{-3}$,
$I=e$ and $\Xc\ll1$, $q=11/6$, then $G\approx6.5\times10^{6} [\Qd
  r/(M_\star e^2)]^{-5/6}$ and $\tc$ can be written more simply as
\begin{equation}
\tc=1.4\times10^{-9} r^{13/3} (dr/r) \Dc \Qd^{5/6} e^{-5/3} M^{-1} M_\star^{-4/3}, \label{eq:tcsimple}
\end{equation}
which is the same as Equation 16 in \cite{Wyatt2008}.

\begin{figure*}
  \centering \includegraphics[trim=0.4cm 0.5cm 0.5cm 0.4cm, clip=true,
    width=0.99\textwidth]{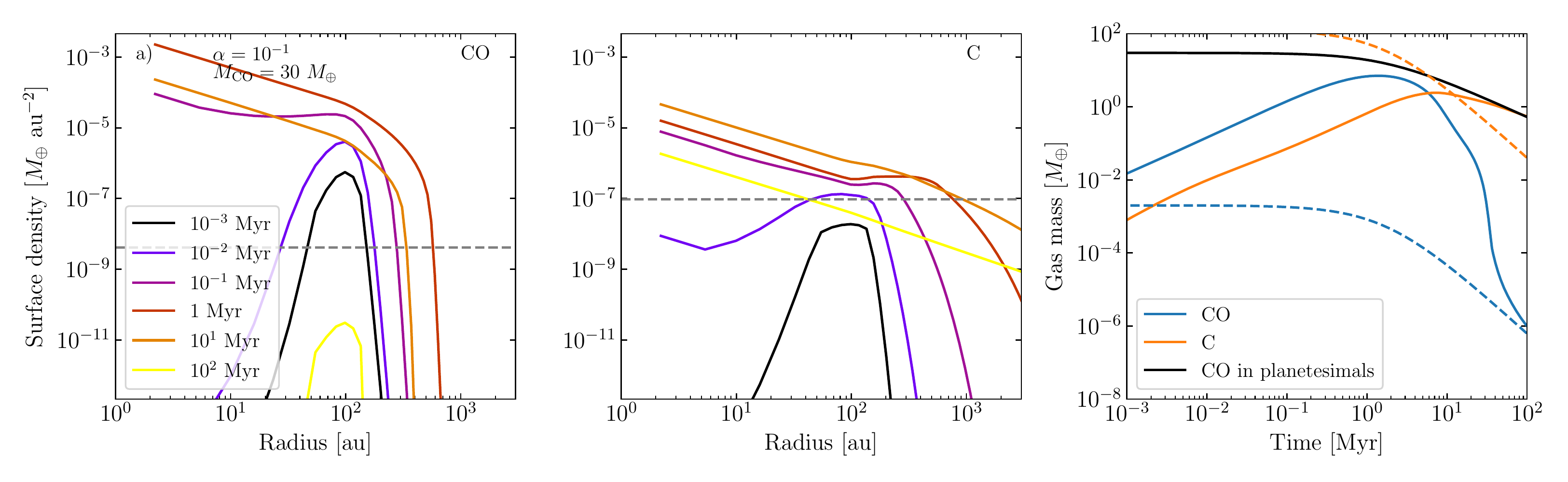}
  \includegraphics[trim=0.4cm 0.5cm 0.5cm 0.4cm, clip=true,
    width=0.99\textwidth]{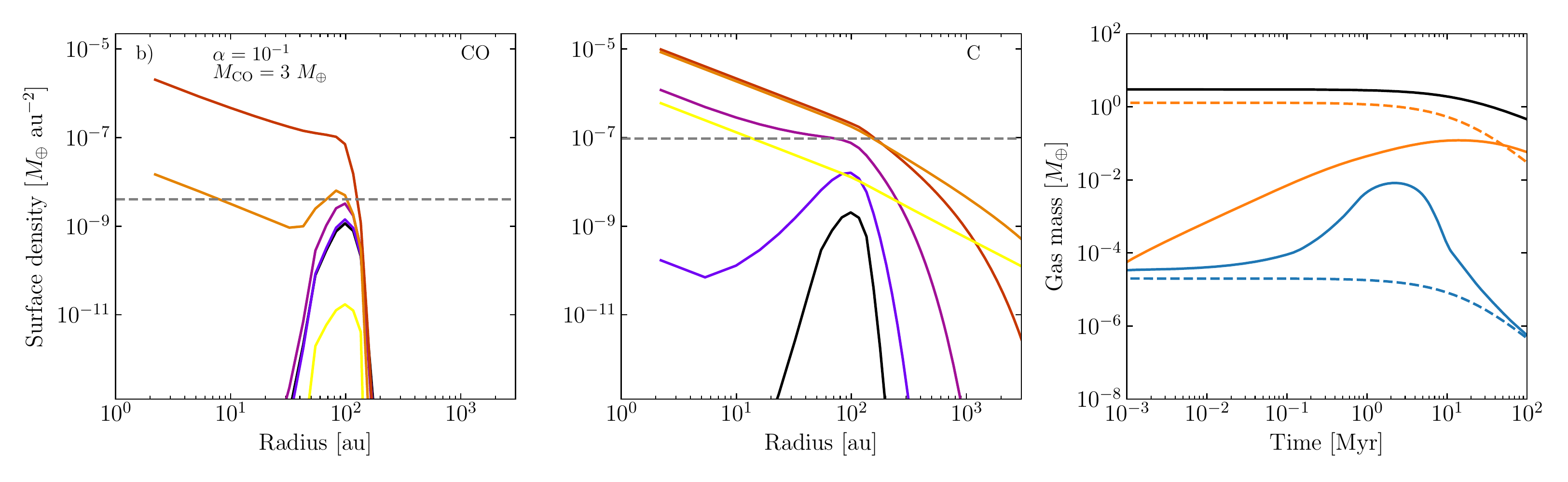}
 \caption{Same as Figure \ref{fig:constMdot}, but with a CO input rate
   decaying in time and high viscosity
   ($\alpha=0.1$). \textbf{\textit{Top:}} initial disc mass of
   300~\Me\ with a 10\% CO fractional mass. \textbf{\textit{Bottom:}}
   initial disc mass of 30~\Me\ with a 10\% CO fractional mass. In the
   right panels, the black line represents the CO mass that is still
   trapped in planetesimals and which decreases in time as the disc
   collisionally evolves. The steady state mass represented by dashed
   lines now decrease with time as the CO mass input rate declines.}
 \label{fig:highalpha}
\end{figure*}

We use Equation \ref{eq:tc} to calculate $\tc$ and $\dot{M}(t)$
assuming an internal density of 2700~kg~cm$^{-3}$, a maximum
planetesimal size $\Dc= 1$~km, eccentricities and inclinations of 0.1,
$\Qd$ of 330~J~kg$^{-1}$, $q=11/6$ (i.e. a size distribution with a
power law index of -3.5) and a fractional disc width of 0.5. Note that
this particular choice of parameters is arbitrary, but in
\S\ref{sec:population} we will fix these parameters to those chosen in
\cite{Wyatt2007Astars} to fit the infrared excess evolution at 24 and
70~$\mu$m in nearby A stars \citep[][]{Rieke2005, Su2006}, which are
the focus of this paper. Finally, we assume a CO mass fraction in
planetesimals of 0.1, consistent with levels inferred in Solar System
comets \citep{Mumma2011}. Using this set of parameters, below we study
the evolution of CO and carbon when the initial planetesimal disc mass
is 300 or 30~\Me\ ($\tc=2$ and 20~Myr, respectively), and assuming
$\alpha=0.1$ (\S\ref{sec:colevol_highalpha}) and $10^{-3}$
(\S\ref{sec:colevol_lowalpha}). This choice of masses is arbitrary,
but it helps to illustrate how important the collisional evolution can
be for the gas evolution.

\begin{figure*}
  \centering
  \includegraphics[trim=0.4cm 0.5cm 0.5cm 0.4cm, clip=true, width=0.99\textwidth]{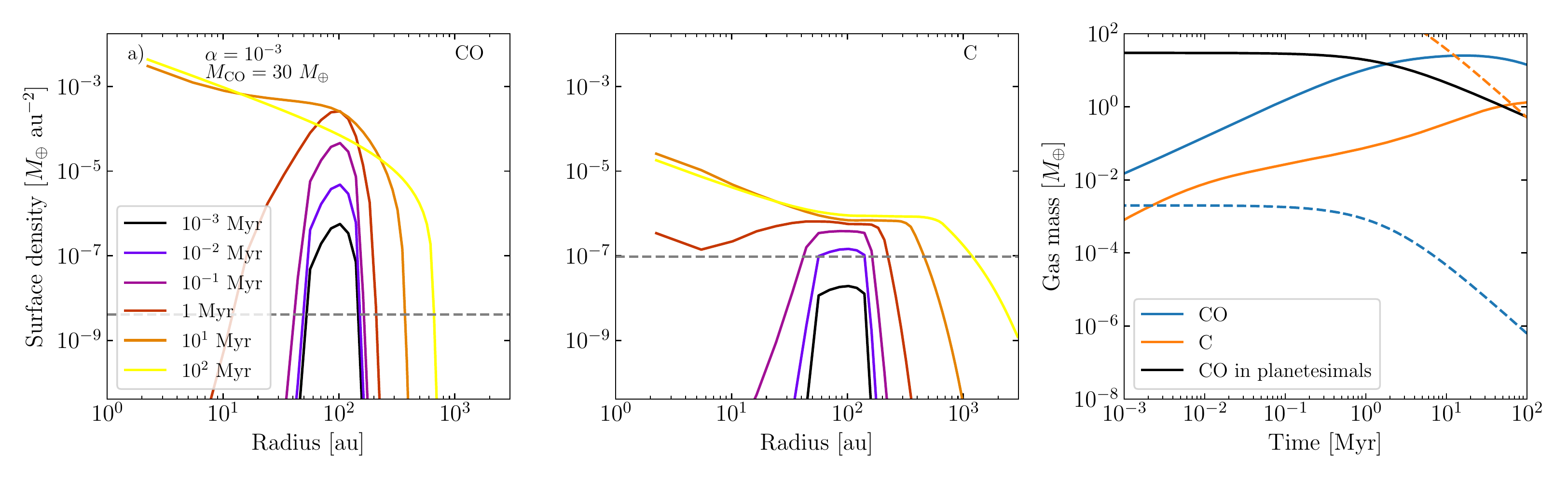}
  \includegraphics[trim=0.4cm 0.5cm 0.5cm 0.4cm, clip=true, width=0.99\textwidth]{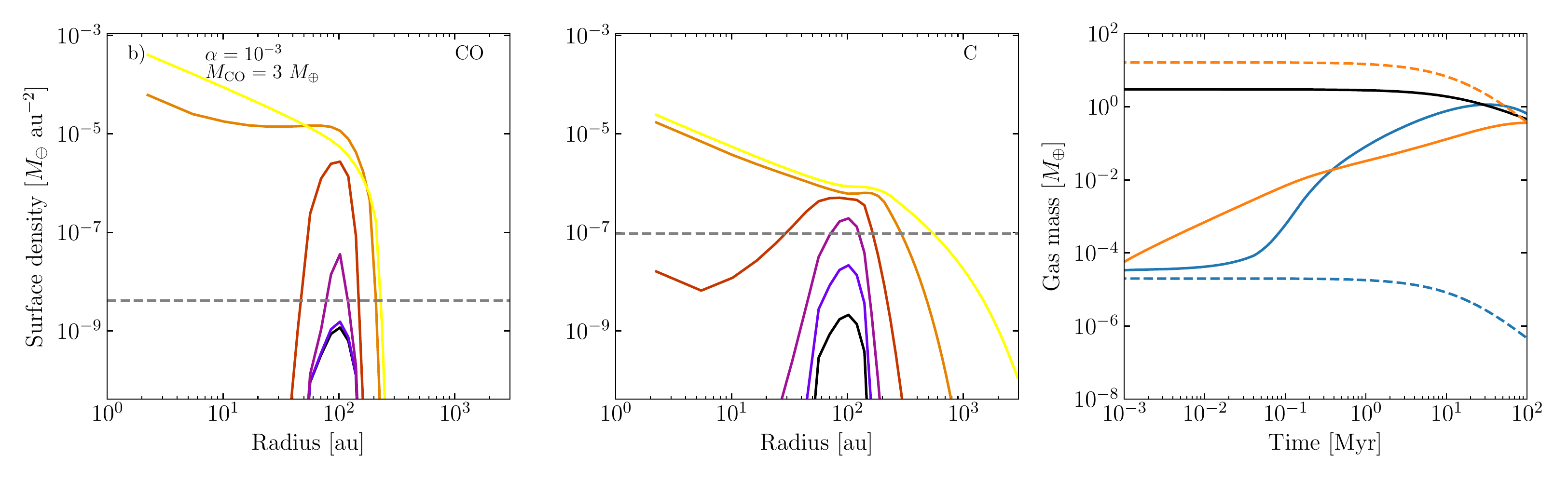}
 \caption{Same as Figure \ref{fig:highalpha}, but with a low
   viscosity ($\alpha=10^{-3}$).}
 \label{fig:lowalpha}
\end{figure*}

\subsubsection{High $\alpha$}
\label{sec:colevol_highalpha}
In Figure \ref{fig:highalpha} we show the gas evolution for two discs
starting with a total mass of 300 and 30~\Me\ (10\% of which is in CO)
in the high viscosity case. In the high initial mass case (top panel),
CO becomes self-shielded before carbon reaches the critical mass since
the initial mass loss rate and CO mass input rate are very high
($\dot{M}_\mathrm{CO}\sim20$~\Me~Myr$^{-1}$). Both CO and carbon
viscously spread within 0.1 Myr, with carbon reaching larger radii,
but not high enough surface densities to shield CO at radii larger
than 500~au. For this initial mass, $\tc=2$~Myr, hence the evolution
after a few Myr is characterised by a declining CO mass input rate
($1/t^2$) which cannot support anymore the high CO surface densities
reached at $t=1$~Myr since the lifetime of CO at the belt location is
about 10~Myr. Therefore, the surface density of CO starts to decline,
but not as steeply as an exponential decline because the more CO is
photodissociated the higher the surface density of carbon which indeed
shields CO (negative feedback). By 10~Myr the carbon surface density
and CO lifetime peak, after which CO declines steeply approaching a
new and less massive quasi equilibrium (see blue dashed line in right
panel). Carbon, on the other hand, decreases slowly with time due to
accretion. This difference in the lifetime of CO vs carbon could be
observed and tested, but at the moment there are not enough
observations of carbon to constrain its lifetime.

If the starting mass is lower (bottom panel), CO is only slightly
self-shielded given the lower CO mass input rate
($\dot{M}_\mathrm{CO}\sim0.2$~\Me~Myr$^{-1}$). After 0.1 Myr, the
carbon surface density is high enough at 100~au to shield CO and thus
the surface density of CO increases until 3~Myr when it reaches its
maximum.
%% , similar to the scenario shown in Figure~\ref{fig:constMdot}b
%% where the CO mass input rate is 0.1~\Me~Myr$^{-1}$ and constant
%% over time.
After a few Myr the CO mass input rate starts declining reducing both
the CO and carbon surface densities (although the total carbon mass
stays almost constant as the gas viscously spreads to larger radii).

An interesting consequence of the nature of how the mass input rate
evolves is that at 100 Myr both scenarios have similar planetesimal
disc masses ($\sim10$~\Me), and thus also fractional luminosities
($\fir$) or dust masses. However, their gas content is not the same
because it depends on the mass loss rate integrated over time and thus
there is some memory from their previous state. For example, if we
compare the carbon final masses or surface densities, we find that in
the high initial mass case they are a factor 10 and 3 larger,
respectively. This is because the viscous evolution is slower than the
collisional evolution as commented above.

Larger initial disc masses would increase this difference on the gas
content even more, while keeping constant the disc mass at 100 Myr,
thus unnoticeable from the amount of dust in the system (assuming gas
and dust do not interact, see \S\ref{dis:gas-dust} for a discussion on
this). On the other hand, much lower initial masses would result in an
evolution with an almost constant $\dot{M}_\mathrm{CO}$ because it
traces $M^2$, where $M$ is constant because the lifetime of the
largest planetesimals is longer than 100~Myr, thus analogous to the
picture presented in Figure~\ref{fig:constMdot}a.

\subsubsection{Low $\alpha$}
\label{sec:colevol_lowalpha}
These differences in the final gas content are more extreme if the
viscosity is lower, i.e. if the viscous evolution is slower, thus
being even more sensitive to the evolutionary history of the
system. As an example, in Figure~\ref{fig:lowalpha} we show the
evolution of two systems with the same initial masses as in
Figure~\ref{fig:highalpha}, but with a lower $\alpha$ value. For both
examples, CO becomes shielded and stays as such beyond 100 Myr since
the viscous timescale in this case is 60~Myr at $\rb$, hence there is
not enough time for the gas to viscously evolve and accrete onto the
star. Therefore, at 100 Myr the differences in gas masses and surface
densities are even larger between the two scenarios. For example,
there is an order of magnitude difference in the final CO and carbon
masses and CO surface density between the two scenarios. The carbon
final surface density, on the other hand, is very similar for both
scenarios, with differences only beyond 200~au.

Here we have found that not only the present mass input rate is
important, but also the input rate in the past since this is likely to
have started at a higher value. This adds a degree of extra complexity
to the results shown in \citet[][see their Figure~19]{Kral2019},
where it was shown that the carbon and CO mass depend on the viscosity
and CO input rate (assumed to be constant). Moreover, the systems
simulated here are far from a quasi steady state between 10--100 Myr,
except for CO if the initial mass was low enough for CO to be
unshielded in which case its evolution would follow the blue dashed
line in Figures \ref{fig:highalpha} and \ref{fig:lowalpha}. But even
in this case, the carbon surface density and mass will be out of
equilibrium as the viscous evolution is slower than the fast decline
rate of $\dot{M}_\mathrm{CO}$. These degeneracies and non-linearity
between the model parameters, initial conditions and subsequent
evolution complicate the interpretation of observations of gas in
debris discs, especially for shielded cases. Nevertheless, population
studies can be used as a tool to constrain some of the model
parameters by comparing observed and model distributions (see
\S\ref{sec:population} below).

\section{Population study for A stars}
\label{sec:population}

In this section we study the gas component of a sample of A stars
through population synthesis. This approach is ideal for degenerate
problems such as the unknown initial disc mass of a particular system,
and can provide constraints on population properties. Our population
synthesis model simulates the collisional evolution of their
planetesimal discs and how the released gas would evolve in those
systems. Here we use the results from \citet[][]{Wyatt2007Astars} that
used the same collisional evolution model to fit the observed
distribution of disc temperatures and evolution of infrared excess (or
$\fir$) around A stars. Wyatt et al. used the distribution of observed
24 and 70~$\mu$m excess as a function of age to constrain the initial
distribution of disc radii (approximating grains by blackbody spheres)
and the maximum $\fir$ that a disc of a given age and radius can
have. The latter is determined by planetesimal intrinsic properties
and level of stirring. In particular, they assume all systems are born
with a debris disc with a random blackbody radius between 3-120~au
with a power-law distribution $N(r)\propto r^{\gamma}$ with
$\gamma=-0.8$ and initial mass following a log-normal distribution
with a mean of $M_\mathrm{mid}$ and a standard deviation of
1.14~dex. Because the initial fractional luminosity of discs is
proportional to $M_\mathrm{mid} \Dc^{-0.5}$ and the maximum fractional
luminosity to $M_\mathrm{mid}\tc\Dc^{-0.5}$, \citet{Wyatt2007Astars}
could constrain these products to $B\equiv M_\mathrm{mid}
\Dc^{-0.5}=1.3$ and $A\equiv\Dc^{1/2}
\Qd^{5/6}e^{-5/3}=7.4\times10^{4}$, respectively, despite the
degeneracy that exists between $\Qd$, $e$, $I$, $\Dc$ and
$M_\mathrm{mid}$.

%% This is because the initial fractional luminosity of
%% discs is proportional to $B=M_\mathrm{mid} \Dc^{-0.5}$ and the maximum
%% fractional luminosity to $A=M\tc\Dc^{-0.5}$.

To study the gas evolution, the main quantity we aim to obtain from
previous dust studies is the mass loss rate since in our model it
determines the gas input rate. However, the collisional evolution
model used here does not fully constrain the total mass and mass loss
rate of discs. This can be shown using Equations \ref{eq:Mt} and
\ref{eq:Mdot}, and the products constrained by
\citet{Wyatt2007Astars}. We find that $\dot{M}\propto B
M_\mathrm{mid}/A$ for $t\ll\tc$ and $A M_\mathrm{mid}/ B$ for
$t\gg\tc$. This means that fixing A and B does not fully constrain
$\dot{M}$, but further assumptions need to be made. Because the gas
evolution is very sensitive to $\dot{M}$, one could try to vary $\Dc$
or $M_\mathrm{mid}$ to fit gas observations, but these parameters are
degenerate with the fraction of CO in planetesimals. For example,
increasing $M_\mathrm{mid}$ would increase the CO mass input rate
generating larger gas masses, which could also be achieved by
increasing the fraction of CO in planetesimals, or alternatively
lowering the viscosity. Note that in principle, under certain
assumptions one can estimate $\dot{M}$ from the disc radius, emitting
area (or fractional luminosity) and the mass of the particles that
contribute the most to the disc's cross sectional area
\citep[e.g.][Appendix B]{Matra2017fomalhaut}, without making
assumptions about the large planetesimals in the disc. We discuss such
an approach in \S\ref{dis:mdot}. In this section we keep
$M_\mathrm{mid}$ fixed to a default value and in \S\ref{dis:mdot} we
show that our choice is roughly consistent with inferred mass loss
rates from the dust emission.

%% , nevertheless,
%% assumes that no mass in the collisional cascade is lost before
%% reaching the blow-out size. In reality, 

%% the initial
%% distribution of disc radii and the products
%% $A=\Dc^{1/2}\Qd^{5/6}e^{-5/3}\approx7.4\times10^{4}$ and
%% $B=M_\mathrm{mid}\Dc^{-1/2}\approx1.3$ are well constrained
%% ($M_\mathrm{mid}$ in units of \Me). This is because assuming dust
%% grains are blackbody spheres, their distance to the star is well
%% constrained by the disc temperatures (or excess at 24 and
%% 70~$\mu$m)can be connected to a disc radius

%% These products set the maximum
%% disc's collisional evolution timescale (and maximum fractional
%% luminosity for a given age and radius), and the distribution of
%% initial fractional luminosities, respectively.

Here we make two subtle changes to the relations that define $A$ and
$B$. First, we introduce a correction factor $\Gamma$ which is the
ratio between the true disc radius over the blackbody radius. This is
necessary since the previous study used the \textit{blackbody radius}
($\rbb$) when calculating collisional timescales (since it is the dust
temperature which is fitted), which usually differs significantly from
the true disc radius being on average underestimated \citep{Booth2013,
  Pawellek2014, Pawellek2015}. The difference is due to small grains
that tend to have higher temperatures than blackbody spheres,
therefore their distance to the star is usually underestimated when
fitting blackbody models to SEDs (note that the model in
\S\ref{sec:collevol} is based on the true radius). Since the initial
fractional luminosity is proportional to $\rb^{-2}$, and the maximum
fractional luminosity to $\rb^{-2}$ and $\tc$ (with $\tc\propto
\rb^{13/3}$), with the introduction of $\Gamma$ we obtain
\begin{eqnarray}
  \Dc^{1/2}\ \Qd^{5/6}\ e^{-5/3}\ \Gamma^{7/3}&=&7.4\times10^{4},\\ \label{eq:A}
  M_\mathrm{mid} \Dc^{-1/2} \Gamma^{-2}&=&1.3 \label{eq:B}
\end{eqnarray}
which is the same as used in \citet{Kains2011}.

Second, the planetesimal strength can be parametrized as a function of
planetesimal size, with $\Qd\propto \Dc^{b_g}$ for planetesimal sizes
above a few hundred meters (gravity dominated). For example, for
compact planetesimals with sizes larger than a few hundred meters,
simulations by \citet{Benz1999} show $\Qd=330 \Dc^{1.36}$. We will
take this value as reference, although $\Qd$ could be significantly
lower if planetesimals are made of aggregates \citep[see][and
  references therein]{Krivov2018}. This change assumes however that
the collisional evolution can still be approximated by $q=11/6$ and
Equation \ref{eq:Mt} where $\tc$ is the lifetime of the largest
body.

%% Second, because the initial distribution of
%% fractional luminosities has to remain the same, any change in
%% $D_{\max}$ translates to a change in $M_\mathrm{mid}$ since
%% $M_\mathrm{mid}D_{\max}^{-1/2}\approx1.3$.

With these changes and a fixed value of $\Gamma$, the general
evolution of fractional luminosities is left with only one degree of
freedom, e.g. fixing $e$ or $M_\mathrm{mid}$ leaves completely
constrained the rest of the parameters. For $\Gamma=1$ and assuming
$e=0.05$, we have the canonical values $\Dc=4.7$ km,
$\Qd=2700$~J~kg$^{-1}$, and $M_\mathrm{mid}$=2.8~\Me, which are
equivalent to the ones used in \citet{Wyatt2007Astars}. For an
arbitrary value of $\Gamma$ and $e$ we have for A stars
\begin{eqnarray}
  %% \Dc&=&1.0 \left(\frac{M_\mathrm{mid}}{2.5\ M_\oplus}\right)^{2} \mathrm{km}, \\
  M_\mathrm{mid} &=& 1.7 \left(\frac{e}{0.01}\right)^{0.5} \left(\frac{\Gamma}{1.7}\right)^{1.2} M_\oplus,\\
  \Dc&=&0.2 \left(\frac{M_\mathrm{mid}}{1.7}\right)^2 \left(\frac{\Gamma}{1.7}\right)^{-4} \mathrm{km}.
  %Qd^{0.26},
  %% e&=&0.05 \left(\frac{M_\mathrm{mid}}{2\ M_\oplus}\right)^{2} \Gamma^{13/5}.
\end{eqnarray}
Here, we assume $\Gamma=1.7$ which is in the middle of the observed
range for A stars \citep[1--2.5,][]{Matra2018mmlaw}, and adopt the
following values as standard: $\Dc=0.02$~km, $\Qd=1.6$~J~kg$^{-1}$,
and $M_\mathrm{mid}$=0.5~\Me\ and e=0.001. Note that there are strong
degeneracies between these parameters, and thus these cannot be
interpreted physically. The values used here are chosen such that the
mass loss rate in the discs that we model are roughly consistent with
the mass loss rate inferred from their fractional luminosity (see
\S\ref{dis:mdot}).

We now proceed with these parameters and generate a random sample of
8,000 A stars. The mass of each star is randomly chosen between 1.7
and 3.4 \Msun\ and its corresponding luminosity estimated using the
mass-luminosity relation. The total disc mass is drawn from a
log-normal distribution centred at $M_\mathrm{mid}$ and with a
standard deviation of 1.14 dex as in \cite{Wyatt2007Astars}. The
planetesimal strength, mean eccentricity and maximum size are set to
the values defined above, and we assume a CO mass fraction of 0.1. The
disc radii are drawn following a power law distribution of exponent
-0.8 and with a minimum and maximum radius of 5.1 to 204~au,
consistent with \citet{Wyatt2007Astars} when $\Gamma$ is set to a
value of 1.7. Finally, we assume exocometary gas release starts at an
age of 3~Myr (i.e. shortly after protoplanetary disc dispersal) and we
evolve systems up to a random age between 3 and 100 Myr to get an age
spread and we keep the value of $\alpha$ fixed to 0.1
(\S\ref{sec:pophigh}) or 0.001 (\S\ref{sec:poplow}).

In the following sections \ref{sec:pophigh} and \ref{sec:poplow}, we
compare the model distributions of CO and carbon masses to
observations of a sample of A stars. To do so, it is important to
first consider biases of any selected sample. We choose to compare our
results with the complete sample of A stars within 150~pc that have
been targeted by ALMA, and that meet the following criteria
\citep[][the most complete study of gas around A stars with debris
  discs]{Moor2017}: i) disc fractional luminosity between
$5\times10^{-4}$ and $10^{-2}$; ii) dust temperature below 140~K; iii)
detection at $\geq 70$~$\mu$m with Spitzer or Herschel; and iv) age
between 10 and 50~Myr. We apply the same filtering process to our
synthesised sample to do a fair comparison between model and
observations. In this way both observed and model samples have the
same biases. In the following sections we present both whole and
filtered model populations, representing them with small dots and blue
contours, respectively\footnote{To increase the number of points of
  this filtered sample we simulated 8,000 extra systems with the same
  initial random distribution, but that passed the Mo\'or et
  al. selection criteria.}.

%% Because \citet{Wyatt2007Astars} used the blackbody radius
%% of the belt, $\rbb$, to compute $\tc$ given the lack of resolved
%% observations, which can differ by a factor of a few to the real radius
%% \citep{Pawellek2014, Pawellek2015}, we follow the same procedure and
%% first draw $\rbb$ from a power law distribution with a minimum and
%% maximum of 3 and 120~au, and an exponent of -0.8. This is used to
%% calculate $\tc$ assuming a fractional disc width of 0.5. We then
%% compute the real belt radius using the empirical relation reported in
%% \citet[for astrosilicates, see Table 8.1]{Pawellekthesis} which links
%% the true radius of a disc and its blackbody radius derived from its
%% SED.

\subsection{High $\alpha$}
\label{sec:pophigh}

\begin{figure}
  \centering \includegraphics[trim=0.2cm 0.2cm 0.6cm 1.0cm, clip=true,
    width=1.0\columnwidth]{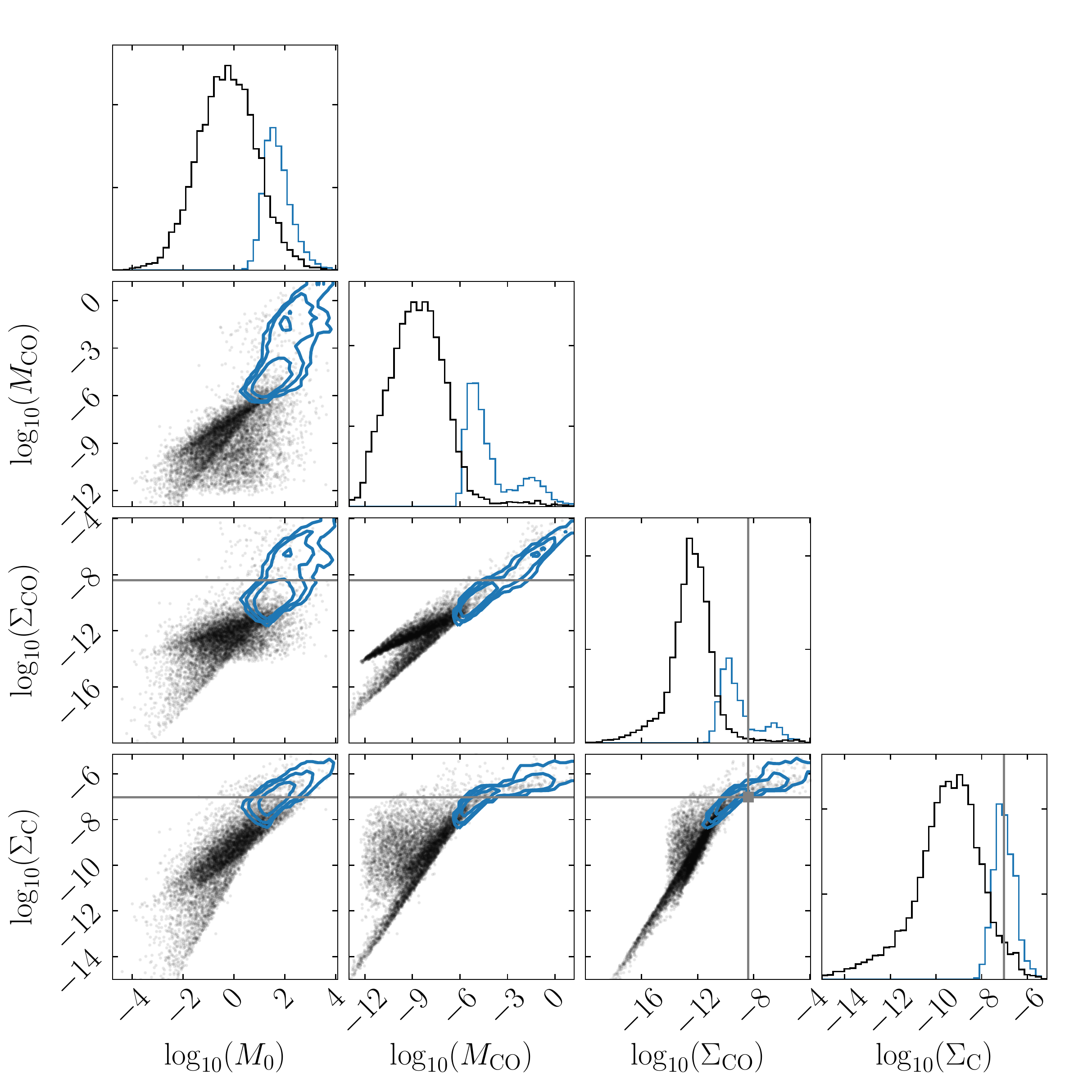}
 \caption{Distribution of the initial disc mass $M_0$, final CO mass
   $M_\mathrm{CO}$ and final surface density of CO and carbon. The
   contours represent regions that enclose the 68\%, 95\% and 99.7\%
   of the distribution of the filtered population. The grey lines
   represent the critical surface density over which CO and carbon can
   shield CO from UV photons. All variables are in units of \Me\ or
   \Me~au$^{-2}$. The marginalised distributions in the top panels are
   shown at an arbitrary scale for display purposes.}
 \label{fig:corner_halpha}
\end{figure}

In Figure~\ref{fig:corner_halpha} we present the distribution of the
initial disc mass ($M_0$), final CO mass contained within 500~au (at a
random age between 3--100~Myr), and the final surface density of CO
and carbon at the belt location, that result from our simulations. We
overlay in blue contours the distribution of systems that would have
been selected by \citet[][]{Moor2017}. We find that the distributions
of some parameters are highly correlated as expected. First, the more
massive the initial disc mass is, the higher the amount of CO and
carbon gas that will be present at a later epoch. For initial disc
masses higher than $\sim10$~\Me, the CO mass distribution of the
filtered sample extends to much higher values as CO becomes
shielded. This is also true for the full population, but it is harder
to see due to the age and radius spread which means that small warm
discs with low initial masses can also reach shielded states for a
short period. On the other hand, the surface density of carbon never
reaches values as high as CO because the more carbon there is, the
more CO is shielded and thus the lower carbon production rate.

The top panel of the second column shows the distribution of CO. While
the bulk of the population is unshielded, the distribution has a tail
extending to larger masses, with a small fraction (2\%) having CO
masses between $10^{-4}-1$~\Me\ \citep[the typical CO masses derived
  for shielded discs,][]{Moor2017} or carbon surface densities at the
belt location above the critical value to shield CO (4\%). We find
that applying the same selection mentioned above filters 99\% of
systems. Of the remaining 1\%, 33\% have CO masses higher than
$10^{-4}$~\Me\ displaying a bimodal distribution, and 49\% have carbon
surface densities above the critical value. These numbers are both
consistent with the number of A stars that meet the selection criteria
(17 of $\sim8,000$ main sequence A stars within 150~pc) and with the
statistics for shielded discs \citep[i.e. 9/17 systems with CO masses
  $\gtrsim 10^{-4}-1$~\Me,][]{Bovy2017, Moor2017}.

\begin{figure}
  \centering \includegraphics[trim=0.2cm 0.4cm 0.2cm 0.4cm, clip=true,
    width=0.95\columnwidth]{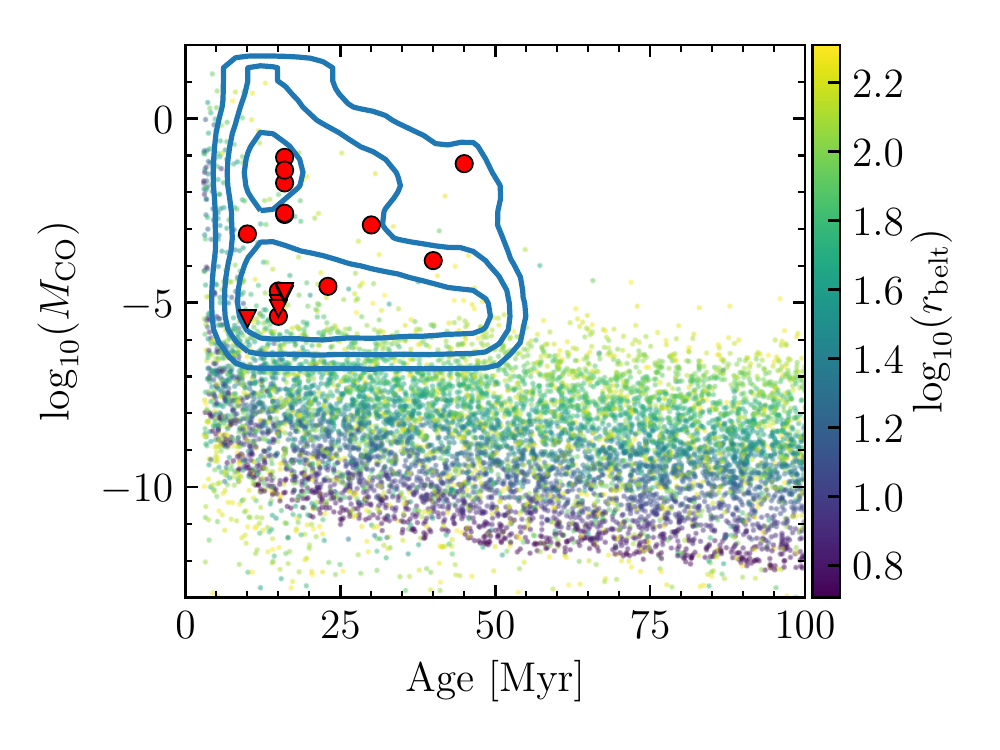}
  \includegraphics[trim=0.2cm 0.4cm 0.2cm 0.4cm, clip=true,
    width=0.95\columnwidth]{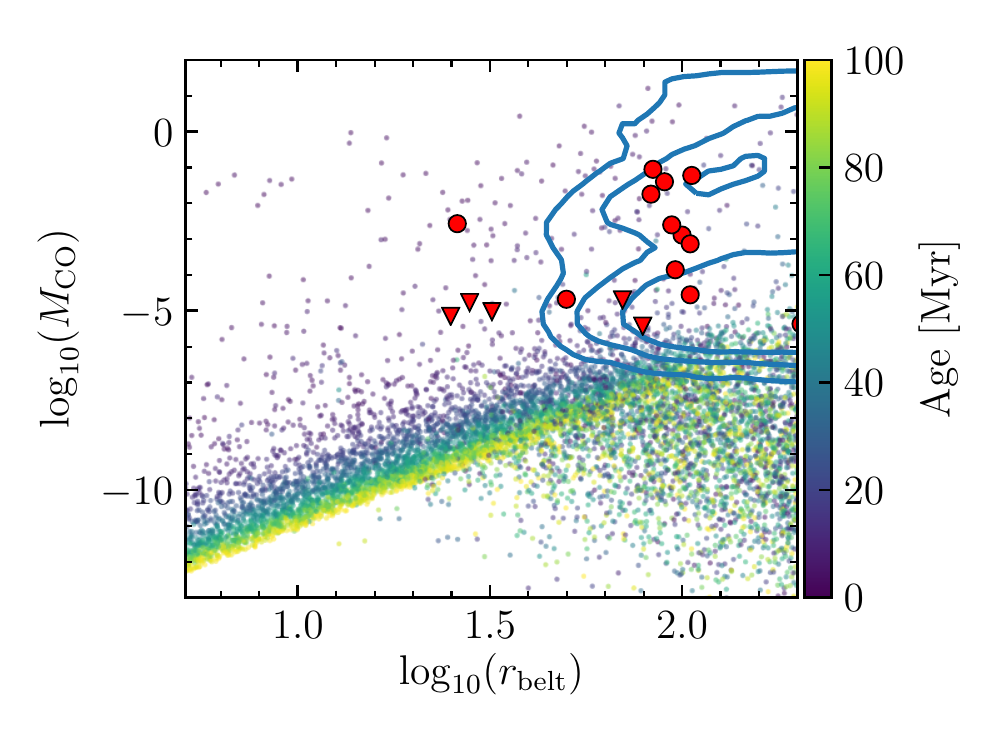}
  \includegraphics[trim=0.2cm 0.4cm 0.2cm 0.4cm, clip=true,
    width=0.95\columnwidth]{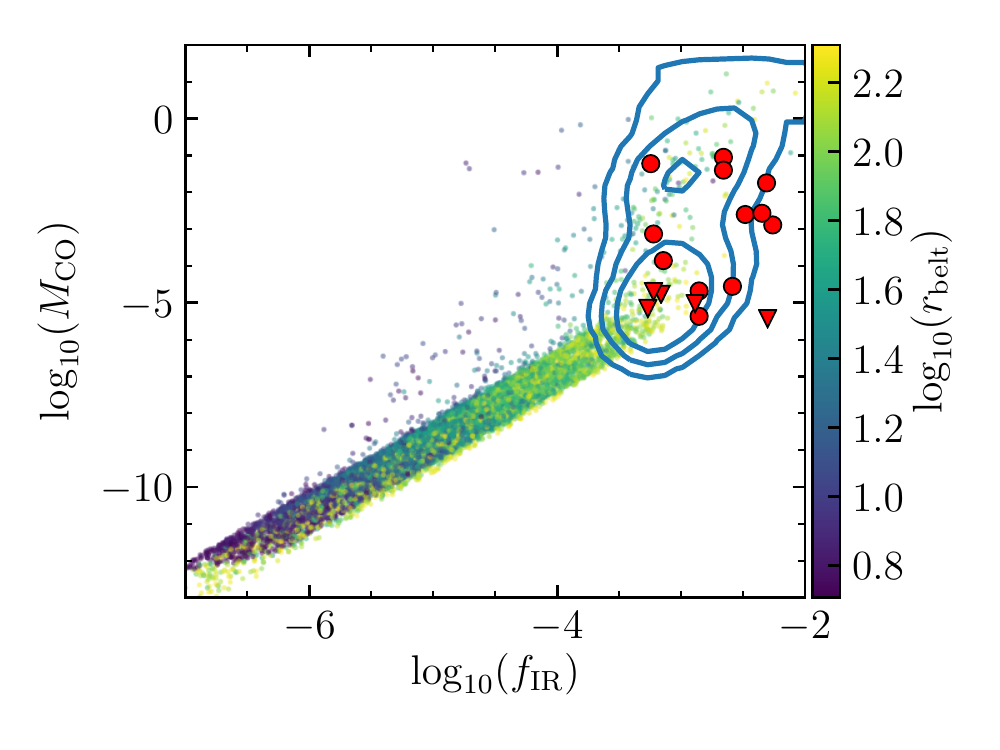}
 \caption{Population synthesis for A stars with
   $\alpha=10^{-1}$. \textbf{\textit{Top:}} CO gas mass
   ($M_\mathrm{CO}$) in \Me\ vs age. \textbf{\textit{Middle:}} CO gas
   mass vs radius. \textbf{\textit{Bottom:}} CO gas mass vs disc
   fractional luminosity. Each point corresponds to a different
   simulation. The contours represent regions that enclose the 68\%,
   95\% and 99.7\% of the distribution of the filtered population. The
   red circles and triangles represent the CO detections and upper
   limits around A stars that meet the same selection criteria as the
   filtered population. }
 \label{fig:2dhists}
\end{figure}

As expected, discs with large CO masses all have carbon surface
densities above which shielding by carbon becomes important (grey
lines). Note that there are no discs that are self-shielded but with
carbon surface densities below the critical value. The reason for this
can be understood by assuming quasi-steady state. Consider a disc with
a constant input rate $\dotCO$. The minimum CO mass input rate such
that CO is self-shielded is $\Sigma_\mathrm{CO,c} \upi
\rb^2/T_\mathrm{ph}$ (assuming CO does not viscously spread and
$\Delta \rb/\rb=0.5$), while the minimum to reach carbon shielding is
$7\Sigma_\mathrm{C,c}\upi \nu_0 \rb$ \citep[where $\nu=\nu_0
  r$,][]{Metzger2012}. By equating these two expressions, we find that
shielding by carbon will be more important in debris discs larger than
\begin{equation}
  \rc=7\ T_\mathrm{ph} \nu_0 \frac{\Sigma_\mathrm{C,c}}{\Sigma_\mathrm{CO,c}},
\end{equation}
which evaluated for $\alpha=0.1$ gives 3~au (lower than the 5~au
minimum disc radius considered here). Therefore, it is very unlikely
to observe a disc in the bottom right corner of the intersection
between the grey lines, i.e. only self-shielded. Lower viscosities
will make $\rc$ even lower. A caveat in this estimate is that for mass
input rates much higher than the minimum to become shielded, at early
times CO might become self-shielded before becoming shielded by carbon
as shown in Figure~\ref{fig:highalpha}.

In order to understand what is the dependence of the CO mass with the
age, belt radius and fractional luminosity of a system,
Figure~\ref{fig:2dhists} shows the scatter of these parameters for
each simulated system and their distribution in blue when applying the
selection criteria commented above. As a comparison, the red circles
and triangles represent the CO detections and upper limits around A
stars that meet the selection criteria \citep[][see Table
  \ref{tab:A}]{Moor2017, Kral2017CO, Kennedy2018, Booth2019,
  Hales2019}. For the non-detections around HD~98363, HD~109832,
HD~143675 and HD~145880 we use the non-LTE tool from \cite{Matra2018}
to compute CO mass upper limits, assuming kinetic temperatures of
150~K and a radial distance of 1.7 their dust blackbody temperatures
(their belts have not been resolved). The top panel shows
$M_\mathrm{CO}$ vs age. This distribution reveals how the fraction of
systems with large CO gas masses decreases with time and by 50~Myr
less than 1\% of the whole population remain with CO masses higher
than $10^{-5}$~\Me. The timescale at which shielded discs disappear is
related to how fast gas viscously evolves, thus a smaller viscosity
would result in long-lasting shielded discs as discussed in
\S\ref{sec:colevol_lowalpha}. Note that the distribution of red dots
is consistent with the blue density map, and we do not expect CO
masses lower than $10^{-7}$~\Me, which depending on line excitation
conditions is at the limit of typical deep ALMA observations
\citep[e.g.][]{Marino2016, Matra2017fomalhaut}, but below the
sensitivity of shallow CO surveys \citep[e.g.][]{Moor2017,
  Kral2017CO}. Hence given the above selection cuts, the distribution
of CO gas masses inferred from observations is consistent with our
model. It is also worth mentioning that the CO mass estimates shown
here have uncertainties that typically are an order of magnitude wide.

The middle panel of Figure~\ref{fig:2dhists} shows the distribution of
$M_\mathrm{CO}$ vs $\rb$. We find that the bimodal distribution of
$M_\mathrm{CO}$ discussed above is present across the whole
distribution of $\rb$, and can be split into three populations, two
with low CO masses and one with shielded and massive CO discs. For
those systems that are unshielded, CO mass seems to increase with
radius up to $\rb\sim50$~au after which the median CO mass decreases
with radius. This is expected given that the CO mass in the unshielded
case is simply proportional to the mass loss rate, which for $t\gg\tc$
is proportional to $\tc\propto r^{13/3}$ (Equation \ref{eq:tcsimple}),
whereas for $t\ll\tc$ is proportional to $1/\tc\propto r^{-13/3}$. The
transition happens where $\tc\sim50$~Myr (mean system age), which
using $M_\mathrm{mid}$ we find this should happen at
$\rb\sim56$~au. The larger dispersion for discs with a large radius is
due to the mass input rate being proportional to $M_0^2$ (for
$t\ll\tc$) which has a large dispersion. The third population is
composed of shielded discs that exists predominantly in young systems
with a high initial mass and with a large disc radius. We find that
the distribution of observed discs matches reasonably well the
expected distribution, with the exception of HD~121191, whose belt has
not been resolved with enough sensitivity \citep[we used 1.7 times its
  blackbody radius, which could be easily an
  underestimate,][]{Moor2017}, thus it could be significantly larger
than the value used here.

%% No discs with large radius ($>100$~au) and
%% CO masses or vice versa have been detected, although only two large
%% discs are present within this sample.

The bottom panel of Figure~\ref{fig:2dhists} shows the distribution of
$M_\mathrm{CO}$ vs fractional luminosity ($\fir$). The fractional
luminosity is calculated using Equations 14 and 15 in
\cite{Wyatt2008}, but using the blackbody radius instead of the true
radius. For $\fir\lesssim10^{-4}$ we find that $M_\mathrm{CO}$ is
roughly proportional to $\fir^2$, a result that is expected since the
CO release rate and mass loss rate in a belt is proportional to the
square of the mass or $\fir^2$. On the other hand, for larger $\fir$
the CO masses grow steeply with fractional luminosity since CO is
shielded in these systems. Overall, the observed distribution (red
dots) is consistent with the filtered model population (blue contours)
given the typical uncertainties in gas masses ($\sim1$~dex). The big
exception is HR~4796. The CO mass upper limit for this system is below
the blue $3\sigma$ confidence region by orders of magnitude. This disc
has one of the highest fractional luminosities and is very narrow,
hence the CO release rate should be high. \cite{Kennedy2018} concluded
that the planetesimals must have a CO+CO$_2$ ice mass fraction below
2\%. This is discussed further in \S\ref{dis:othergas}.

\begin{figure}
  \centering \includegraphics[trim=0.2cm 0.4cm 0.2cm 0.4cm, clip=true,
    width=1.0\columnwidth]{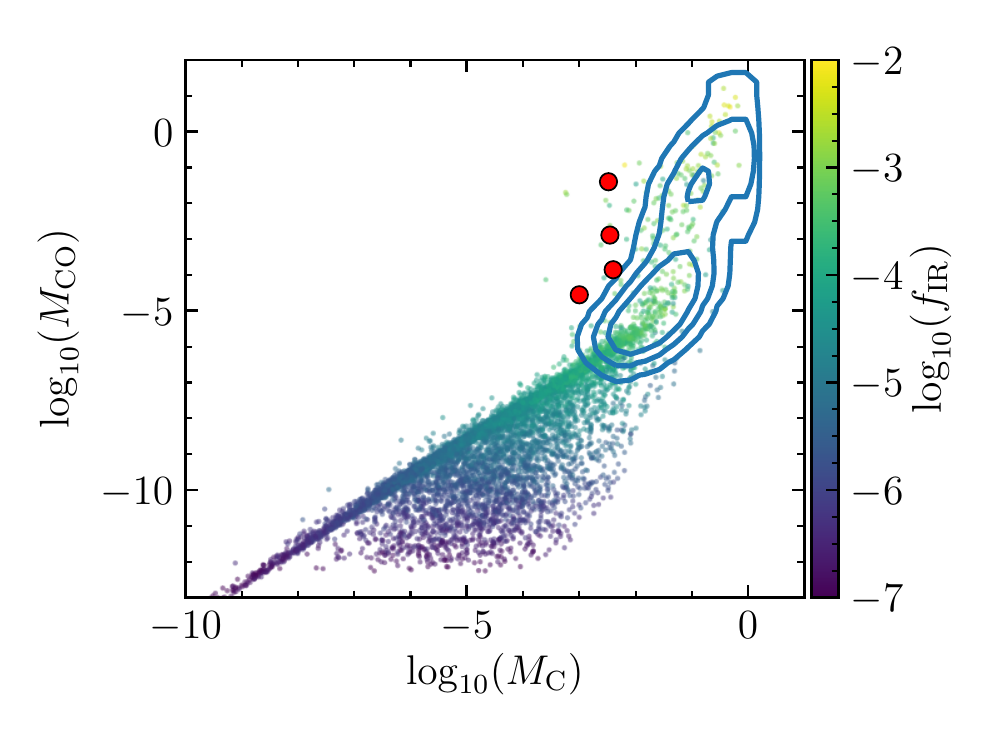}
  \includegraphics[trim=0.2cm 0.4cm 0.2cm 0.4cm, clip=true,
    width=1.0\columnwidth]{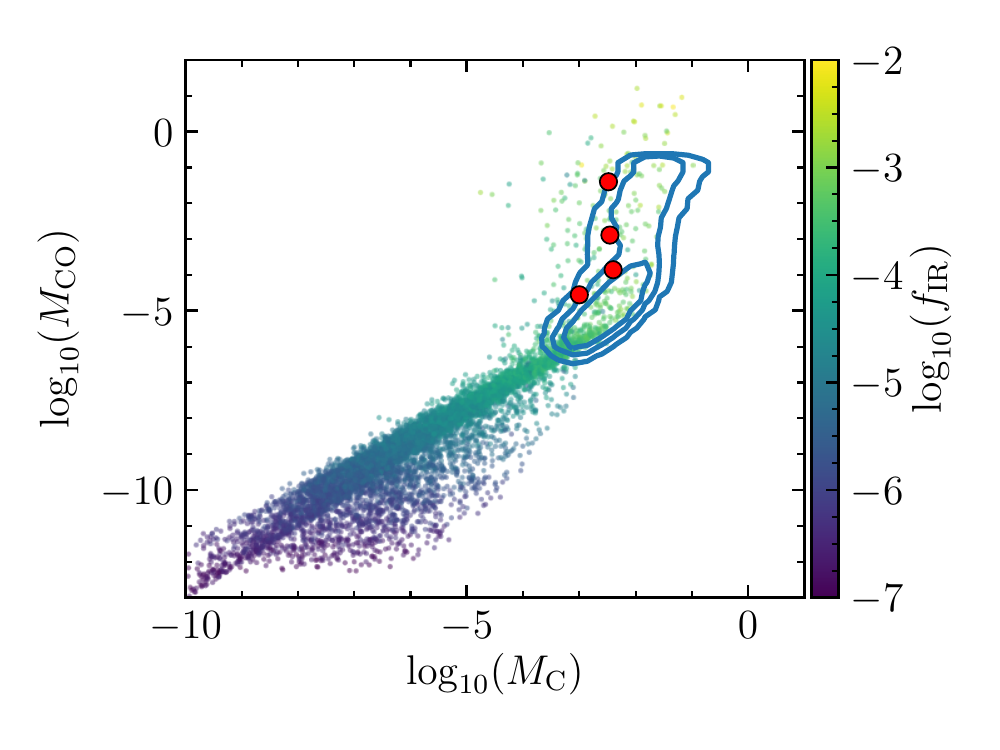}
  \caption{Population synthesis model for the carbon and CO masses
    around A stars assuming a high viscosity
    ($\alpha=0.1$). \textbf{\textit{Top:}} The carbon model mass
    represents all mass within 500~au. \textbf{\textit{Bottom:}} The
    carbon model mass is calculated as $\Sigma_{\mathrm{C},\rb} \upi
    \rb^2$. Each point corresponds to a different simulation. The
    contours represent regions that enclose the 68\%, 95\% and 99.7\%
    of the distribution of the filtered population. As a comparison,
    the red circles represent the CO and carbon detections around A
    stars that meet the same selection criteria as the filtered
    population.}
 \label{fig:carbon}
\end{figure}

Finally, in the top panel of Figure~\ref{fig:carbon} we compare the
carbon and CO masses of our models (contained within 500~au) with the
estimated CO and neutral carbon mass in four discs \citep[$\beta$~Pic,
  49~Ceti, HD~131835, HD32297,][]{Higuchi2017, Higuchi2019,
  Cataldi2018, Kral2019, Cataldi2019}. Overall we find that the
observed carbon masses are lower than predicted by our model. This has
been highlighted recently by \cite{Cataldi2018, Cataldi2019} and used
to argue in favour of a scenario in which gas was released only
recently in the $\beta$~Pic and HD~32297 systems. This comparison is
difficult since most of the carbon mass resides at large radii where
typically ALMA observations are less sensitive (due to the primary
beam and low surface densities). This is why we report gas masses as
the total masses contained within 500~au, as a proxy for the
\textit{observable} mass. Note that this affects more the carbon mass
than the CO mass, since carbon is always distributed up to larger
radii. In steady state we expect the total carbon mass to be
proportional to $\rmax^{1/2}$ for $\rmax>\rb$ (Equation \ref{eq:mc}),
so the observable mass could vary by a factor of a few if we used a
different definition. This together with the low number statistics,
unknown ionisation fraction (see \ref{dis:ionisation}) and uncertain
excitation conditions for the observed lines, hinders any strong
conclusions on the model success at reproducing observations. If the
carbon mass was indeed lower than predicted by our model, this could
be due to a higher viscosity, however this would lower the CO mass of
shielded discs in our model (lower shielding and shorter viscous
timescale) making it inconsistent with observations. Alternatively we
can estimate an observable carbon mass as the product between the
carbon surface density at the belt location times the area of the
belt, i.e. $\Sigma_{\mathrm{C},\rb} \upi \rb^2$ (for a fractional
width of 0.5). This is shown in the bottom panel of
Figure~\ref{fig:carbon}. Using this definition for carbon mass, we
find a better agreement with observations. This illustrates that
simple comparisons between model and observations are difficult. We
identify the gas surface density at the belt location as a better
quantity to test shielding models, since it is the vertical column
densities of carbon and CO that set the CO photodissociation timescale
in this context. Note that estimating the neutral carbon surface
density from current observations is very difficult since the observed
systems were edge-on ($\beta$~Pic and HD32297) or marginally resolved
(HD~131835). Moreover, the observed single carbon line in some of
these systems is close to being optically thick and degenerate with
excitation conditions, hindering determining the total carbon mass
present (49~Ceti).

\subsection{Low $\alpha$}
\label{sec:poplow}

\begin{figure}
  \centering
  \includegraphics[trim=0.2cm 0.4cm 0.2cm 0.4cm, clip=true,
    width=0.95\columnwidth]{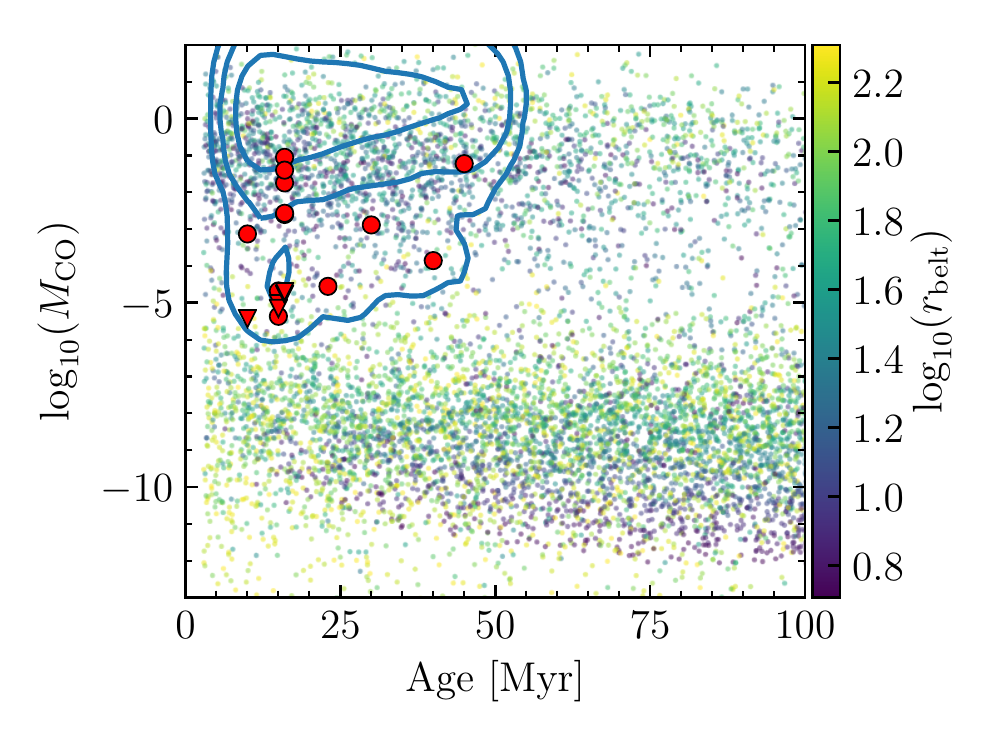}
  \includegraphics[trim=0.2cm 0.4cm 0.2cm 0.4cm, clip=true,
    width=.95\columnwidth]{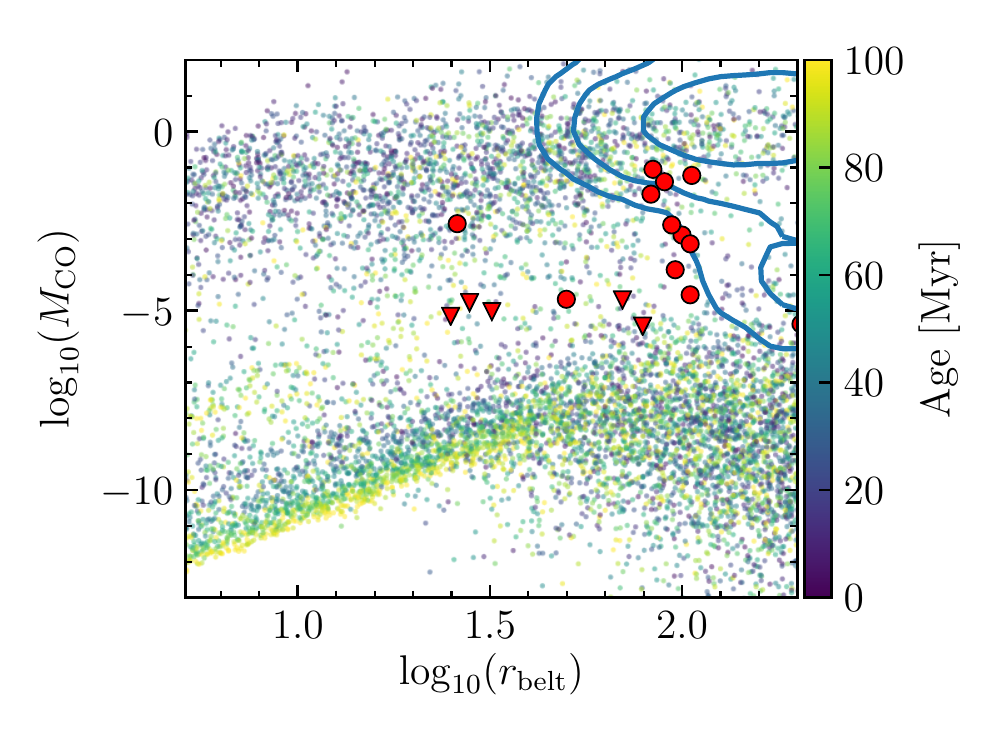}
  \includegraphics[trim=0.2cm 0.4cm 0.2cm 0.4cm, clip=true,
    width=.95\columnwidth]{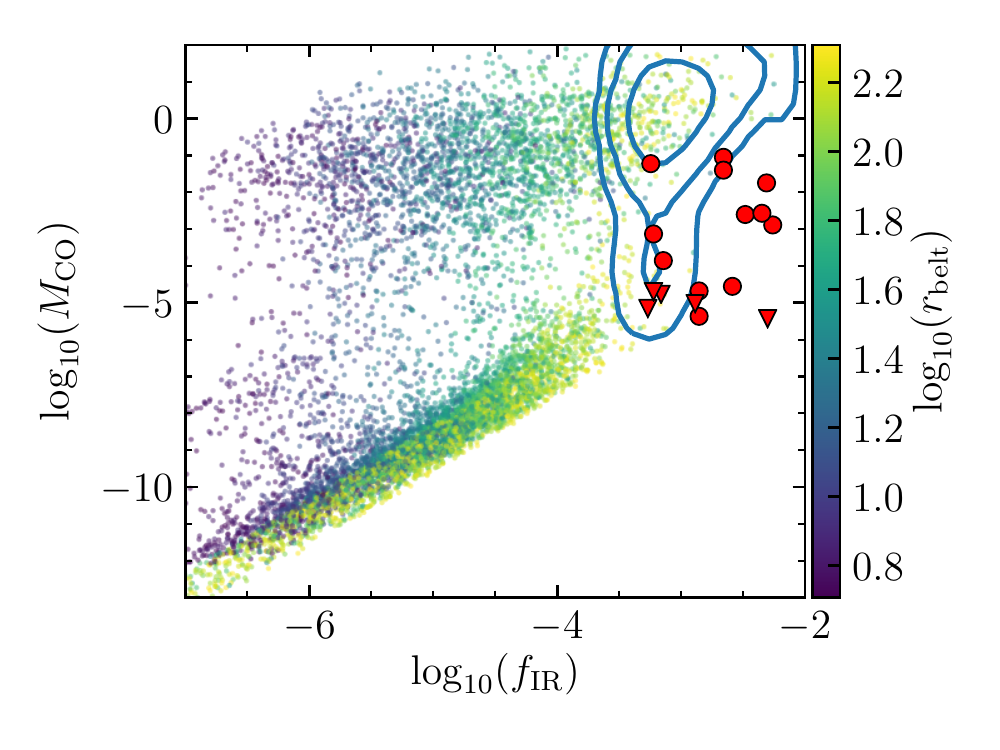}
  \caption{Population synthesis for A stars with
    $\alpha=10^{-3}$. \textbf{\textit{Top:}} CO gas mass
    ($M_\mathrm{CO}$) in \Me\ vs age. \textbf{\textit{Middle:}} CO gas
    mass vs radius. \textbf{\textit{Bottom:}} CO gas mass vs disc
    fractional luminosity. Each point corresponds to a different
    simulation. The contours represent regions that enclose the 68\%,
    95\% and 99.7\% of the distribution of the filtered
    population. The red circles and triangles represent the CO
    detections and upper limits around A stars that meet the same
    selection criteria as the filtered population.}
 \label{fig:2dhists_lowalpha}
\end{figure}

Using the same collisional evolution model as in the previous section,
we use a lower $\alpha=10^{-3}$ and make predictions for the mass of
CO and carbon for a population of systems around A stars. Figure
\ref{fig:2dhists_lowalpha} shows the results for the CO mass as a
function of age (top), $\rb$ (middle), and $\fir$ (bottom). We find a
stronger bimodal distribution compared to the results with a high
$\alpha$. A larger fraction of systems become shielded and stay almost
constant in time because the viscous evolution timescale for these
discs is 100 times longer. Moreover, the CO masses of shielded discs
are much higher because accretion is slower, and inconsistent with the
observations (compare the blue contours with the red markers). In
fact, even discs with low fractional luminosities which would have
been undetectable with Spitzer or Herschel have high levels of
shielded CO which hints at their high initial masses and fast
collisional evolution. Because viscosity is low, CO gas takes much
longer to disappear compared to solids. Therefore, the model in this
paper disfavours low viscosities. Nevertheless, unbiased searches for
CO have not been done around nearby A stars, thus we cannot rule out
the presence of massive gas discs around stars without detectable
debris discs.

Because the solid mass loss rate and thus the CO mass input rate are
not fully constrained by the collisional evolution model used here, it
might still be possible to fine tune parameters such that with a low
viscosity the predicted CO masses are lower. We showed above that the
mass input rate of the population is proportional to $M_\mathrm{mid}$
or $\Dc^{0.5}$ (and constants), so in order to reduce the mass input
rate by a factor of 100 to counteract the 100 times lower viscosity,
we would need to reduce $\Dc$ by a factor of $10^{4}$, and thus also
decrease $M_\mathrm{mid}$ by a factor 100 and increase $\Qd$ by a
factor 250. Such a large strength for such small bodies is unrealistic
\citep{Benz1999}, thus we conclude that $\alpha$ values as low as
$10^{-3}$ are unlikely to explain the observed properties of gas in
high fractional luminosity discs. Moreover, high gas masses are only
found in systems younger than 50~Myr \citep{Greaves2016}, thus
favouring a viscous timescale shorter than 50~Myr at the belt
location.

Another source of degeneracy is the assumed CO fraction in
planetesimals. A lower CO abundance would lower the input rate of CO,
which combined with a low viscosity could still fit the observed
distribution of CO masses. This is not explored in this paper, but in
\S\ref{sec:alpha} we discuss avenues to break these degeneracies using
independent observations.

%% small would contradict there is still
%% have very low fractional
%% luminosities. This means that the debris discs around this system
%% changing alpha and dmax at the same time, but Mdot has to be similar
%% to Matra, so it cannot be.

\section{Solar-type stars} 
\label{sec:fgk}
Exocometary gas is not only found around A stars, but also around
lower mass stars, from F to M \citep{Marino2016, Marino2017etacorvi,
  Lieman-Sifry2016, Matra2019twa7}, however high CO mass discs are
only found around A stars so far. This could be due to A stars being
born with more massive discs on average, which was hinted by
\cite{Matra2019twa7} based on a higher fractional luminosity in the
more luminous stars that have been observed with ALMA. Here we aim to
show that shielded discs are unlikely to be found around FGK stars
simply because their planetesimal belts are born less massive than
their A stars counterpart (or equivalently have a lower $\dot{M}$).

We take the same approach as for A stars, and we simulate a population
of FGK stars following the results by \cite{Sibthorpe2018} that used
both Spitzer and Herschel FIR photometry of 275 FGK stars to fit a
collisional evolution model, similar to \cite{Wyatt2007Astars}. That
study found best-fit parameters $A=5.5\times10^{5}$, $B=0.1$ and
$\gamma=-1.7$. We assume the same planetesimal strength relation as
before, eccentricities of 0.001 and $\Gamma=3.1$ \citep[the average
  true to blackbody radius ratio for FGK stars resolved by
  ALMA,][]{Matra2018mmlaw}. Fixing these parameters we have
$\Dc=0.03$~km and $M_\mathrm{mid}=0.16$~\Me\ (vs 0.02~km and
0.5~\Me\ for A stars), i.e. discs around FGK stars are born on average
with a lower mass and lower fractional luminosities. We can anticipate
that FGK stars will have lower mass CO discs, and for the same
$\alpha$ value used here (0.1) viscosities will be slightly lower (a
factor $\sim2$) due to the different stellar masses and
luminosities. We simulate then 8,000 FGK stars with stellar masses
between 0.56 and 1.6~\Msun, blackbody radius between 1 and 1000~au
(i.e. real radius between 3.1 to 3100~au) and ages between 3 and
100~Myr. Note that this large upper limit on the disc radius beyond
500~au is unrealistic given the observed population of debris and
protoplanetary discs. Moreover, temperatures beyond 500~au might be
below the CO sublimation temperature and thus CO would not be
released. We keep this large upper limit because if not there is no
guarantee that the model distribution would still fit the distribution
of fractional luminosities of FGK stars. Nevertheless, because of the
low value of $\gamma$ only a very small fraction of simulated systems
have radii larger than 500~au. Therefore, this unrealistic choice of
upper boundary in the radius distribution does not affect our
conclusions.

%% Despite the
%% larger value of the parameter $A$, the collisional evolution of discs
%% with the same true radius and mass is faster for FGK stars compared to
%% A stars because the $\Gamma$ assumed here is a factor 1.8 larger (see
%% Equation~\ref{eq:A}). 

\begin{figure}
  \centering \includegraphics[trim=0.2cm 0.4cm 0.2cm 0.4cm, clip=true,
    width=.95\columnwidth]{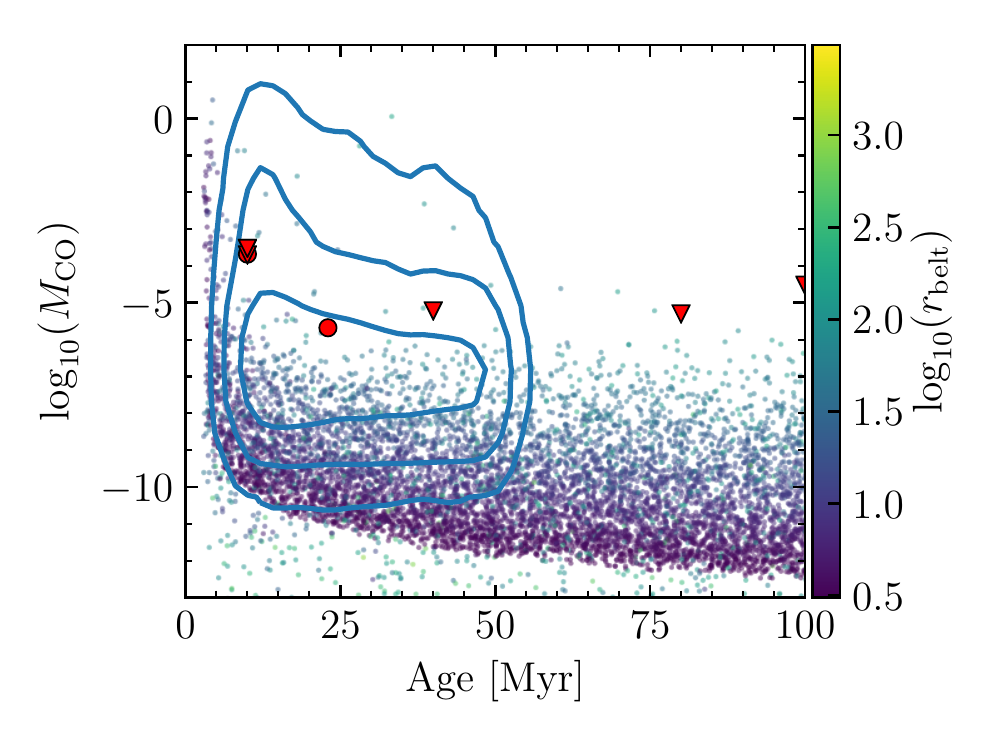}
  \includegraphics[trim=0.2cm 0.4cm 0.2cm 0.4cm, clip=true,
    width=.95\columnwidth]{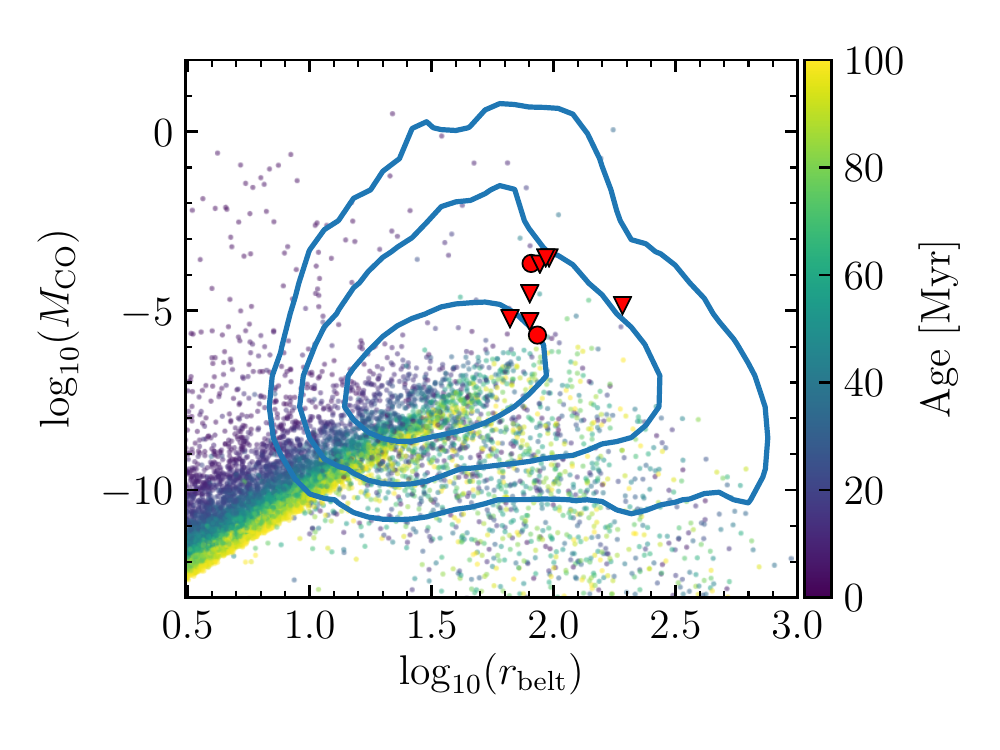}
  \includegraphics[trim=0.2cm 0.4cm 0.2cm 0.4cm, clip=true,
    width=.95\columnwidth]{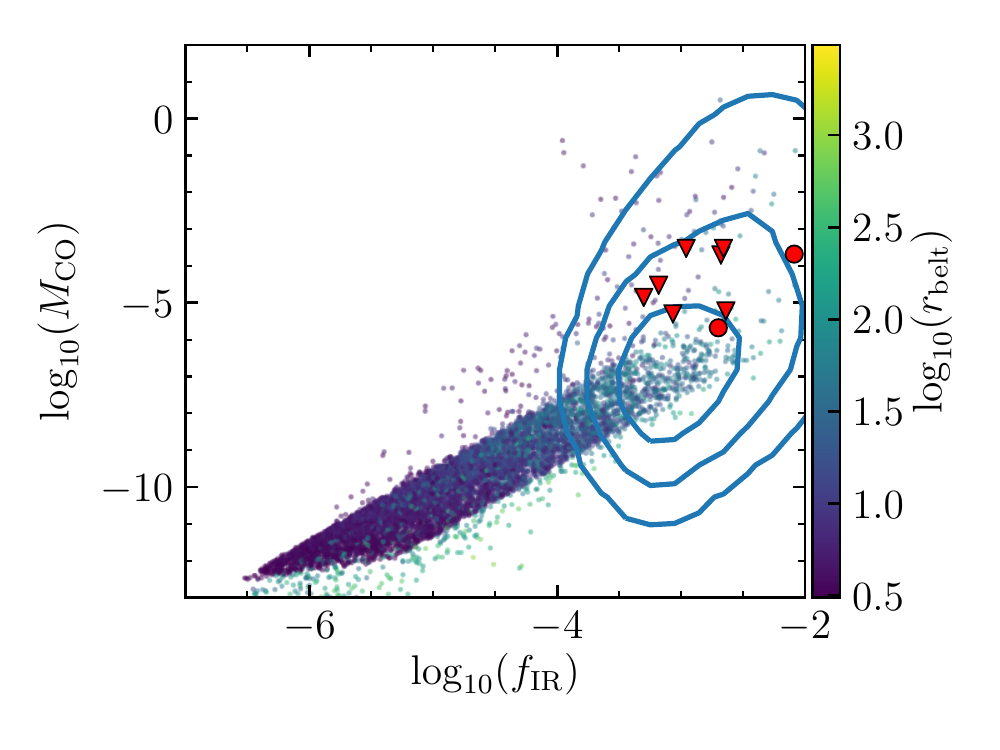}
  \caption{Population synthesis for FGK stars with
    $\alpha=10^{-1}$. \textbf{\textit{Top:}} CO gas mass
    ($M_\mathrm{CO}$) in \Me\ vs age. \textbf{\textit{Middle:}} CO gas
    mass vs radius. \textbf{\textit{Bottom:}} CO gas mass vs disc
    fractional luminosity. Each point corresponds to a different
    simulation. The contours represent regions that enclose the 68\%,
    95\% and 99.7\% of the distribution of the filtered
    population. The red circles and triangles represent the CO
    detections and upper limits around FGK stars that meet the same
    selection criteria as the filtered population. }
 \label{fig:2dhists_fgk}
\end{figure}

Similar to the previous figures, we present in Figure
\ref{fig:2dhists_fgk} the distribution of CO masses vs age, belt
radius, and fractional luminosity. We find that overall the CO masses
of the population are much lower, even after applying the same
selection cuts as for A stars (blue contours, 2\% of the whole
population). This is because for a fixed fractional luminosity, the
gas release rate is higher for more luminous stars
\citep{Matra2019twa7}. We find that only 1\% of the whole fraction has
CO masses above $10^{-4}\ M_\oplus$, most of which are young systems
with small discs. This fraction only grows to 4\% for the filtered
population. When we look instead at the carbon surface density at the
belt location, we find that only 3\% is above the critical value to
shield CO, and this number increases only to 11\% in the filtered
population. Note that the filtered population is consistent with
observations of FGK stars with bright debris discs represented by red
circles and triangles (see Table \ref{tab:FGK}). Our model then
explains why we do not find hybrid/shielded discs around FGK stars as
they are not as common given the collisional evolution models that fit
their infrared excesses.

\section{Discussion}
\label{sec:discussion}

\subsection{Caveats}

In this section we discuss some of the assumptions made in our
modelling, when these are justified and how replacing these by more
complex models could have an impact on some of our conclusions.

\subsubsection{Other volatiles}
\label{dis:othergas}
In this paper, we have focused on the release of CO only. Other
volatiles that are abundant in Solar System comets such as H$_2$O,
CO$_2$, CH$_4$, HCN, CN, C$_2$H$_6$, NH$_3$, H$_2$S, etc. could also
be released in collisions that expose the interior of
planetesimals. This will only happen if temperatures are high enough
for thermal desorption (before or after collisions) or if there is a
strong UV radiation so photodesorption is effective
\citep{Grigorieva2007}. Assuming planetesimals at large radii are
similar in composition to Solar System comets, we expect that only
water and CO$_2$ to be as abundant or more than CO so their effect
could be important in the model presented here. We expect that water
thermal desorption will only be important interior to 10--20~au where
blackbody temperatures are higher than 150~K around A stars, although
photodesorption could act instead as the release mechanism at large
radii. This could mean that the total amount of gas is higher by a
factor of a few, but this would not affect the evolution of CO and
carbon since the viscous evolution equations are scale independent for
the mass and these molecules cannot act as shielding agents because
they photodissociate at (UV) wavelengths longer than CO
photodissociates and C ionises. The only noticeable effect on the gas
would be that the oxygen to carbon ratio in the disc would be higher,
which was used as an argument in \cite{Kral2016} to infer that water
is also being released around $\beta$~Pic, although not detected
\citep{Cavallius2019}. On the other hand, including CO$_2$ would be
equivalent to increasing the fraction of CO in planetesimals, since
CO$_2$ is photodissociated into CO and oxygen in a timescale shorter
than CO. Even if neutral carbon is present, it cannot shield CO$_2$ by
more than a factor 0.33 \citep{Rollins2012}, thus whether the CO in
the model was released directly from collisions or is a result of
CO$_2$ photodissociation has little effect. The only effect would be
that the oxygen to carbon ratio in the disc would be higher, and even
more if water is released too. A potential noticeable effect of
including other volatile species could be that the higher gas density
could increase the gas drag on the dust (see discussion in
\S~\ref{dis:gas-dust}).

One effect that could be important and neglected here, is that if
temperatures of the largest planetesimals are higher than the water
sublimation temperature, volatiles might not remain trapped in the
interiors and planetesimals could devolatise on short timescales
compared to their collisional lifetime. Therefore, any conclusion on
the evolution of gas in discs within 20~au must be taken with caution
since gas released might not be controlled by collisional processes
\citep[e.g.][]{Marino2017etacorvi}.

Moreover, in this paper we have assumed a constant CO abundance in
planetesimals. This could not be the case since we expect that the CO
abundances will depend both on the CO abundance in their progenitors
protoplanetary discs, but also on the disc temperature where
planetesimals formed. If midplane temperatures at formation were close
or higher than $\sim20$~K (CO freeze out temperature), then
planetesimals might lack CO. The best example for this possibility is
HR~4796, which given that it is an A0 star and its belt is at only
79~au, it could well be that its planetesimals were formed within the
CO snowline and thus poor in CO. In fact, HR~4796 is the system with
the highest predicted midplane temperature at the belt location
\citep[see snowline locations in Figure 1 in][]{Matra2018mmlaw}, or
smallest radius compared to the CO snowline given its central
star. Computing the effect of this temperature dependence on the CO
abundance in planetesimals is difficult since it would involve
knowledge of the temperatures in the parent protoplanetary disc. In
addition, even if planetesimals were CO poor, they could contain large
fractions of CO$_2$ which has a higher freeze out
temperature. Therefore CO$_2$, which photodissociates very quickly and
cannot be fully shielded by carbon, could be released producing large
quantities of CO gas.

\subsubsection{Carbon ionisation fraction}
\label{dis:ionisation}

Here we discuss the effect of carbon ionisation on the results
presented in this paper. This can be important since it is the carbon
ionization continuum that can shield CO and explain the observed
massive CO discs around A stars. In our model we neglect the presence
of ionized carbon, or in other words we assume that the ionisation
fraction of carbon is much lower than 1 (i.e. $\lesssim0.3$). While
carbon ionisation can be significant \citep[e.g. in discs with low CO
  masses like $\beta$~Pic
  C$^0$/C$\sim\hbox{0.015--0.2}$,][]{Cataldi2014, Cataldi2018}, it has
only been constrained by observations in one shielded disc
\citep[lower than 0.4 in HD~131835,][]{Kral2019}. Here we aim to
show that the ionization fraction is lower than 0.3 for shielded
discs, i.e. for discs with a carbon surface density higher than its
critical surface density of $10^{-7}$~$M_\oplus$~au$^{-2}$.

\begin{figure}
  \centering \includegraphics[trim=0.0cm 0.0cm 0.0cm 0.0cm, clip=true,
    width=1.0\columnwidth]{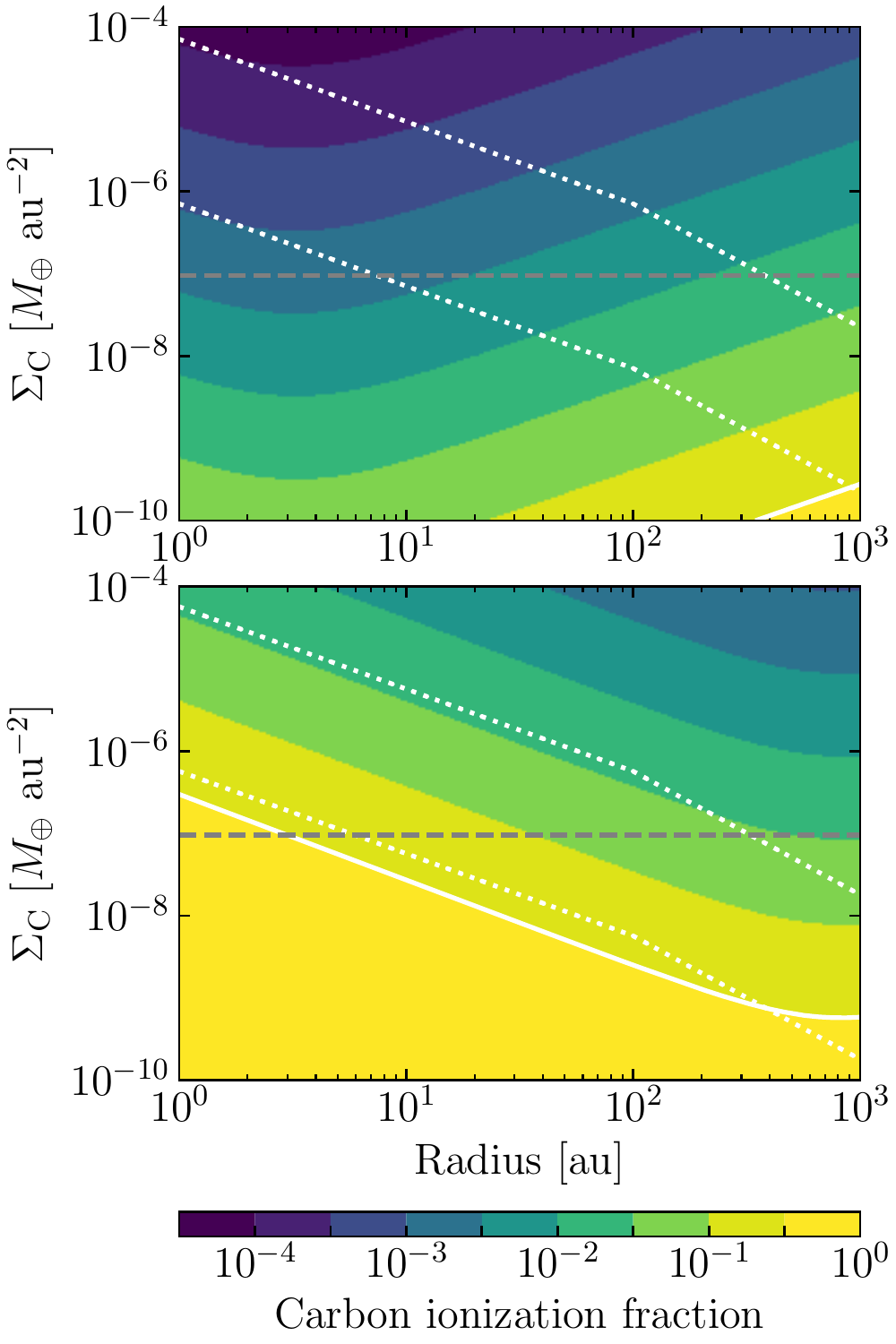}
 \caption{Carbon ionisation fraction as a function of radius and
   surface density around an A9V (top) and A0V star (bottom). The
   horizontal grey dashed line represents the critical density at
   which neutral carbon starts shielding CO. The white continuous line
   represents the 0.3 ionization level. The white dotted lines
   represent the expected surface density for a disc with a high
   viscosity ($\alpha=0.1$) where carbon is being input at 100~au at a
   rate of $10^{-1}$ and $10^{-3}$~$M_\oplus$~Myr$^{-1}$.}
 \label{fig:ionisation}
\end{figure}

We calculate the ionisation fraction of carbon using Equation 15 in
\cite{Kral2017CO}, taking into account the stellar and interstellar
radiation field \citep{Draine1978, vanDishoeck2006} and assuming an
optically thin medium (worst case scenario). In Figure
\ref{fig:ionisation} we present the ionisation fraction as a function
of radius and carbon surface density for an A9V (top) and A0V star
(bottom). The grey dashed horizontal line represents the critical
surface density of neutral carbon for shielding and the continuous
white contour represents a ionization fraction of 0.3. Overlaid in
white dotted lines we plot the surface density expected for carbon in
steady state \citep{Metzger2012} if it is being input at 100~au at a
rate of $10^{-1}$ and $10^{-3}$~$M_\oplus$~Myr$^{-1}$ and with
$\alpha=0.1$. We find that the ionization fraction is always lower
than 0.3 when densities are higher than the carbon critical
density. Only within 3~au for the A0V star, the ionisation fraction is
higher than 0.3 for surface densities above the critical value for
shielding. Therefore we conclude that for our 1D model neglecting
carbon ionisation is a valid assumption for shielded discs.

For low mass gaseous discs, the ionisation fraction can be close to
one, specially for discs around early A type stars, although the
ionisation fractions are likely lower when taking into account the
optical depth in the UV. This high ionisation fraction has little
effect on the lifetime of CO since it would be vertically unshielded
even if all carbon were neutral. Therefore, the evolution of these
discs is not dependant on the level of ionisation. Note however that
for those discs, the mass of neutral carbon that our model predicts
could be overestimated.

A caveat in this calculation is that we are computing the ionisation
fraction in the disc midplane, while the carbon atoms that could
shield CO must lie in upper layers. Further work in 2D including the
vertical dimension are necessary to estimate more precisely the
ionisation fraction of carbon in the upper layers, as well as the
vertically dependent UV radiation impinging the CO molecules
\citep[e.g.][]{Kral2019}. Moreover, here we only considered the
photospheric emission, while these stars can have significant
additional chromospheric emission in the UV \citep{Matra2018}. These
considerations require of detailed modelling of the stellar emission
and are beyond the scope of this paper.

\subsubsection{CO photodissociation including stellar UV}
\label{dis:photodissociation}

\begin{figure}
  \centering \includegraphics[trim=0.3cm 0.4cm 0.2cm 0.2cm, clip=true,
    width=1.0\columnwidth]{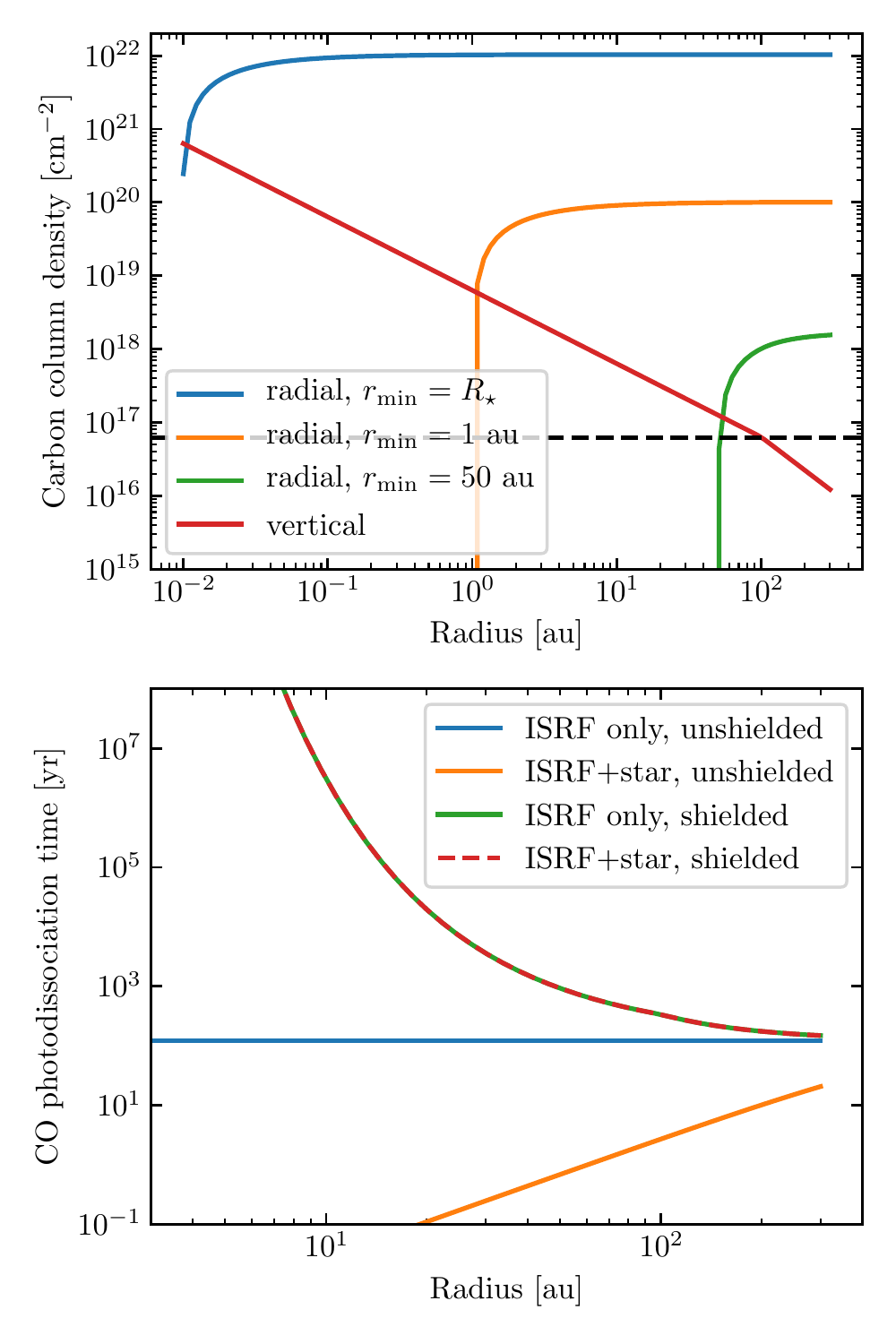}
  \caption{\textbf{\textit{Top:}} Column density of carbon in the
    radial (towards the star) and vertical (from the midplane, red
    line) directions. The radial component is shown for three
    different cases in which the disc inner edge is at the stellar
    surface (2.1$R_\odot$), 1~au and 50~au (blue, orange and green
    lines). The horizontal dashed black line represents the critical
    column density of carbon above which which it can effectively
    shield the CO. \textbf{\textit{Bottom:}} CO photodissociation
    timescale assuming a carbon disc extending down to the stellar
    surface and calculated in four different scenarios: radiation
    dominated by the ISRF and neglecting shielding (blue), radiation
    dominated by the ISRF and considering shielding (green), radiation
    from the ISRF and an A0 star and neglecting shielding (orange
    line), and radiation from the ISRF and an A0 star and considering
    shielding (red dashed line).}
 \label{fig:shielding}
\end{figure}

Throughout this paper we have assumed that the photodissociation of CO
is controlled purely by UV radiation impinging CO molecules in the
vertical direction. This is only valid when the stellar UV flux is
lower than the ISRF (e.g. for gas at 100~au around stars with
luminosities below $\sim20\ L_\odot$) or if CO is shielded in the
radial direction by carbon. We are interested in the second case since
this could be common among massive discs independently of the stellar
type. In order to check this we first compare the radial column
density of neutral carbon in the radial (towards the star) and
vertical direction (from the midplane), assuming the following: i) a
disc surface density corresponding to an accretion disc with an input
source at a belt radius of 100~au \citep{Metzger2012}; ii) a disc
inner edge at $R_\star$ (2.1$R_\odot$), 1~au and 50~au; iii) a surface
density at 100~au equal to the carbon critical surface density
($10^{-7}$~$M_\oplus$~au$^{-2}$); iv) a disc scale height of
$0.05r$. Figure~\ref{fig:shielding} shows these column densities as a
function of radius. We find that the radial component is always orders
of magnitude larger than the vertical one, even when using a disc
inner edge of 50~au.

Then, in the case of a disc extending inwards until reaching the
stellar surface, we calculate the CO photodissociation rate
\citep{Matra2018} around a A0 star considering shielding by carbon
only. In the bottom panel of Figure~\ref{fig:shielding}, we compare
the photodissociation timescale of four different cases or
assumptions: ISRF only and no shielding; the ISRF plus the stellar
radiation and no shielding; the ISRF only and vertical shielding (as
in this paper); and the ISRF and stellar radiation and both vertical
and radial shielding. We find that as expected, when neglecting carbon
shielding (orange line), the photodissociation timescale is much lower
than 120~yr. However, when considering carbon shielding (red and green
lines) the photodissociation timescale is primarily set by the ISRF
and vertical shielding. In fact, including the stellar contribution
and the carbon shielding along the radial direction does not change
significantly the CO photodissociation rate. We find this is true at
100~au even for low neutral carbon surface densities of
$3\times10^{-12}$~$M_\oplus$~au$^{-2}$, i.e. for all discs in the
filtered model population and most of the discs that we model (see
bottom right corner in Figure \ref{fig:corner_halpha}). Therefore we
conclude that for the filtered population it is safe to neglect the
stellar UV contribution to the CO photodissociation.

\subsection{Breaking degeneracies}

\subsubsection{Measuring $\alpha$}
\label{sec:alpha}

As commented in \S\ref{sec:poplow}, in our population synthesis model
a degeneracy exists between the kinematic viscosity (or $\alpha$) and
the fraction of CO in planetesimals. A potential way to break this is
by estimating the level of turbulence in the disc through line
observations and thus constraining $\alpha$ independently. Generally
speaking, the local line width of an emission line in velocity units
(without taking into account Doppler shifts due to Keplerian rotation)
is
\begin{equation}
  \Delta v_\mathrm{line}=\sqrt{2 k_\mathrm{B} T / (\mu m_p) +
    v_\mathrm{turb}^{2}} ,
\end{equation}
where $T$ is the gas temperature, $\mu$ the mean molecular weight,
$m_p$ the proton mass, and $v_\mathrm{turb}$ is the turbulent
linewidth\footnote{Note that $\Delta v_\mathrm{line}$ is defined as
  $\sigma_\mathrm{line}/\sqrt{2}$, with $\sigma_\mathrm{line}$ being
  the standard deviation of the line profile}. The turbulent line
width can be approximated by $\sqrt{\alpha} c_s$, thus obtaining
$\Delta v_\mathrm{line}=c_s\sqrt{2+\alpha}$, where $c_s$ is the sound
speed. Therefore, if we can constrain $c_s$ from the disc scale height
or temperature (e.g. through line excitation temperature, line
brightness from optically thick emission or dust observations), then
$\alpha$ could be constrained independently from our population
synthesis model, allowing to constrain the fraction of CO in
planetesimals. This idea is not new, and multiple studies exist in
protoplanetary discs trying to constrain $\alpha$ through line
observations \citep[e.g.][]{Hughes2011}. The difference here is that a
larger $\alpha$ value compared to protoplanetary discs should be
easier to observe and the vertical dependence of the temperature
should be weaker since these discs are optically thin.

In order to constrain $\alpha$ to values near 0.1, it would be
necessary to constrain the line FWHM with a precision below 1\%,
which is achievable with high S/N observations that can resolve the
line intrinsic widths. The precision or fractional uncertainty on the
line FWHM can be written as \citep{Lenz1992}
\begin{equation}
  \frac{\sigma_\mathrm{FWHM}}{\mathrm{FWHM}} = 1.4 \sqrt{ \frac{ \Delta
      v}{ \mathrm{FWHM} }} \frac{1}{\mathrm{(S/N)}_0},
\end{equation}
where $\Delta v$ is the spectral resolution and $\mathrm{(S/N)}_0$ is
the S/N at the line peak. Since the peak S/N typically scales as the
square root of the spectral resolution for well resolved lines, it can
be shown that the expression above is equivalent to
\begin{equation}
\frac{\sigma_\mathrm{FHM}}{\mathrm{FWHM}} = \frac{0.24}{\mathrm{(S/N)}} ,
\end{equation}
where S/N is the integrated line signal-to-noise. This form shows that
as long as the line is well resolved, the precision to measure the
linewidth does not depend on the spectral resolution, but only on the
integrated line S/N (which does not depend on $\Delta v$ for ALMA
observations). The above expression then implies that in order to
constrain the linewidth with a precision of 0.2\%, it is necessary
to achieve an integrated line S/N greater than 100. Although such high
S/N might be difficult to achieve over a single beam, azimuthal
averaging with Keplerian masking can be employed
\citep[e.g.][]{Matra2017fomalhaut, Teague2018} and thus obtaining high
precision measurements of the intrinsic linewidth and constraints on
$\alpha$.

%% One caviat in the above discussion is that in order for that method to work

%%  For example, let's imagine we want to determine the CO 3-2
%% linewidth with a precission of 0.2\% in order to constrain $\alpha$
%% for a gas at a temperature of 50 K (intrinsic line FWHM of 290
%% m/s). Using ALMA's highest spectral resolution in band 7 of 27 m/s we
%% would need an overall S/N of

%% where $\Delta v$ is the spectral resolution and S/N is the integrated
%% line signal-to-noise. This form shows how important is the chosen
%% spectral resolution in order to determine the linewidth with a high
%% precission since for typical observations of resolved lines, the
%% integrated line S/N does no depend on the spectral resolution. For
%% example, let's imagine we want to determine the CO 3-2 linewidth with
%% a precission of 0.2\% in order to constrain $\alpha$ for a gas at a
%% temperature of 50 K (intrinsic line FWHM of 290 m/s). Using ALMA's
%% highest spectral resolution in band 7 of 27 m/s we would need an
%% overall S/N of 

%% For example, 
%% integrated line S/N of 

%% the
%% above expression is equivalent to It seems then that with Thus with
%% $N_line$

%% For example, with a disc
%% temperature of 50 K and a mean molecular weight of 14 (gas dominated
%% by atomic carbon and oxygen) the sounds speed is $\sim170$~m/s.

%% check shear.

%% We identify a way

\subsubsection{Measuring the CO fraction in planetesimals}
\label{dis:cofraction}
The degeneracies between $\alpha$ and fraction of CO in planetesimals
($f_\mathrm{CO}$) could also be broken by estimating $f_\mathrm{CO}$
in unshielded discs. Because the CO mass in those discs only depends
on the photodissociation timescale ($t_\mathrm{ph}$) and the product
between the mass loss rate and $f_\mathrm{CO}$,
i.e. $M_\mathrm{CO}=t_\mathrm{ph} f_\mathrm{CO} \dotCO$, by estimating
the UV field at the CO location and mass loss rate from the disc
fractional luminosity it is possible to derive $f_\mathrm{CO}$ as
shown by \cite{Marino2016} and \cite{Matra2017fomalhaut}. The main
obstacle for determining $f_\mathrm{CO}$ with this method are the
uncertainties on the CO masses due to NLTE effects. Observations of
multiple transitions could help improve these mass estimates
\citep{Matra2017betapic}.

%% \subsection{Hydrodynamic limit}

%% The timescale thus to

%% the and thus decoupled from the
%% gas. We find though that sub-$\mu$m grains can

%% is such that the carbon surface density is five times its critical value to shield 

%% is worth comparing the timescale of collisions of grains
%% of a given size $D$ with the stopping time due to gas drag. In Figure
%% x we show the Stoke number as a function of radius and

%% ... This was found also in
%% \cite{Kral2019}....

%% stoke numbers for constant Mdot

%% stoke number for micron size grains to check blow out

\subsection{Resolved observations of atomic carbon}
\label{dis:carbon}

An important test of the model presented here and previous viscous
evolution modelling is to check whether the atomic gas is more spread
than CO and planetesimal belts, with a surface density characteristic
of an accretion disc and thus decreasing with radius. So far, the four
resolved observations of atomic carbon have not found that, but rather
inner cavities in the neutral carbon distribution \citep{Kral2019,
  Cataldi2018, Cataldi2019, Higuchi2019}. This could be due to low
viscosities, but that would contradict our findings that high
viscosities are preferred. Alternatively, it could be that carbon is
significantly ionised within a few tens of au, although even the
ionised carbon in $\beta$~Pic was found to have a large inner cavity
\citep{Cataldi2014}. Thus, it might be that something is preventing
carbon from spreading inwards. As commented before, around A stars
carbon can become unbound due to radiation pressure, but we expect
both self-shielding and collisions with bound oxygen atoms to prevent
this. If radiation pressure was stronger than previously estimated
though \citep[e.g.][]{Kral2017CO}, this could explain the gas cavities
observed so far. In fact, there is already evidence that the stellar
UV flux where most of the stellar absorption occurs could be
underestimated around these young A stars \citep[e.g.][]{Deleuil2001,
  Bouret2002}. \cite{Matra2018} compared observed and photospheric
model UV spectra for $\beta$~Pic finding it was underestimated by
orders of magnitude. If both carbon and oxygen are loosely bound, then
this could explain why carbon is not forming an accretion disc.

A counter argument for this possibility is that if carbon and oxygen
were not spreading inwards due to radiation pressure and instead
accumulating near the planetesimal belt, then the carbon surface
densities should be even higher, which is inconsistent with the
observed carbon masses and analysis presented here. Only if radiation
pressure can remove carbon from the system or at least beyond the
planetesimal belt this could solve this apparent
inconsistency. Alternatively, the gas distribution could still have
large cavities if the gas was released recently in a giant collision,
as suggested by \cite{Dent2014} and \cite{Cataldi2018, Cataldi2019},
or the disc was stirred only recently (in less than a viscous
timescale) and thus gas release has only been in place for a small
fraction of the ages of these systems. This however raises even more
questions regarding the frequency of these events since we would also
expect to observe systems where gas has viscously spread filling the
cavities with gas.

\subsection{Interaction with planets}

The gas evolution modelled here does not take into account any
possible interaction with planets. These planets might be massive and
thus significantly affect the gas distribution and evolution, or small
enough such that any exocometary accreted gas could dramatically
change the mass and composition of their atmospheres. Because the
accretion of gas onto low mass planets requires detailed modelling, we
leave this topic for future work. Here we focus on the effects that
massive planets could have on the gas distribution, which can be
studied using simple relations. For a more detailed discussion on
planet disc interactions see Kral et al. submitted.

Similarly to the study of planet disc interactions in protoplanetary
discs, there is a critical planet mass above which a planet could
create a deep gap in the gas distribution around its orbit and affect
the gas transport processes in the system \citep[e.g.][]{
  Crida2006}. Such a planet can reduce the gas flow from the
planetesimal belt (exterior to the planet's orbit) into the inner
regions of the system, creating a cavity \citep{Lubow1999,
  Lubow2006}. This scenario is of particular interest here as it could
solve the tension between the few carbon observations that show
cavities and the viscous evolution models of exocometary gas in the
literature \citep[as suggested by][]{Cataldi2019}. Thus, the question
is how massive must a planet be to reduce the gas flow. Both analytic
and numerical studies have shown that the accretion rate onto the star
can be a factor 4--10 times lower than the accretion rate outside of
the orbit of the planet for planet-to-star mass ratios of
$5\times10^{-5}-10^{-3}$ \citep{Lubow2006}. In this case, most of the
gas flowing through the planet's orbit will be accreted by the
planet. One of the basic requirements to create such gaps and affect
the radial flow is that the Planet's Hill radius must be larger than
the disc scale height. Defining the vertical aspect ratio of the gas
disc as $h=H/r$, where $H$ is the scale height of the gaseous disc, we
find that in order to explain the observed cavities planets must be
more massive than
\begin{equation}
  \Mp > 50\  \left(\frac{h}{0.03}\right)^3 \left(\frac{M_\star}{2\ M_\odot}\right)  M_\oplus.
\end{equation}
This minimum planet mass is thus very sensitive to the disc scale
height. Assuming blackbody temperatures for the gas and a mean
molecular weight of 14 (i.e. gas is dominated by carbon and oxygen) we
find $h\approx0.03$ at 40~au. Gas temperatures could be much lower
\citep[as suggested by the low excitation temperature of CO in some
  discs,][]{Kospal2013, Flaherty2016, Matra2017betapic} and the mean
molecular weight larger if dominated by CO, which would make $h$ and
the minimum planet mass smaller. So far, 49~Ceti is the only example
there is where $h$ has been constrained for the gas. \cite{Hughes2017}
constrained $h$ to be smaller than 0.04 \footnote{calculated at 40~au
  using their best fit parameters} and therefore planets with masses
similar or larger than Neptune could explain the observed carbon
cavities. Note that planets with masses below a few Jupiter masses are
currently undetectable through direct methods \citep[e.g. direct
  imaging,][]{Bowler2018}, thus the non-detections of planets within
these cavities are still consistent with this scenario. Moreover, we
can hypothesise that the inner edges of the observed planetesimal
discs are truncated by planets, similar to the Kuiper belt in the
Solar System. These planets could also be the ones responsible for
stirring the orbits of planetesimals igniting a collisional cascade
and the gas release \citep[e.g.][]{Mustill2009}. Therefore the radial
distribution of gas could provide important constraints to the
presence of planets that are undetectable with current
instrumentation.

These massive planets decreasing the inward flow of gas would accrete
most of it \citep[e.g.][]{Machida2010}. We can calculate the planet
accretion rate using Equation~\ref{eq:vr} and the analytic solution
for the surface density of gas by \cite{Metzger2012} for an $\alpha$
disc model. In steady state and assuming only a small fraction of the
gas flow is able to cross the gap, we find $\dot{M}_\mathrm{p}=3\upi
\nu \Sigma$, which for a given disc is independent of radius as long
as the planet is interior to the planetesimal belt where gas is being
released. Evaluating this expression we find
\begin{equation}
  \begin{split}
    \dot{M}_\mathrm{p}=10^{-2}  \left(\frac{\alpha}{0.1}\right) \left(\frac{\Sigma_{\rm G}(\rb)}{10^{-7} M_\oplus\ \mathrm{au}^{-2}}\right)  \left(\frac{\rb}{100\ \mathrm{au}}\right)  \\ \left(\frac{\mu}{14}\right)^{-1}  \left(\frac{M_\star}{2\ M_\odot}\right)^{-1/2} \left(\frac{L_\star}{16\ L_\odot}\right)^{1/4}    M_\oplus\ \mathrm{Myr}^{-1}.
  \end{split}
\end{equation}
Therefore, over 10-100~Myr of evolution we do not expect that the gas
accretion will significantly change the mass of such a
planet. Nevertheless, because the gas will be dominated by carbon and
oxygen (in contrast to protoplanetary disc gas dominated by H$_2$) the
abundance of these elements relative to hydrogen could change
significantly. Even the abundance of carbon relative to oxygen (the
C/O ratio) could be affected since the accreted gas could have a low
C/O if originating from CO$_2$ or H$_2$O outgassed molecules compared
to gas accreted during the protoplanetary disc phase with a C/O ratio
close to one at tens of au \citep[e.g.][]{Oberg2011}. Finally, the
accretion luminosity of such an accreting planet will be
$\sim10^{-9}\ L_\odot$, assuming a planet mass of 50~\Me\ and a radius
of 6 Earth radii. Therefore it is unlikely that exocometary gas
accretion will increase the intrinsic luminosity of these young
planets \citep[][]{Mordasini2009}.

\subsection{Other gas removal processes}
\label{dis:removal}
 
If gas were not being removed through viscous accretion onto the star
or planets, then some other mechanism must be at play. We identify two
potential candidates: radiation pressure acting on the gas and unbound
grains pushing the gas out \citep{Tazaki2015, Kuiper2018}. For
example, radiation pressure could remove mass in the form of winds
launched from the surface of the disc where densities are low enough
that atomic carbon is unshielded from stellar radiation. We have
already discussed in \S\ref{dis:carbon} that radiation pressure on
carbon could be underestimated by current models
\citep[e.g.][]{Kral2017CO} and thus this scenario cannot be ruled out
yet. If so, the $\alpha$ value derived here gives a sense for the
timescales involved. We found that observations were best matched with
$\alpha=0.1$, which leads to a viscous timescale of 0.6~Myr or
$\sim700$ orbits for gas released at 100~au around a 2~$M_\odot$
star. This means that the mechanism is not efficient enough to remove
the bulk of the gas mass on a dynamical timescale, which is expected
given the large optical depth in the radial direction at the
frequencies where gas absorbs and low dust densities. Detailed
modelling of these processes with more accurate stellar spectra are
required to test these mechanisms.

\subsection{Gas - dust interactions}
\label{dis:gas-dust}

So far through this paper we have neglected how gas could affect the
dynamics of dust; particularly, how gas could damp the eccentricity of
small dust grains in high eccentricity orbits or unbound trajectories
due to radiation pressure, and how gas could drag small dust interior
or exterior to the planetesimal belt on timescales shorter than
collisions or P-R drag. Here we aim to quantitatively check if gas
could affect the dust dynamics in the context of shielded discs
\citep[see also][]{Kral2019}.

\begin{figure}
  \centering \includegraphics[trim=0.0cm 0.0cm 0.0cm 0.0cm, clip=true,
    width=1.0\columnwidth]{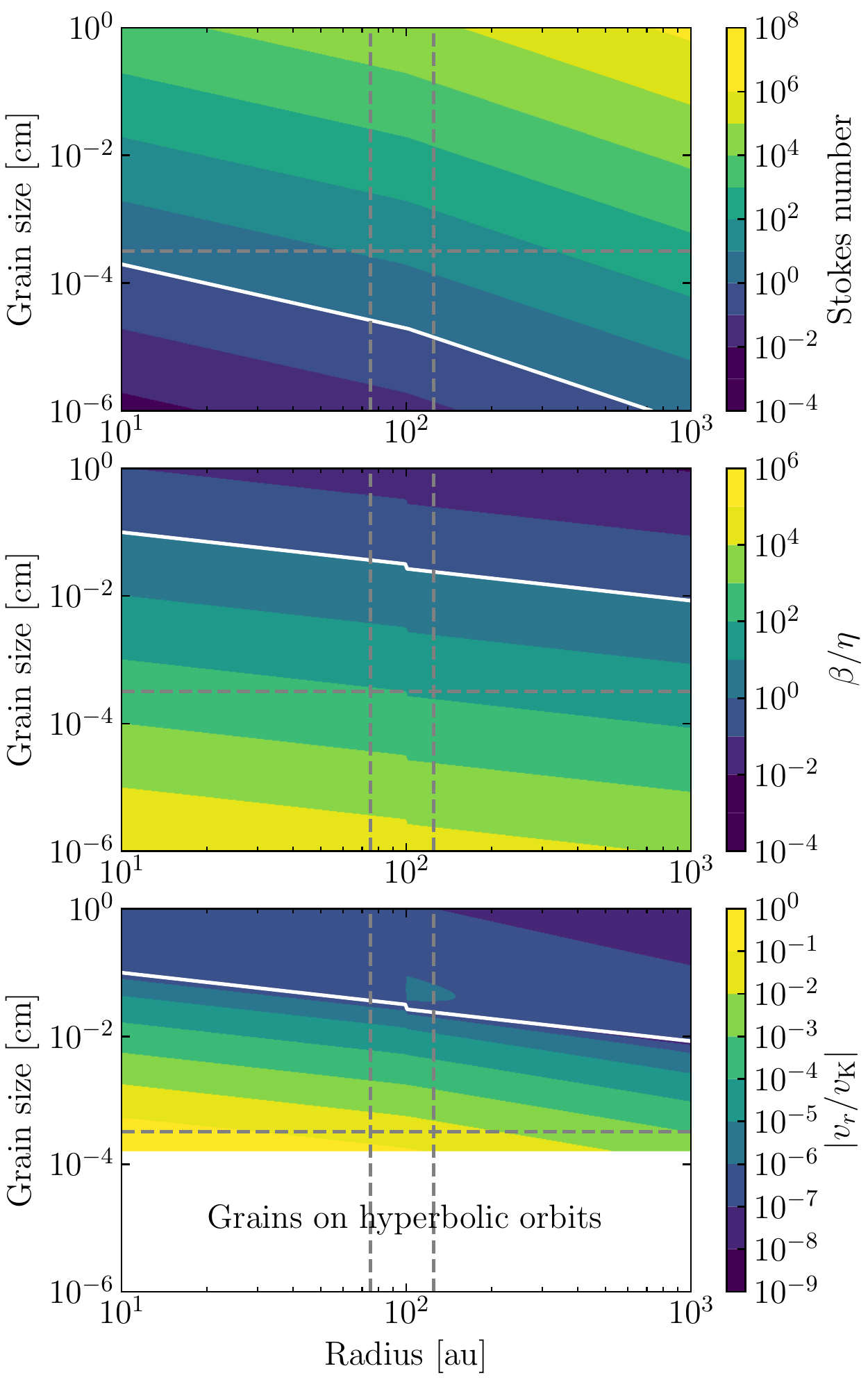}
 \caption{Effect of gas on dust as a function of grain size and
   stellar distance assuming a steady state surface density of gas
   when released at 100 au, reaching a local surface density of
   $10^{-6}$~\Me~au$^{-2}$. \textit{\textbf{Top:}} Dimensionless
   stopping time of Stokes number in the subsonic regime. The white
   line represents a Stokes number of one. \textit{\textbf{Middle:}}
   Ratio between the $\beta$ (the radiation to gravitational force
   ratio) and $\eta$ the ratio between the pressure gradient force and
   gravitational force. If larger than 1, then grains will be pushed
   outwards and vice versa. \textit{\textbf{Bottom:}} Modulus of the
   radial velocity of dust grains over the Keplerian velocity. In the
   three panels, the horizontal dashed line represents the blowout
   size for a A3V star and the vertical dashed lines represent the
   inner and outer edge of a belt at 100~au with a fractional width of
   0.5.}
 \label{fig:dust}
\end{figure}

First, to understand whether gas drag could be important we compute
the dimensionless stopping time (or Stokes number)\footnote{Using the
  subsonic stopping time.} of grains of different sizes and at
different radii around an A3V star (Figure~\ref{fig:dust} top
panel). We assume a steady state surface density of gas
\citep{Metzger2012} for a system where gas is released from a
planetesimal belt at 100~au and the surface density of gas at 100~au
is $10^{-6}$~\Me~au$^{-2}$---a typical surface density for the
predicted distribution of shielded discs (see
Figure~\ref{fig:corner_halpha}). We find that all grains above the
blowout size (dashed horizontal line) have Stokes numbers larger than
10 near 100~au where dust is released (in between the vertical dashed
lines), and thus we do not expect grains to be significantly damped
before experiencing disrupting collisions. In particular, mm-sized
grains have Stokes numbers of $\sim10^{4}$, thus are well decoupled
from the gas. Nevertheless, we find that sub-$\mu$m grains below the
blowout size have Stokes number close and below unity, and thus these
unbound grains could remain in the system for longer timescales due to
gas drag. The dynamics of such grains has been studied by
\cite{Lecavelier1998}, finding that unbound dust grains can be heavily
decelerated by gas drag creating a stationary outflow. Therefore, gas
rich debris discs might have more massive halos of small grains which
can be traced in scattered light
\citep[e.g. HD32297,][]{Schneider2005, Bhowmik2019}.

Even though grains above the blowout size have larger Stokes numbers,
they could still systematically migrate in or out due to a difference
in their orbital speed. Ignoring Poynting-Robertson drag (PR drag),
the azimuthal velocity of a dust grain on a circular orbit is
\begin{equation}
  v_{\phi}=v_\mathrm{K}(1-\beta)^{1/2},
\end{equation}
where $\beta$ is the ratio between the radiation and gravitational
forces \citep{Burns1979}, thus small grains in circular orbits will
have sub-Keplerian speeds. Similarly, because of pressure support gas
will also orbit at a sub-Keplerian speed that is
\begin{eqnarray}
  v_\mathrm{g}=v_\mathrm{K}(1-\eta)^{1/2},\\
  \eta= - \frac{1}{r\Omega_\mathrm{K}^2 \rho_\mathrm{g}}\frac{dP_\mathrm{g}}{dr},
\end{eqnarray}
where $\rho_\mathrm{g}$ and $P_\mathrm{g}$ are the gas density and
pressure, respectively, and $\eta$ the ratio between the pressure
gradient force and gravitational force. If $\beta>\eta$, dust grains
will orbit slower than the gas, and thus gas drag will increase their
angular momentum producing an outward migration
\citep{Takeuchi2001}. In the middle plot of Figure~\ref{fig:dust} we
show the ratio $\beta/\eta$ for different grain sizes at different
radii. We calculate $\beta$ assuming blackbody grains with an internal
density of 2.7~g~cm$^{-3}$ and $\eta$ according to blackbody
temperatures and the analytic expression for the gas surface density
profile. We find that grains below 1~mm have $\beta>\eta$ and thus
will experience outward migration, while larger grains will tend to
migrate inwards. This will only happen if the timescales are shorter
or comparable to the collisional lifetime of these grains. Using
Equation 26 in \cite{Takeuchi2001}, which assumes circular orbits, we
compute the radial velocity of grains under the influence of gas and
PR drag too (Figure~\ref{fig:dust} bottom panel). We find that for
small $\mu$m-sized grains their radial speed could be a few percent of
the Keplerian speed, and thus could exit the planetesimal belt within
a few orbits.

\begin{figure}
  \centering \includegraphics[trim=0.2cm 0.2cm 0.5cm 0.3cm, clip=true,
    width=1.0\columnwidth]{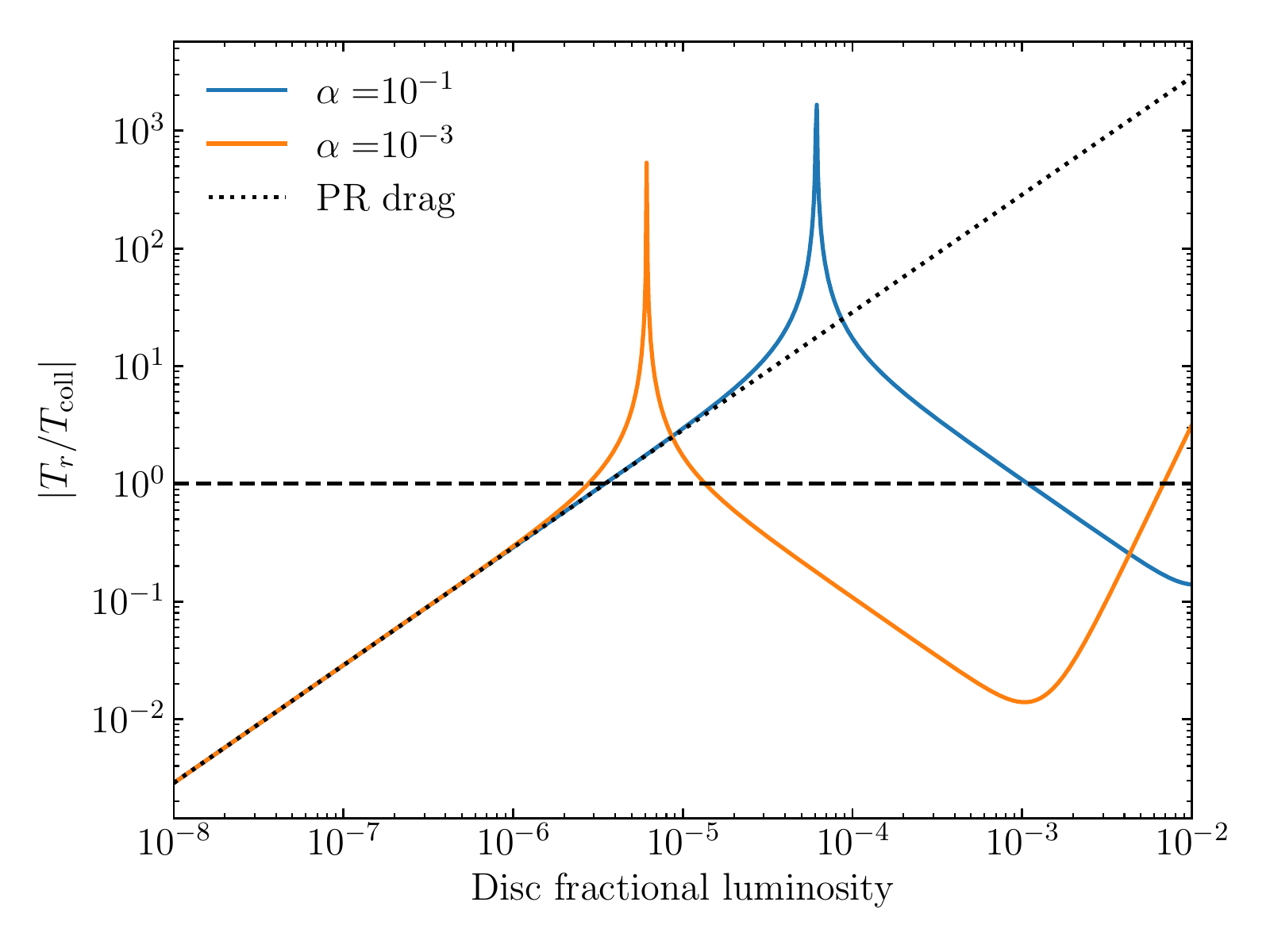}
 \caption{Radial migration timescale over collisional timescales for
   dust grains at the bottom of the collisional cascade. The blue and
   orange lines represent a case with an $\alpha$ viscosity value of
   $10^{-3}$ and 0.1, respectively. The black dotted line represents a
   case when neglecting gas drag and considering PR drag only.}
 \label{fig:radial_velocity}
\end{figure}

Small grains will only migrate out of the belt efficiently if they can
do so on a timescale shorter than their collisional timescale. In
Figure \ref{fig:radial_velocity} we compare the timescale at which a
small grain at the bottom of the collisional cascade on a circular
orbit would exit the belt, assuming a disc width of 50~au, with their
collisional timescale as a function of the disc fractional
luminosity. For this, we use equations B5 and B6 in
\cite{Matra2017fomalhaut} to calculate the mass loss rate and
collisional timescales. We also assume a CO mass fraction of 0.1
inside planetesimals, which sets the gas release rate and thus the gas
surface density which will affect the dust. We find that for
$\alpha=0.1$ (blue line) and fractional luminosities larger than
$10^{-3}$, grains can exit the belt on timescales shorter than their
collisional lifetime (blue line is below the dashed black line). For
smaller $\fir$ drift timescales due to gas drag become too long, and
actually, for $\fir<6\times10^{-5}$ PR drag dominates as a drag force
(see the blue line converging to the black dotted line). Nevertheless,
PR drag timescales will be longer than collisional timescales for
$\fir>4\times10^{-6}$, which is at the limit of detectability with
current instrumentation \citep[see][for a detailed discussion on PR
  drag]{Wyatt2005prdrag}. For smaller $\alpha$'s, gas drag could
become important even in less massive planetesimal discs since gas can
accumulate for longer and thus build larger gas surface densities. For
$\alpha=10^{-3}$, small dust in discs with $\fir>2\times10^{-5}$ could
efficiently migrate outwards with the fastest drift achieved with
$\fir\sim10^{-3}$, which is when the Stokes number for grains with
$\beta=0.5$ is equal to 1. Therefore, we conclude that outward
migration of small dust grains due to gas drag in exocometary gaseous
discs could be an efficient process, increasing the cross sectional
area of disc halos \citep[e.g. HD32297,][]{Schneider2005,
  Bhowmik2019}.

%% - peak pr drag uprise for f>10-3 due to small St

%% This timescale is comparable to their collisional
%% timescale for belts with fractional luminosities between
%% $10^{-3}$--$10^{-2}$, therefore a fraction of small grains could
%% efficiently migrate outwards resulting in a radial segregation of dust
%% grains \citep{Takeuchi2001}. This was suggested by \cite{Kral2019}
%% as a mechanism that could explain the multiple rings seen in
%% HD~131835.

Based on the derived Stokes numbers we find that photoelectric
instability \citep{Lyra2013photoelectric, Richert2017} should not play
a major role affecting the distribution of gas and dust. This is
because the Stokes numbers are significantly larger than unity for all
grains sizes above the blowout size, and the dust-to-gas ratio is
expected to be much lower than unity when considering only grains
smaller than 10~$\mu$m as in \citet[see their Figure 2]{Richert2017}.

Finally, it is possible that the overall gas surface densities are
higher than assumed here if other volatiles (e.g. water) are being
released as well. Assuming a water ice fraction of 0.5 in
planetesimals, the gas surface densities would be a factor 5 larger
than assumed. Therefore, dust grains close to the blow-out size would
have lower Stokes numbers in the range 1-10 \citep[as suggested
  in][]{Bhowmik2019}, and migrate outwards even faster. Note that the
smaller Stokes number could be enough to trigger the photoelectric
instability, although the higher gas surface density would also
decrease the dust-to-gas ratio. Detailed simulations are necessary to
assess whether this instability could act on timescales shorter than
the age of these systems with these dust-to-gas ratios and Stokes
numbers.

\subsection{Mass loss rate inferred from the SED}
\label{dis:mdot}

In this paper we have shown that overall CO gas observations around A
stars can be explained with our population synthesis model presented
in \S\ref{sec:population}. Nevertheless, this is not entirely
satisfactory since our model has three main free parameters to fit one
main observable, the observed CO gas mass. Moreover, these parameters
(the viscosity, the fraction of CO in planetesimals, and the maximum
planetesimal size or initial median disc mass) are degenerate. In
\S\ref{sec:alpha} and \S\ref{dis:cofraction} we discuss how the first
two could be constrained through independent methods and thus break
degeneracies. The maximum planetesimal size is however unconstrained
by the collisional model used here. Namely, given a disc radius and
fractional luminosity, the total disc mass and mass loss rate are
unconstrained. \cite{Matra2017fomalhaut} showed however that the mass
loss rate at the bottom of the collisional cascade can indeed be
estimated analytically based on the disc cross sectional area and
minimum grain size. The reason for this is that the lifetime or
collisional rate of the smallest grains (near the blowout size) is
dominated by grains of similar size which is not the case at the top
of the size distribution. This slight but significant difference is
not accounted for in the collisional model used here which assumes a
single power law size distribution.

\begin{figure}
  \centering
  \includegraphics[trim=0.0cm 0.0cm 0.0cm 0.0cm, clip=true,
    width=1.0\columnwidth]{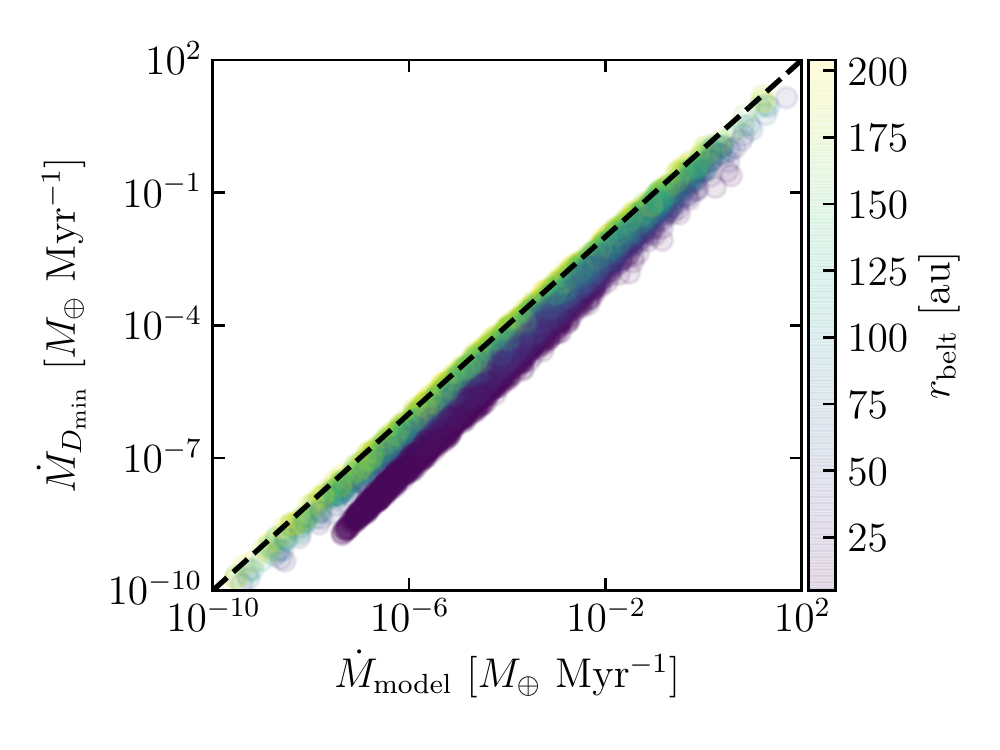}
 \caption{Mass loss rate used in our modelling based on planetesimal
   parameters and total mass (x axis) vs mass loss rate derived from
   our model fractional luminosity as in \citet[][y
     axis]{Matra2017fomalhaut}. Each circle corresponds to a different
   simulation presented in \S\ref{sec:pophigh}, and are colour coded
   according to their belt radius, with purple circles representing
   small belts and yellow circles large belts. The dashed line
   represents the y=x curve, i.e. points along the dashed line have
   consistent mass loss rates.}
 \label{fig:mdot}
\end{figure}

In Figure~\ref{fig:mdot} we compare the mass loss rate used in our
model (\S\ref{sec:population}) vs the mass loss rate that would be
inferred from the model fractional luminosity using Equation B6 in
\cite{Matra2017fomalhaut}. We find that for the chosen parameters
($e=0.001$, $\Dc=0.02$~km, $\Qd=1.6$~J~kg$^{-1}$ and
$M_\mathrm{mid}=0.5$~\Me) the mass loss rate in our model roughly
agrees with the one calculated considering the bottom of the
collisional cascade ($\dot{M}_{D_\mathrm{min}}$), except for belts
with radius smaller than $\sim50$~au. The dependence on belt radius is
due to both analytic expressions having a different dependence on
$\rb$. The belt radius sets the Keplerian and relative velocities,
which are important for determining the minimum planetesimal size that
disrupt the planetesimals at the top of the size distribution, whereas
at the bottom of the size distribution that size is simply set by the
blowout size. In order to reconcile the mass loss rate from small
grains and the largest planetesimals, more detailed modelling is
necessary taking into account, for example, the size dependent
planetesimal strength and wavy patterns in the size distribution
\citep{Krivov2006, Thebault2007}.

\subsection{Hydrodynamic limit}

Finally, here we discuss whether it is reasonable to treat the
evolution of exocometary gaseous discs with the standard viscous
evolution equations used here. While these equations are normally
valid for the evolution of protoplanetary discs, the gas densities in
the exocometary gas context can be much lower and thus we may approach
the limit at which these equations are valid. To check this, we
compare the disc scale height ($H=c_s/\Omega_\mathrm{K}$) with the
mean free path in the gas, $l$ \citep[see][for a specific discussion
  related to $\beta$~Pic's gaseous disc]{Kral2016}. We find that at
100~au, the mean free path of gas (not ionised) in the midplane will
be smaller than $H$ as long as the gas surface density is larger than
$\sim10^{-9}$~\Me~au$^{-2}$. If the gas is highly ionised as expected
for low density discs, the mean free path will be orders of magnitude
smaller due to the large cross section of ionized species, and
therefore we estimate that for surface densities larger than
$\sim10^{-17}$~\Me~au$^{-2}$ our treatment is valid. This surface
density is much lower than the typical gas surface densities of
shielded discs ($\gtrsim10^{-7}$~\Me~au$^{-2}$) and the unshielded gas
discs detected around A stars ($M_\mathrm{gas}>10^{-5}$~\Me). Only for
the lower tail of the population distribution of surface densities
around A stars (see Figure \ref{fig:corner_halpha}), the mean free
path of molecules and atoms could be comparable or larger than $H$ and
thus the evolution of those discs with extreme low gas masses must be
taken with caution. Note that this critical surface density is not
very sensitive to the choice of $H$ since $l$ is proportional to $H$.

\section{Conclusions}
\label{sec:conclusions}

In this paper we have presented a new numerical model to explain the
presence of gas found around nearby young stars with debris discs,
with an emphasis on A stars. This gas originates in the interior of
planetesimals, and mutual collisions release both dust and gas. Our
model (\S\ref{sec:model}) solves for the viscous evolution of the gas
(CO, carbon and oxygen) in 1D as it is released by planetesimals (in
the form of CO), taking into account the CO photodissociation,
self-shielding and shielding by neutral carbon. Our model has one
significant addition compared to previous work, it takes into account
the time dependent gas release rate due to the disc collisional
evolution.

With this model we found that the present gas mass of a disc is highly
dependant on the assumed initial disc mass in the form of
planetesimals. A system with a constant gas input rate can have a gas
mass orders of magnitude lower compared to a system that started with
a higher mass and collisionally evolved to an equal planetesimal disc
mass. Because of the nature of collisional evolution of planetesimal
discs, dust levels do not depend strongly on the initial disc mass and
therefore it is a degenerate problem to try to infer the initial disc
mass from dust levels. This means that while a disc might contain low
quantities of dust in a collisionally evolved disc, it might have a
massive gas disc due to a large initial disc mass. These
considerations must be taken into account when analysing gas
observations and comparing it with dust levels. However, if CO is
unshielded, its short lifetime ($\sim120$~yr) means that its mass will
be set by the present mass loss rate. Conversely, atomic carbon has a
longer lifetime as it will viscously evolve on longer timescales which
will depend on the kinematic viscosity.

We used this new numerical code to produce the first population
synthesis model for exocometary gas around A stars. Population
synthesis studies are ideal to deal with degenerate problems and
provide constraints on population properties rather than individual
systems. Informed by previous studies that fit the dust evolution, we
generated samples of $10^4$ stars hosting planetesimal discs with
random radius, mass and stellar properties. We evolved these systems
up to a random age between 3--100~Myr and compare the final
distribution to the observed distribution by applying the same
selection filters to our model population. We found that our model can
reproduce the distribution of CO masses well when viscosities are high
($\alpha\approx0.1$), even producing a bimodal distribution with a
large population of unshielded CO discs and a small fraction of
shielded ones. The shielded population is nevertheless common in our
models among systems with bright debris discs, which explains why
shielded discs are common among the sample of 17 A stars with bright
and cold debris discs. The high $\alpha$ value implies that gas needs
to be lost on timescales of $\sim1-10$~Myr. Otherwise, carbon readily
accumulates shielding CO and producing a large population of shielded
and massive CO discs, even around systems with low planetesimal disc
masses, which is inconsistent with observations.

Although we focused on A stars, we also tested our model against
observations of FGK stars. While massive and shielded CO gas discs are
found around A stars, observations of FGK stars have showed that these
have low CO gas masses, with only a few detections. Using our model we
showed that this is consistent with the collisional evolution studies
of discs around FGK stars compared to A stars. Their lower initial
disc masses means that gas release rates are lower. Our model
quantitatively shows that massive CO gas discs should be very rare
around FGK stars if they have similar viscosity levels compared to A
stars.

An important test of viscous evolution models of gas in debris discs,
is whether gas is observed to extend inwards and accrete onto the
star. In \S\ref{sec:discussion} we discussed recent observations of
neutral carbon around a few systems which seem to indicate that carbon
has not spread inwards forming an accretion disc as expected in a
viscous evolution scenario. This might suggest that either gas release
started recently or is lost through another mechanism, e.g. radiation
pressure. Nevertheless, our modelling constrains the gas loss
timescale to be $\sim1-10$~Myr (viscous timescale), which could
provide insights into alternative mass loss mechanisms. More resolved
observations of carbon are needed to conclude.

A potential explanation for the observed cavities in the distribution
of carbon could be the presence of planets. A planet orbiting the
system interior to the planetesimal belt could block the inward flow
of gas if more massive that $\gtrsim50$~\Me, accreting most of the
exocometary gas flowing in. In this scenario, the gas distribution and
dynamics could provide constraints on the mass and location of planets
at tens of au and interior to massive planetesimal belts.

Finally, we also discussed if gas could affect the dynamics of dust
grains. We found that dust should not be heavily affected by gas
drag. Only small sub-$\mu$m grains that are unbound due to radiation
pressure could feel a strong gas drag that could slow their unbound
trajectories. However, we find that the predicted gas densities could
be high enough to make $\mu$m-sized grains to effectively migrate out
on timescales comparable to their collisional timescales.

\section*{Acknowledgements}

We thank Richard Booth for valuable input and advice in setting up the
numerical code to solve for the viscous evolution of gas considering
multiple species. We also thank the anonymous referee for a
constructive and thorough review. T.H. acknowledges support from the
European Research Council under the Horizon 2020 Framework Program via
the ERC Advanced Grant Origins 83 24 28. L.M. acknowledges support
from the Smithsonian Institution as a Submillimeter Array (SMA)
Fellow.

%% thanks to Richard for discussion and help 

%% Simulations in this paper made use
%% of the REBOUND code which can be downloaded freely at
%% http://github.com/hannorein/rebound.

%%%%%%%%%%%%%%%%%%%%%%%%%%%%%%%%%%%%%%%%%%%%%%%%%%



\bibliographystyle{mnras} \bibliography{SM_pformation} % if your bibtex file is called example.bib

\newcommand{\noop}[1]{}
\begin{thebibliography}{}
\makeatletter
\relax
\def\mn@urlcharsother{\let\do\@makeother \do\$\do\&\do\#\do\^\do\_\do\%\do\~}
\def\mn@doi{\begingroup\mn@urlcharsother \@ifnextchar [ {\mn@doi@}
  {\mn@doi@[]}}
\def\mn@doi@[#1]#2{\def\@tempa{#1}\ifx\@tempa\@empty \href
  {http://dx.doi.org/#2} {doi:#2}\else \href {http://dx.doi.org/#2} {#1}\fi
  \endgroup}
\def\mn@eprint#1#2{\mn@eprint@#1:#2::\@nil}
\def\mn@eprint@arXiv#1{\href {http://arxiv.org/abs/#1} {{\tt arXiv:#1}}}
\def\mn@eprint@dblp#1{\href {http://dblp.uni-trier.de/rec/bibtex/#1.xml}
  {dblp:#1}}
\def\mn@eprint@#1:#2:#3:#4\@nil{\def\@tempa {#1}\def\@tempb {#2}\def\@tempc
  {#3}\ifx \@tempc \@empty \let \@tempc \@tempb \let \@tempb \@tempa \fi \ifx
  \@tempb \@empty \def\@tempb {arXiv}\fi \@ifundefined
  {mn@eprint@\@tempb}{\@tempb:\@tempc}{\expandafter \expandafter \csname
  mn@eprint@\@tempb\endcsname \expandafter{\@tempc}}}

\bibitem[\protect\citeauthoryear{{Bath} \& {Pringle}}{{Bath} \&
  {Pringle}}{1981}]{Bath1981}
{Bath} G.~T.,  {Pringle} J.~E.,  1981, \mn@doi [\mnras]
  {10.1093/mnras/194.4.967}, \href
  {https://ui.adsabs.harvard.edu/abs/1981MNRAS.194..967B} {194, 967}

\bibitem[\protect\citeauthoryear{{Benz} \& {Asphaug}}{{Benz} \&
  {Asphaug}}{1999}]{Benz1999}
{Benz} W.,  {Asphaug} E.,  1999, \mn@doi [\icarus] {10.1006/icar.1999.6204},
  \href {http://adsabs.harvard.edu/abs/1999Icar..142....5B} {142, 5}

\bibitem[\protect\citeauthoryear{{Beust} \& {Morbidelli}}{{Beust} \&
  {Morbidelli}}{1996}]{Beust1996}
{Beust} H.,  {Morbidelli} A.,  1996, \mn@doi [\icarus]
  {10.1006/icar.1996.0056}, \href
  {http://adsabs.harvard.edu/abs/1996Icar..120..358B} {120, 358}

\bibitem[\protect\citeauthoryear{{Beust}, {Vidal-Madjar}, {Ferlet}  \&
  {Lagrange-Henri}}{{Beust} et~al.}{1990}]{Beust1990}
{Beust} H.,  {Vidal-Madjar} A.,  {Ferlet} R.,   {Lagrange-Henri} A.~M.,  1990,
  \aap, \href {http://adsabs.harvard.edu/abs/1990A%26A...236..202B} {236, 202}

\bibitem[\protect\citeauthoryear{{Bhowmik} et~al.,}{{Bhowmik}
  et~al.}{2019}]{Bhowmik2019}
{Bhowmik} T.,  et~al., 2019, arXiv e-prints, \href
  {https://ui.adsabs.harvard.edu/abs/2019arXiv190808511B} {p. arXiv:1908.08511}

\bibitem[\protect\citeauthoryear{{Booth} et~al.,}{{Booth}
  et~al.}{2013}]{Booth2013}
{Booth} M.,  et~al., 2013, \mn@doi [\mnras] {10.1093/mnras/sts117}, \href
  {http://adsabs.harvard.edu/abs/2013MNRAS.428.1263B} {428, 1263}

\bibitem[\protect\citeauthoryear{{Booth}, {Clarke}, {Madhusudhan}  \&
  {Ilee}}{{Booth} et~al.}{2017}]{Booth2017gas}
{Booth} R.~A.,  {Clarke} C.~J.,  {Madhusudhan} N.,   {Ilee} J.~D.,  2017,
  \mn@doi [\mnras] {10.1093/mnras/stx1103}, \href
  {https://ui.adsabs.harvard.edu/abs/2017MNRAS.469.3994B} {469, 3994}

\bibitem[\protect\citeauthoryear{{Booth} et~al.,}{{Booth}
  et~al.}{2019}]{Booth2019}
{Booth} M.,  et~al., 2019, \mn@doi [\mnras] {10.1093/mnras/sty2993}, \href
  {http://adsabs.harvard.edu/abs/2019MNRAS.482.3443B} {482, 3443}

\bibitem[\protect\citeauthoryear{{Bouret}, {Deleuil}, {Lanz}, {Roberge},
  {Lecavelier des Etangs}  \& {Vidal-Madjar}}{{Bouret}
  et~al.}{2002}]{Bouret2002}
{Bouret} J.~C.,  {Deleuil} M.,  {Lanz} T.,  {Roberge} A.,  {Lecavelier des
  Etangs} A.,   {Vidal-Madjar} A.,  2002, \mn@doi [\aap]
  {10.1051/0004-6361:20020741}, \href
  {https://ui.adsabs.harvard.edu/abs/2002A&A...390.1049B} {390, 1049}

\bibitem[\protect\citeauthoryear{{Bovy}}{{Bovy}}{2017}]{Bovy2017}
{Bovy} J.,  2017, \mn@doi [\mnras] {10.1093/mnras/stx1277}, \href
  {http://adsabs.harvard.edu/abs/2017MNRAS.470.1360B} {470, 1360}

\bibitem[\protect\citeauthoryear{{Bowler} \& {Nielsen}}{{Bowler} \&
  {Nielsen}}{2018}]{Bowler2018}
{Bowler} B.~P.,  {Nielsen} E.~L.,  2018, {Occurrence Rates from Direct Imaging
  Surveys}.
p.~155, \mn@doi{10.1007/978-3-319-55333-7_155}

\bibitem[\protect\citeauthoryear{{Burns}, {Lamy}  \& {Soter}}{{Burns}
  et~al.}{1979}]{Burns1979}
{Burns} J.~A.,  {Lamy} P.~L.,   {Soter} S.,  1979, \mn@doi [\icarus]
  {10.1016/0019-1035(79)90050-2}, \href
  {http://adsabs.harvard.edu/abs/1979Icar...40....1B} {40, 1}

\bibitem[\protect\citeauthoryear{{Cataldi} et~al.,}{{Cataldi}
  et~al.}{2014}]{Cataldi2014}
{Cataldi} G.,  et~al., 2014, \mn@doi [\aap] {10.1051/0004-6361/201323126},
  \href {http://adsabs.harvard.edu/abs/2014A%26A...563A..66C} {563, A66}

\bibitem[\protect\citeauthoryear{{Cataldi} et~al.,}{{Cataldi}
  et~al.}{2018}]{Cataldi2018}
{Cataldi} G.,  et~al., 2018, \mn@doi [\apj] {10.3847/1538-4357/aac5f3}, \href
  {http://adsabs.harvard.edu/abs/2018ApJ...861...72C} {861, 72}

\bibitem[\protect\citeauthoryear{{Cataldi} et~al.,}{{Cataldi}
  et~al.}{2019}]{Cataldi2019}
{Cataldi} G.,  et~al., 2019, arXiv e-prints, \href
  {http://adsabs.harvard.edu/abs/2019arXiv190407215C} {}

\bibitem[\protect\citeauthoryear{{Cavallius}, {Cataldi}, {Brandeker},
  {Olofsson}, {Larsson}  \& {Liseau}}{{Cavallius} et~al.}{2019}]{Cavallius2019}
{Cavallius} M.,  {Cataldi} G.,  {Brandeker} A.,  {Olofsson} G.,  {Larsson} B.,
   {Liseau} R.,  2019, arXiv e-prints, \href
  {https://ui.adsabs.harvard.edu/abs/2019arXiv190611106C} {p. arXiv:1906.11106}

\bibitem[\protect\citeauthoryear{{Crida}, {Morbidelli}  \& {Masset}}{{Crida}
  et~al.}{2006}]{Crida2006}
{Crida} A.,  {Morbidelli} A.,   {Masset} F.,  2006, \mn@doi [\icarus]
  {10.1016/j.icarus.2005.10.007}, \href
  {https://ui.adsabs.harvard.edu/abs/2006Icar..181..587C} {181, 587}

\bibitem[\protect\citeauthoryear{{Deleuil} et~al.,}{{Deleuil}
  et~al.}{2001}]{Deleuil2001}
{Deleuil} M.,  et~al., 2001, \mn@doi [\apjl] {10.1086/323005}, \href
  {https://ui.adsabs.harvard.edu/abs/2001ApJ...557L..67D} {557, L67}

\bibitem[\protect\citeauthoryear{{Dent} et~al.,}{{Dent}
  et~al.}{2014}]{Dent2014}
{Dent} W.~R.~F.,  et~al., 2014, \mn@doi [Science] {10.1126/science.1248726},
  \href {http://adsabs.harvard.edu/abs/2014Sci...343.1490D} {343, 1490}

\bibitem[\protect\citeauthoryear{{Dominik} \& {Decin}}{{Dominik} \&
  {Decin}}{2003}]{Dominik2003}
{Dominik} C.,  {Decin} G.,  2003, \mn@doi [\apj] {10.1086/379169}, \href
  {http://adsabs.harvard.edu/abs/2003ApJ...598..626D} {598, 626}

\bibitem[\protect\citeauthoryear{{Draine}}{{Draine}}{1978}]{Draine1978}
{Draine} B.~T.,  1978, \mn@doi [\apjs] {10.1086/190513}, \href
  {http://adsabs.harvard.edu/abs/1978ApJS...36..595D} {36, 595}

\bibitem[\protect\citeauthoryear{{Ferlet}, {Vidal-Madjar}  \& {Hobbs}}{{Ferlet}
  et~al.}{1987}]{Ferlet1987}
{Ferlet} R.,  {Vidal-Madjar} A.,   {Hobbs} L.~M.,  1987, \aap, \href
  {http://adsabs.harvard.edu/abs/1987A%26A...185..267F} {185, 267}

\bibitem[\protect\citeauthoryear{{Fern{\'a}ndez}, {Brandeker}  \&
  {Wu}}{{Fern{\'a}ndez} et~al.}{2006}]{Fernandez2006}
{Fern{\'a}ndez} R.,  {Brandeker} A.,   {Wu} Y.,  2006, \mn@doi [\apj]
  {10.1086/500788}, \href {http://adsabs.harvard.edu/abs/2006ApJ...643..509F}
  {643, 509}

\bibitem[\protect\citeauthoryear{{Flaherty} et~al.,}{{Flaherty}
  et~al.}{2016}]{Flaherty2016}
{Flaherty} K.~M.,  et~al., 2016, \mn@doi [\apj] {10.3847/0004-637X/818/1/97},
  \href {https://ui.adsabs.harvard.edu/abs/2016ApJ...818...97F} {818, 97}

\bibitem[\protect\citeauthoryear{{Greaves} et~al.,}{{Greaves}
  et~al.}{2016}]{Greaves2016}
{Greaves} J.~S.,  et~al., 2016, \mn@doi [\mnras] {10.1093/mnras/stw1569}, \href
  {http://adsabs.harvard.edu/abs/2016MNRAS.461.3910G} {461, 3910}

\bibitem[\protect\citeauthoryear{{Grigorieva}, {Th{\'e}bault}, {Artymowicz}  \&
  {Brandeker}}{{Grigorieva} et~al.}{2007}]{Grigorieva2007}
{Grigorieva} A.,  {Th{\'e}bault} P.,  {Artymowicz} P.,   {Brandeker} A.,  2007,
  \mn@doi [\aap] {10.1051/0004-6361:20077686}, \href
  {http://adsabs.harvard.edu/abs/2007A%26A...475..755G} {475, 755}

\bibitem[\protect\citeauthoryear{{Hales}, {Gorti}, {Carpenter}, {Hughes}  \&
  {Flaherty}}{{Hales} et~al.}{2019}]{Hales2019}
{Hales} A.~S.,  {Gorti} U.,  {Carpenter} J.~M.,  {Hughes} M.,   {Flaherty} K.,
  2019, \mn@doi [\apj] {10.3847/1538-4357/ab211e}, \href
  {https://ui.adsabs.harvard.edu/abs/2019ApJ...878..113H} {878, 113}

\bibitem[\protect\citeauthoryear{{Higuchi} et~al.,}{{Higuchi}
  et~al.}{2017}]{Higuchi2017}
{Higuchi} A.~E.,  et~al., 2017, \mn@doi [\apjl] {10.3847/2041-8213/aa67f4},
  \href {http://adsabs.harvard.edu/abs/2017ApJ...839L..14H} {839, L14}

\bibitem[\protect\citeauthoryear{{Higuchi} et~al.,}{{Higuchi}
  et~al.}{2019}]{Higuchi2019}
{Higuchi} A.~E.,  et~al., 2019, arXiv e-prints, \href
  {https://ui.adsabs.harvard.edu/abs/2019arXiv190807032H} {p. arXiv:1908.07032}

\bibitem[\protect\citeauthoryear{{Hudson}}{{Hudson}}{1971}]{Hudson1971}
{Hudson} R.~D.,  1971, \mn@doi [Reviews of Geophysics and Space Physics]
  {10.1029/RG009i002p00305}, \href
  {http://adsabs.harvard.edu/abs/1971RvGSP...9..305H} {9, 305}

\bibitem[\protect\citeauthoryear{Hughes, Wilner, Andrews, Qi  \&
  Hogerheijde}{Hughes et~al.}{2011}]{Hughes2011}
Hughes a.~M.,  Wilner D.~J.,  Andrews S.~M.,  Qi C.,   Hogerheijde M.~R.,
  2011, \mn@doi [The Astrophysical Journal] {10.1088/0004-637X/727/2/85}, 727,
  85

\bibitem[\protect\citeauthoryear{{Hughes} et~al.,}{{Hughes}
  et~al.}{2017}]{Hughes2017}
{Hughes} A.~M.,  et~al., 2017, \mn@doi [\apj] {10.3847/1538-4357/aa6b04}, \href
  {https://ui.adsabs.harvard.edu/abs/2017ApJ...839...86H} {839, 86}

\bibitem[\protect\citeauthoryear{{Iglesias} et~al.,}{{Iglesias}
  et~al.}{2018}]{Iglesias2018}
{Iglesias} D.,  et~al., 2018, \mn@doi [\mnras] {10.1093/mnras/sty1724}, \href
  {https://ui.adsabs.harvard.edu/abs/2018MNRAS.480..488I} {480, 488}

\bibitem[\protect\citeauthoryear{{Kains}, {Wyatt}  \& {Greaves}}{{Kains}
  et~al.}{2011}]{Kains2011}
{Kains} N.,  {Wyatt} M.~C.,   {Greaves} J.~S.,  2011, \mn@doi [\mnras]
  {10.1111/j.1365-2966.2011.18566.x}, \href
  {http://adsabs.harvard.edu/abs/2011MNRAS.414.2486K} {414, 2486}

\bibitem[\protect\citeauthoryear{{Kennedy}, {Marino}, {Matr{\`a}}, {Pani{\'c}},
  {Wilner}, {Wyatt}  \& {Yelverton}}{{Kennedy} et~al.}{2018}]{Kennedy2018}
{Kennedy} G.~M.,  {Marino} S.,  {Matr{\`a}} L.,  {Pani{\'c}} O.,  {Wilner} D.,
  {Wyatt} M.~C.,   {Yelverton} B.,  2018, \mn@doi [\mnras]
  {10.1093/mnras/sty135}, \href
  {http://adsabs.harvard.edu/abs/2018MNRAS.475.4924K} {475, 4924}

\bibitem[\protect\citeauthoryear{{Kiefer}, {Lecavelier des Etangs}, {Boissier},
  {Vidal-Madjar}, {Beust}, {Lagrange}, {H{\'e}brard}  \& {Ferlet}}{{Kiefer}
  et~al.}{2014}]{Kiefer2014b}
{Kiefer} F.,  {Lecavelier des Etangs} A.,  {Boissier} J.,  {Vidal-Madjar} A.,
  {Beust} H.,  {Lagrange} A.-M.,  {H{\'e}brard} G.,   {Ferlet} R.,  2014,
  \mn@doi [\nat] {10.1038/nature13849}, \href
  {http://adsabs.harvard.edu/abs/2014Natur.514..462K} {514, 462}

\bibitem[\protect\citeauthoryear{{K{\'o}sp{\'a}l} et~al.,}{{K{\'o}sp{\'a}l}
  et~al.}{2013}]{Kospal2013}
{K{\'o}sp{\'a}l} {\'A}.,  et~al., 2013, \mn@doi [\apj]
  {10.1088/0004-637X/776/2/77}, \href
  {http://adsabs.harvard.edu/abs/2013ApJ...776...77K} {776, 77}

\bibitem[\protect\citeauthoryear{{Kral} \& {Latter}}{{Kral} \&
  {Latter}}{2016}]{Kral2016b}
{Kral} Q.,  {Latter} H.,  2016, \mn@doi [\mnras] {10.1093/mnras/stw1429}, \href
  {http://adsabs.harvard.edu/abs/2016MNRAS.461.1614K} {461, 1614}

\bibitem[\protect\citeauthoryear{{Kral}, {Th{\'e}bault}  \& {Charnoz}}{{Kral}
  et~al.}{2013}]{Kral2013}
{Kral} Q.,  {Th{\'e}bault} P.,   {Charnoz} S.,  2013, \mn@doi [\aap]
  {10.1051/0004-6361/201321398}, \href
  {http://adsabs.harvard.edu/abs/2013A%26A...558A.121K} {558, A121}

\bibitem[\protect\citeauthoryear{{Kral}, {Wyatt}, {Carswell}, {Pringle},
  {Matr{\`a}}  \& {Juh{\'a}sz}}{{Kral} et~al.}{2016}]{Kral2016}
{Kral} Q.,  {Wyatt} M.,  {Carswell} R.~F.,  {Pringle} J.~E.,  {Matr{\`a}} L.,
  {Juh{\'a}sz} A.,  2016, \mn@doi [\mnras] {10.1093/mnras/stw1361}, \href
  {http://adsabs.harvard.edu/abs/2016MNRAS.461..845K} {461, 845}

\bibitem[\protect\citeauthoryear{{Kral}, {Matr{\`a}}, {Wyatt}  \&
  {Kennedy}}{{Kral} et~al.}{2017}]{Kral2017CO}
{Kral} Q.,  {Matr{\`a}} L.,  {Wyatt} M.~C.,   {Kennedy} G.~M.,  2017, \mn@doi
  [\mnras] {10.1093/mnras/stx730}, \href
  {http://adsabs.harvard.edu/abs/2017MNRAS.469..521K} {469, 521}

\bibitem[\protect\citeauthoryear{{Kral}, {Marino}, {Wyatt}, {Kama}  \&
  {Matr{\`a}}}{{Kral} et~al.}{2019}]{Kral2019}
{Kral} Q.,  {Marino} S.,  {Wyatt} M.~C.,  {Kama} M.,   {Matr{\`a}} L.,  2019,
  \mn@doi [\mnras] {10.1093/mnras/sty2923}, \href
  {https://ui.adsabs.harvard.edu/abs/2019MNRAS.489.3670K} {489, 3670}

\bibitem[\protect\citeauthoryear{{Krivov}, {L{\"o}hne}  \& {Srem{\v
  c}evi{\'c}}}{{Krivov} et~al.}{2006}]{Krivov2006}
{Krivov} A.~V.,  {L{\"o}hne} T.,   {Srem{\v c}evi{\'c}} M.,  2006, \mn@doi
  [\aap] {10.1051/0004-6361:20064907}, \href
  {http://adsabs.harvard.edu/abs/2006A%26A...455..509K} {455, 509}

\bibitem[\protect\citeauthoryear{{Krivov}, {Ide}, {L{\"o}hne}, {Johansen}  \&
  {Blum}}{{Krivov} et~al.}{2018}]{Krivov2018}
{Krivov} A.~V.,  {Ide} A.,  {L{\"o}hne} T.,  {Johansen} A.,   {Blum} J.,  2018,
  \mn@doi [\mnras] {10.1093/mnras/stx2932}, \href
  {http://adsabs.harvard.edu/abs/2018MNRAS.474.2564K} {474, 2564}

\bibitem[\protect\citeauthoryear{{Kuiper} \& {Hosokawa}}{{Kuiper} \&
  {Hosokawa}}{2018}]{Kuiper2018}
{Kuiper} R.,  {Hosokawa} T.,  2018, \mn@doi [\aap]
  {10.1051/0004-6361/201832638}, \href
  {https://ui.adsabs.harvard.edu/abs/2018A&A...616A.101K} {616, A101}

\bibitem[\protect\citeauthoryear{{Lecavelier Des Etangs}, {Vidal-Madjar}  \&
  {Ferlet}}{{Lecavelier Des Etangs} et~al.}{1998}]{Lecavelier1998}
{Lecavelier Des Etangs} A.,  {Vidal-Madjar} A.,   {Ferlet} R.,  1998, \aap,
  \href {https://ui.adsabs.harvard.edu/abs/1998A%26A...339..477L} {339, 477}

\bibitem[\protect\citeauthoryear{{Lenz} \& {Ayres}}{{Lenz} \&
  {Ayres}}{1992}]{Lenz1992}
{Lenz} D.~D.,  {Ayres} T.~R.,  1992, \mn@doi [\pasp] {10.1086/133096}, \href
  {https://ui.adsabs.harvard.edu/abs/1992PASP..104.1104L} {104, 1104}

\bibitem[\protect\citeauthoryear{{Lieman-Sifry}, {Hughes}, {Carpenter},
  {Gorti}, {Hales}  \& {Flaherty}}{{Lieman-Sifry}
  et~al.}{2016}]{Lieman-Sifry2016}
{Lieman-Sifry} J.,  {Hughes} A.~M.,  {Carpenter} J.~M.,  {Gorti} U.,  {Hales}
  A.,   {Flaherty} K.~M.,  2016, \mn@doi [\apj] {10.3847/0004-637X/828/1/25},
  \href {http://adsabs.harvard.edu/abs/2016ApJ...828...25L} {828, 25}

\bibitem[\protect\citeauthoryear{{L{\"o}hne}, {Krivov}  \&
  {Rodmann}}{{L{\"o}hne} et~al.}{2008}]{Lohne2008}
{L{\"o}hne} T.,  {Krivov} A.~V.,   {Rodmann} J.,  2008, \mn@doi [\apj]
  {10.1086/524840}, \href {http://adsabs.harvard.edu/abs/2008ApJ...673.1123L}
  {673, 1123}

\bibitem[\protect\citeauthoryear{{Lubow} \& {D'Angelo}}{{Lubow} \&
  {D'Angelo}}{2006}]{Lubow2006}
{Lubow} S.~H.,  {D'Angelo} G.,  2006, \mn@doi [\apj] {10.1086/500356}, \href
  {https://ui.adsabs.harvard.edu/abs/2006ApJ...641..526L} {641, 526}

\bibitem[\protect\citeauthoryear{{Lubow}, {Seibert}  \& {Artymowicz}}{{Lubow}
  et~al.}{1999}]{Lubow1999}
{Lubow} S.~H.,  {Seibert} M.,   {Artymowicz} P.,  1999, \mn@doi [\apj]
  {10.1086/308045}, \href
  {https://ui.adsabs.harvard.edu/abs/1999ApJ...526.1001L} {526, 1001}

\bibitem[\protect\citeauthoryear{{Lynden-Bell} \& {Pringle}}{{Lynden-Bell} \&
  {Pringle}}{1974}]{Lynden-Bell1974}
{Lynden-Bell} D.,  {Pringle} J.~E.,  1974, \mnras, \href
  {http://adsabs.harvard.edu/abs/1974MNRAS.168..603L} {168, 603}

\bibitem[\protect\citeauthoryear{{Lyra} \& {Kuchner}}{{Lyra} \&
  {Kuchner}}{2013}]{Lyra2013photoelectric}
{Lyra} W.,  {Kuchner} M.,  2013, \mn@doi [\nat] {10.1038/nature12281}, \href
  {http://adsabs.harvard.edu/abs/2013Natur.499..184L} {499, 184}

\bibitem[\protect\citeauthoryear{{MacGregor} et~al.,}{{MacGregor}
  et~al.}{2018}]{Macgregor2018}
{MacGregor} M.~A.,  et~al., 2018, \mn@doi [\apj] {10.3847/1538-4357/aaec71},
  \href {https://ui.adsabs.harvard.edu/abs/2018ApJ...869...75M} {869, 75}

\bibitem[\protect\citeauthoryear{{Machida}, {Kokubo}, {Inutsuka}  \&
  {Matsumoto}}{{Machida} et~al.}{2010}]{Machida2010}
{Machida} M.~N.,  {Kokubo} E.,  {Inutsuka} S.-I.,   {Matsumoto} T.,  2010,
  \mn@doi [\mnras] {10.1111/j.1365-2966.2010.16527.x}, \href
  {https://ui.adsabs.harvard.edu/abs/2010MNRAS.405.1227M} {405, 1227}

\bibitem[\protect\citeauthoryear{{Marino} et~al.,}{{Marino}
  et~al.}{2016}]{Marino2016}
{Marino} S.,  et~al., 2016, \mn@doi [\mnras] {10.1093/mnras/stw1216}, \href
  {http://adsabs.harvard.edu/abs/2016MNRAS.460.2933M} {460, 2933}

\bibitem[\protect\citeauthoryear{{Marino} et~al.,}{{Marino}
  et~al.}{2017}]{Marino2017etacorvi}
{Marino} S.,  et~al., 2017, \mn@doi [\mnras] {10.1093/mnras/stw2867}, \href
  {http://adsabs.harvard.edu/abs/2017MNRAS.465.2595M} {465, 2595}

\bibitem[\protect\citeauthoryear{{Marino} et~al.,}{{Marino}
  et~al.}{2018}]{Marino2018hd107}
{Marino} S.,  et~al., 2018, \mn@doi [\mnras] {10.1093/mnras/sty1790}, \href
  {http://adsabs.harvard.edu/abs/2018MNRAS.479.5423M} {479, 5423}

\bibitem[\protect\citeauthoryear{{Marino}, {Yelverton}, {Booth}, {Faramaz},
  {Kennedy}, {Matr{\`a}}  \& {Wyatt}}{{Marino} et~al.}{2019}]{Marino2019}
{Marino} S.,  {Yelverton} B.,  {Booth} M.,  {Faramaz} V.,  {Kennedy} G.~M.,
  {Matr{\`a}} L.,   {Wyatt} M.~C.,  2019, \mn@doi [\mnras]
  {10.1093/mnras/stz049}, \href
  {https://ui.adsabs.harvard.edu/abs/2019MNRAS.484.1257M} {484, 1257}

\bibitem[\protect\citeauthoryear{{Matr{\`a}} et~al.,}{{Matr{\`a}}
  et~al.}{2017a}]{Matra2017betapic}
{Matr{\`a}} L.,  et~al., 2017a, \mn@doi [\mnras] {10.1093/mnras/stw2415}, \href
  {http://adsabs.harvard.edu/abs/2017MNRAS.464.1415M} {464, 1415}

\bibitem[\protect\citeauthoryear{{Matr{\`a}} et~al.,}{{Matr{\`a}}
  et~al.}{2017b}]{Matra2017fomalhaut}
{Matr{\`a}} L.,  et~al., 2017b, \mn@doi [\apj] {10.3847/1538-4357/aa71b4},
  \href {http://adsabs.harvard.edu/abs/2017ApJ...842....9M} {842, 9}

\bibitem[\protect\citeauthoryear{{Matr{\`a}}, {Wilner}, {{\"O}berg}, {Andrews},
  {Loomis}, {Wyatt}  \& {Dent}}{{Matr{\`a}} et~al.}{2018a}]{Matra2018}
{Matr{\`a}} L.,  {Wilner} D.~J.,  {{\"O}berg} K.~I.,  {Andrews} S.~M.,
  {Loomis} R.~A.,  {Wyatt} M.~C.,   {Dent} W.~R.~F.,  2018a, \mn@doi [\apj]
  {10.3847/1538-4357/aaa42a}, \href
  {http://adsabs.harvard.edu/abs/2018ApJ...853..147M} {853, 147}

\bibitem[\protect\citeauthoryear{{Matr{\`a}}, {Marino}, {Kennedy}, {Wyatt},
  {{\"O}berg}  \& {Wilner}}{{Matr{\`a}} et~al.}{2018b}]{Matra2018mmlaw}
{Matr{\`a}} L.,  {Marino} S.,  {Kennedy} G.~M.,  {Wyatt} M.~C.,  {{\"O}berg}
  K.~I.,   {Wilner} D.~J.,  2018b, \mn@doi [\apj] {10.3847/1538-4357/aabcc4},
  \href {http://adsabs.harvard.edu/abs/2018ApJ...859...72M} {859, 72}

\bibitem[\protect\citeauthoryear{{Matr{\`a}}, {{\"O}berg}, {Wilner}, {Olofsson}
   \& {Bayo}}{{Matr{\`a}} et~al.}{2019}]{Matra2019twa7}
{Matr{\`a}} L.,  {{\"O}berg} K.~I.,  {Wilner} D.~J.,  {Olofsson} J.,   {Bayo}
  A.,  2019, \mn@doi [\aj] {10.3847/1538-3881/aaff5b}, \href
  {http://adsabs.harvard.edu/abs/2019AJ....157..117M} {157, 117}

\bibitem[\protect\citeauthoryear{{Metzger}, {Rafikov}  \&
  {Bochkarev}}{{Metzger} et~al.}{2012}]{Metzger2012}
{Metzger} B.~D.,  {Rafikov} R.~R.,   {Bochkarev} K.~V.,  2012, \mn@doi [\mnras]
  {10.1111/j.1365-2966.2012.20895.x}, \href
  {http://adsabs.harvard.edu/abs/2012MNRAS.423..505M} {423, 505}

\bibitem[\protect\citeauthoryear{{Montgomery} \& {Welsh}}{{Montgomery} \&
  {Welsh}}{2012}]{Montgomery2012}
{Montgomery} S.~L.,  {Welsh} B.~Y.,  2012, \mn@doi [\pasp] {10.1086/668293},
  \href {https://ui.adsabs.harvard.edu/abs/2012PASP..124.1042M} {124, 1042}

\bibitem[\protect\citeauthoryear{{Mo{\'o}r} et~al.,}{{Mo{\'o}r}
  et~al.}{2011}]{Moor2011a}
{Mo{\'o}r} A.,  et~al., 2011, \mn@doi [\apjl] {10.1088/2041-8205/740/1/L7},
  \href {http://adsabs.harvard.edu/abs/2011ApJ...740L...7M} {740, L7}

\bibitem[\protect\citeauthoryear{{Mo{\'o}r} et~al.,}{{Mo{\'o}r}
  et~al.}{2015}]{Moor2015gas}
{Mo{\'o}r} A.,  et~al., 2015, \mn@doi [\apj] {10.1088/0004-637X/814/1/42},
  \href {http://adsabs.harvard.edu/abs/2015ApJ...814...42M} {814, 42}

\bibitem[\protect\citeauthoryear{{Mo{\'o}r} et~al.,}{{Mo{\'o}r}
  et~al.}{2017}]{Moor2017}
{Mo{\'o}r} A.,  et~al., 2017, \mn@doi [\apj] {10.3847/1538-4357/aa8e4e}, \href
  {http://adsabs.harvard.edu/abs/2017ApJ...849..123M} {849, 123}

\bibitem[\protect\citeauthoryear{{Mo{\'o}r} et~al.,}{{Mo{\'o}r}
  et~al.}{2019}]{Moor2019}
{Mo{\'o}r} A.,  et~al., 2019, arXiv e-prints, \href
  {https://ui.adsabs.harvard.edu/abs/2019arXiv190809685M} {p. arXiv:1908.09685}

\bibitem[\protect\citeauthoryear{{Mordasini}, {Alibert}  \& {Benz}}{{Mordasini}
  et~al.}{2009}]{Mordasini2009}
{Mordasini} C.,  {Alibert} Y.,   {Benz} W.,  2009, \mn@doi [\aap]
  {10.1051/0004-6361/200810301}, \href
  {http://adsabs.harvard.edu/abs/2009A%26A...501.1139M} {501, 1139}

\bibitem[\protect\citeauthoryear{{Mumma} \& {Charnley}}{{Mumma} \&
  {Charnley}}{2011}]{Mumma2011}
{Mumma} M.~J.,  {Charnley} S.~B.,  2011, \mn@doi [\araa]
  {10.1146/annurev-astro-081309-130811}, \href
  {http://adsabs.harvard.edu/abs/2011ARA%26A..49..471M} {49, 471}

\bibitem[\protect\citeauthoryear{{Mustill} \& {Wyatt}}{{Mustill} \&
  {Wyatt}}{2009}]{Mustill2009}
{Mustill} A.~J.,  {Wyatt} M.~C.,  2009, \mn@doi [\mnras]
  {10.1111/j.1365-2966.2009.15360.x}, \href
  {http://adsabs.harvard.edu/abs/2009MNRAS.399.1403M} {399, 1403}

\bibitem[\protect\citeauthoryear{{{\"O}berg}, {Murray-Clay}  \&
  {Bergin}}{{{\"O}berg} et~al.}{2011}]{Oberg2011}
{{\"O}berg} K.~I.,  {Murray-Clay} R.,   {Bergin} E.~A.,  2011, \mn@doi [\apj]
  {10.1088/2041-8205/743/1/L16}, \href
  {https://ui.adsabs.harvard.edu/abs/2011ApJ...743L..16O} {743, L16}

\bibitem[\protect\citeauthoryear{{Olofsson} et~al.,}{{Olofsson}
  et~al.}{2016}]{Olofsson2016}
{Olofsson} J.,  et~al., 2016, \mn@doi [\aap] {10.1051/0004-6361/201628196},
  \href {https://ui.adsabs.harvard.edu/abs/2016A&A...591A.108O} {591, A108}

\bibitem[\protect\citeauthoryear{{Pawellek} \& {Krivov}}{{Pawellek} \&
  {Krivov}}{2015}]{Pawellek2015}
{Pawellek} N.,  {Krivov} A.~V.,  2015, \mn@doi [\mnras]
  {10.1093/mnras/stv2142}, \href
  {http://adsabs.harvard.edu/abs/2015MNRAS.454.3207P} {454, 3207}

\bibitem[\protect\citeauthoryear{{Pawellek}, {Krivov}, {Marshall},
  {Montesinos}, {{\'A}brah{\'a}m}, {Mo{\'o}r}, {Bryden}  \& {Eiroa}}{{Pawellek}
  et~al.}{2014}]{Pawellek2014}
{Pawellek} N.,  {Krivov} A.~V.,  {Marshall} J.~P.,  {Montesinos} B.,
  {{\'A}brah{\'a}m} P.,  {Mo{\'o}r} A.,  {Bryden} G.,   {Eiroa} C.,  2014,
  \mn@doi [\apj] {10.1088/0004-637X/792/1/65}, \href
  {http://adsabs.harvard.edu/abs/2014ApJ...792...65P} {792, 65}

\bibitem[\protect\citeauthoryear{{Rebollido} et~al.,}{{Rebollido}
  et~al.}{2018}]{Rebollido2018}
{Rebollido} I.,  et~al., 2018, \mn@doi [\aap] {10.1051/0004-6361/201732329},
  \href {https://ui.adsabs.harvard.edu/abs/2018A&A...614A...3R} {614, A3}

\bibitem[\protect\citeauthoryear{{Richert}, {Lyra}  \& {Kuchner}}{{Richert}
  et~al.}{2018}]{Richert2017}
{Richert} A.~J.~W.,  {Lyra} W.,   {Kuchner} M.~J.,  2018, \mn@doi [\apj]
  {10.3847/1538-4357/aaadaa}, \href
  {http://adsabs.harvard.edu/abs/2018ApJ...856...41R} {856, 41}

\bibitem[\protect\citeauthoryear{{Rieke} et~al.,}{{Rieke}
  et~al.}{2005}]{Rieke2005}
{Rieke} G.~H.,  et~al., 2005, \mn@doi [\apj] {10.1086/426937}, \href
  {http://adsabs.harvard.edu/abs/2005ApJ...620.1010R} {620, 1010}

\bibitem[\protect\citeauthoryear{{Riviere-Marichalar}
  et~al.,}{{Riviere-Marichalar} et~al.}{2012}]{Riviere-Marichalar2012}
{Riviere-Marichalar} P.,  et~al., 2012, \mn@doi [\aap]
  {10.1051/0004-6361/201219745}, \href
  {http://adsabs.harvard.edu/abs/2012A%26A...546L...8R} {546, L8}

\bibitem[\protect\citeauthoryear{{Roberge} et~al.,}{{Roberge}
  et~al.}{2013}]{Roberge2013}
{Roberge} A.,  et~al., 2013, \mn@doi [\apj] {10.1088/0004-637X/771/1/69}, \href
  {http://adsabs.harvard.edu/abs/2013ApJ...771...69R} {771, 69}

\bibitem[\protect\citeauthoryear{{Rollins} \& {Rawlings}}{{Rollins} \&
  {Rawlings}}{2012}]{Rollins2012}
{Rollins} R.~P.,  {Rawlings} J.~M.~C.,  2012, \mn@doi [\mnras]
  {10.1111/j.1365-2966.2012.22121.x}, \href
  {http://adsabs.harvard.edu/abs/2012MNRAS.427.2328R} {427, 2328}

\bibitem[\protect\citeauthoryear{{Schneider}, {Silverstone}  \&
  {Hines}}{{Schneider} et~al.}{2005}]{Schneider2005}
{Schneider} G.,  {Silverstone} M.~D.,   {Hines} D.~C.,  2005, \mn@doi [\apjl]
  {10.1086/452631}, \href
  {https://ui.adsabs.harvard.edu/abs/2005ApJ...629L.117S} {629, L117}

\bibitem[\protect\citeauthoryear{{Sepulveda} et~al.,}{{Sepulveda}
  et~al.}{2019}]{Sepulveda2019}
{Sepulveda} A.~G.,  et~al., 2019, arXiv e-prints, \href
  {https://ui.adsabs.harvard.edu/abs/2019arXiv190608797S} {p. arXiv:1906.08797}

\bibitem[\protect\citeauthoryear{{Sibthorpe}, {Kennedy}, {Wyatt}, {Lestrade},
  {Greaves}, {Matthews}  \& {Duch{\^e}ne}}{{Sibthorpe}
  et~al.}{2018}]{Sibthorpe2018}
{Sibthorpe} B.,  {Kennedy} G.~M.,  {Wyatt} M.~C.,  {Lestrade} J.~F.,  {Greaves}
  J.~S.,  {Matthews} B.~C.,   {Duch{\^e}ne} G.,  2018, \mn@doi [\mnras]
  {10.1093/mnras/stx3188}, \href
  {https://ui.adsabs.harvard.edu/abs/2018MNRAS.475.3046S} {475, 3046}

\bibitem[\protect\citeauthoryear{{Slettebak}}{{Slettebak}}{1975}]{Slettebak1975}
{Slettebak} A.,  1975, \mn@doi [\apj] {10.1086/153493}, \href
  {http://adsabs.harvard.edu/abs/1975ApJ...197..137S} {197, 137}

\bibitem[\protect\citeauthoryear{{Stevenson}}{{Stevenson}}{1990}]{Stevenson1990}
{Stevenson} D.~J.,  1990, \mn@doi [\apj] {10.1086/168282}, \href
  {https://ui.adsabs.harvard.edu/abs/1990ApJ...348..730S} {348, 730}

\bibitem[\protect\citeauthoryear{{Su} et~al.,}{{Su} et~al.}{2006}]{Su2006}
{Su} K.~Y.~L.,  et~al., 2006, \mn@doi [\apj] {10.1086/508649}, \href
  {http://adsabs.harvard.edu/abs/2006ApJ...653..675S} {653, 675}

\bibitem[\protect\citeauthoryear{{Takeuchi} \& {Artymowicz}}{{Takeuchi} \&
  {Artymowicz}}{2001}]{Takeuchi2001}
{Takeuchi} T.,  {Artymowicz} P.,  2001, \mn@doi [\apj] {10.1086/322252}, \href
  {https://ui.adsabs.harvard.edu/abs/2001ApJ...557..990T} {557, 990}

\bibitem[\protect\citeauthoryear{{Tazaki} \& {Nomura}}{{Tazaki} \&
  {Nomura}}{2015}]{Tazaki2015}
{Tazaki} R.,  {Nomura} H.,  2015, \mn@doi [\apj] {10.1088/0004-637X/799/2/119},
  \href {https://ui.adsabs.harvard.edu/abs/2015ApJ...799..119T} {799, 119}

\bibitem[\protect\citeauthoryear{{Teague}, {Bae}, {Bergin}, {Birnstiel}  \&
  {Foreman-Mackey}}{{Teague} et~al.}{2018}]{Teague2018}
{Teague} R.,  {Bae} J.,  {Bergin} E.~A.,  {Birnstiel} T.,   {Foreman-Mackey}
  D.,  2018, \mn@doi [\apjl] {10.3847/2041-8213/aac6d7}, \href
  {http://adsabs.harvard.edu/abs/2018ApJ...860L..12T} {860, L12}

\bibitem[\protect\citeauthoryear{{Th{\'e}bault} \& {Augereau}}{{Th{\'e}bault}
  \& {Augereau}}{2007}]{Thebault2007}
{Th{\'e}bault} P.,  {Augereau} J.-C.,  2007, \mn@doi [\aap]
  {10.1051/0004-6361:20077709}, \href
  {http://adsabs.harvard.edu/abs/2007A%26A...472..169T} {472, 169}

\bibitem[\protect\citeauthoryear{{Turner}, {Willacy}, {Bryden}  \&
  {Yorke}}{{Turner} et~al.}{2006}]{Turner2006}
{Turner} N.~J.,  {Willacy} K.,  {Bryden} G.,   {Yorke} H.~W.,  2006, \mn@doi
  [\apj] {10.1086/499486}, \href
  {https://ui.adsabs.harvard.edu/abs/2006ApJ...639.1218T} {639, 1218}

\bibitem[\protect\citeauthoryear{{Visser}, {van Dishoeck}  \& {Black}}{{Visser}
  et~al.}{2009}]{Visser2009}
{Visser} R.,  {van Dishoeck} E.~F.,   {Black} J.~H.,  2009, \mn@doi [\aap]
  {10.1051/0004-6361/200912129}, \href
  {http://adsabs.harvard.edu/abs/2009A%26A...503..323V} {503, 323}

\bibitem[\protect\citeauthoryear{{Welsh} \& {Montgomery}}{{Welsh} \&
  {Montgomery}}{2018}]{Welsh2018}
{Welsh} B.~Y.,  {Montgomery} S.~L.,  2018, \mn@doi [\mnras]
  {10.1093/mnras/stx2800}, \href
  {https://ui.adsabs.harvard.edu/abs/2018MNRAS.474.1515W} {474, 1515}

\bibitem[\protect\citeauthoryear{{Wyatt}}{{Wyatt}}{2005}]{Wyatt2005prdrag}
{Wyatt} M.~C.,  2005, \mn@doi [\aap] {10.1051/0004-6361:20042073}, \href
  {https://ui.adsabs.harvard.edu/abs/2005A&A...433.1007W} {433, 1007}

\bibitem[\protect\citeauthoryear{{Wyatt}}{{Wyatt}}{2008}]{Wyatt2008}
{Wyatt} M.~C.,  2008, \mn@doi [\araa] {10.1146/annurev.astro.45.051806.110525},
  \href {http://adsabs.harvard.edu/abs/2008ARA%26A..46..339W} {46, 339}

\bibitem[\protect\citeauthoryear{{Wyatt} \& {Dent}}{{Wyatt} \&
  {Dent}}{2002}]{Wyatt2002}
{Wyatt} M.~C.,  {Dent} W.~R.~F.,  2002, \mn@doi [\mnras]
  {10.1046/j.1365-8711.2002.05533.x}, \href
  {http://adsabs.harvard.edu/abs/2002MNRAS.334..589W} {334, 589}

\bibitem[\protect\citeauthoryear{{Wyatt}, {Smith}, {Greaves}, {Beichman},
  {Bryden}  \& {Lisse}}{{Wyatt} et~al.}{2007a}]{Wyatt2007hotdust}
{Wyatt} M.~C.,  {Smith} R.,  {Greaves} J.~S.,  {Beichman} C.~A.,  {Bryden} G.,
   {Lisse} C.~M.,  2007a, \mn@doi [\apj] {10.1086/510999}, \href
  {http://adsabs.harvard.edu/abs/2007ApJ...658..569W} {658, 569}

\bibitem[\protect\citeauthoryear{{Wyatt}, {Smith}, {Su}, {Rieke}, {Greaves},
  {Beichman}  \& {Bryden}}{{Wyatt} et~al.}{2007b}]{Wyatt2007Astars}
{Wyatt} M.~C.,  {Smith} R.,  {Su} K.~Y.~L.,  {Rieke} G.~H.,  {Greaves} J.~S.,
  {Beichman} C.~A.,   {Bryden} G.,  2007b, \mn@doi [\apj] {10.1086/518404},
  \href {http://adsabs.harvard.edu/abs/2007ApJ...663..365W} {663, 365}

\bibitem[\protect\citeauthoryear{{Zuckerman} \& {Song}}{{Zuckerman} \&
  {Song}}{2012}]{Zuckerman2012}
{Zuckerman} B.,  {Song} I.,  2012, \mn@doi [\apj] {10.1088/0004-637X/758/2/77},
  \href {http://adsabs.harvard.edu/abs/2012ApJ...758...77Z} {758, 77}

\bibitem[\protect\citeauthoryear{{Zuckerman}, {Forveille}  \&
  {Kastner}}{{Zuckerman} et~al.}{1995}]{Zuckerman1995}
{Zuckerman} B.,  {Forveille} T.,   {Kastner} J.~H.,  1995, \mn@doi [\nat]
  {10.1038/373494a0}, \href {http://adsabs.harvard.edu/abs/1995Natur.373..494Z}
  {373, 494}

\bibitem[\protect\citeauthoryear{{van Dishoeck} \& {Black}}{{van Dishoeck} \&
  {Black}}{1988}]{vanDishoeck1988}
{van Dishoeck} E.~F.,  {Black} J.~H.,  1988, \mn@doi [\apj] {10.1086/166877},
  \href {http://adsabs.harvard.edu/abs/1988ApJ...334..771V} {334, 771}

\bibitem[\protect\citeauthoryear{van Dishoeck, Jonkheid  \& van Hemert}{van
  Dishoeck et~al.}{2006}]{vanDishoeck2006}
van Dishoeck E.~F.,  Jonkheid B.,   van Hemert M.~C.,  2006, \mn@doi [Faraday
  Discussions] {10.1039/b517564j}, 133, 231

\makeatother
\end{thebibliography}

%%%%%%%%%%%%%%%%%%%%%%%%%%%%%%%%%%%%%%%%%%%%%%%%%%

%%%%%%%%%%%%%%%%% APPENDICES %%%%%%%%%%%%%%%%%%%%%

\appendix

\section{Gas observations used in this work}

In tables \ref{tab:A} and \ref{tab:FGK} we show the properties of each
of the observed systems that were used to compare to our population
synthesis model. Table~\ref{tab:A} only shows A stars with ages
between 10-50~Myr and debris discs that have fractional luminosities
between $5\times10^{-4}-0.01$ and dust temperatures below 140~K.
Table~\ref{tab:FGK} shows a similar filtered sample, but without an
age restriction. Note that there has not been a complete survey for
FGK stars with a selection criteria similar to \cite{Moor2017}, and
thus Table \ref{tab:FGK} is likely incomplete.

\begin{table*}
 %% \centering
 \begin{adjustbox}{max width=1.0\textwidth}
   \begin{threeparttable}
     \caption{Properties of A stars with high fractional luminosity
       discs ($5\times10^{-4}-0.01$), ages ranging between 10-50~Myr
       and dust temperatures lower than 140~K. The upper limits
       correspond to 99.7\% confidence levels}
     \label{tab:A}
     \begin{tabular}{lccccccc} 
       \hline
       \hline
       Name   & Age [Myr] & $L_\star$ [$L_\odot$] & $f_\mathrm{IR}$  & $\rb$ [au] & $M_\mathrm{CO}$ [\Me] & $M_\mathrm{C}$ [\Me]  & ref \\
       \hline
       $\beta$~Pic & 23  &      8.7             & $2.6\times10^{-3}$ & 105   &    $2.8\times10^{-5}$ & $1.0\times10^{-3}$   & 1, 2\\
       49~Ceti     & 40	 &	16              & $7.2\times10^{-4}$ & 96   &    $1.4\times10^{-4}$ & $4.0\times10^{-3}$   & 3, 4\\	
       HD~21997	   & 45   &	9.9             & $5.6\times10^{-4}$ & 106   &	 $6\times10^{-2}$    &  ---                & 5 \\
       HD~32297	   & 30   &	8.2             & $5.5\times10^{-3}$ & 100   &    $1.3\times10^{-3}$  &	$3.5\times10^{-3}$ & 6, 7, 8, 9 \\
       HD~95086	   & 15   &     6.1             & $1.4\times10^{-3}$ &  204  &	  $4.3\times10^{-6}$   & ---                  & 10  \\
       HD~98363	   & 15   &     11              & $1.3\times10^{-3}$ & 32$^{\star}$&$<9.5\times10^{-6}$ & --- & 11, 12 \\
       HD~109832   & 15   &     5.3             & $5.4\times10^{-4}$ & 25$^{\star}$&$<7.\times10^{-6}$ & --- &	11, 12 \\
       HD~110058   & 15   &     5.9             & $1.4\times10^{-3}$ & 50  &	  $2.1\times10^{-5}$  & --- & 7, 11, 13 \\
       HD~121191   & 16	  &     7.2             & $4.5\times10^{-3}$ & 26$^{\star}$ &	$2.7\times10^{-3}$& ---   & 11 \\
       HD~121617   & 16   &     17              & $4.9\times10^{-3}$ &  83   &    $1.8\times10^{-2}$  &  ---		   & 11 \\
       HD~131488   & 16   &     13              & $2.2\times10^{-3}$ &  84   &	 $8.9\times10^{-2}$  &	---                 & 11 \\
       HD~131835   & 16	  &     11              & $2.2\times10^{-3}$ &  90   &	 $4.\times10^{-2}$   &	$3.3\times10^{-3}$ &  13, 14, 15, 16	\\
       HD~138813   & 10	  &     17              & $6.\times10^{-4}$  &  105  &    $7.4\times10^{-4}$  &  ---                 & 11, 13, 16 \\
       HD~143675   & 16   &     8.9	        &$6.9\times10^{-4}$   & 28$^{\star}$&$<1.7\times10^{-5}$ & --- & 11, 12 \\
       HD~145880   & 16   &     19	        & $1.7\times10^{-3}$  & 70$^{\star}$&  $<2.\times10^{-5}$	& --- & 11, 12 \\
       HD~156623   & 16   &     13              & $3.3\times10^{-3}$ &  94   &	  $2.5\times10^{-3}$  & ---                 & 13, 16  \\
       HR~4796	   & 10   &     26.             & $4.8\times10^{-3}$  &	79 & $<3.7\times10^{-6}$ & --- & 19 \\
       \hline
     \end{tabular}
     \begin{tablenotes}
     \item \textbf{References used in this table:} (1):
       \cite{Matra2017betapic}; (2): \cite{Cataldi2018}; (3):
       \cite{Hughes2017}; (4): \cite{Higuchi2017}; (5):
       \cite{Kospal2013}; (6): \cite{Greaves2016}; (7):
       \cite{Kral2017CO}; (8): \cite{Macgregor2018}; (9):
       \cite{Cataldi2019}; (10): \cite{Booth2019}; (11):
       \cite{Moor2017}; (12): upper limits derived in this work; (13):
       \cite{Lieman-Sifry2016}; (14): \cite{Moor2015gas}; (15):
       \cite{Kral2019}; (16) \cite{Hales2019}; (17):
       \cite{Kennedy2018}. $\star$: The belt radius is calculated as
       $r_\mathrm{BB}\times1.7$.
     \end{tablenotes}
   \end{threeparttable}  
  \end{adjustbox}
\end{table*}

\begin{table*}
 %% \centering
 \begin{adjustbox}{max width=1.0\textwidth}
   \begin{threeparttable}
     \caption{Properties of FGK stars with high fractional luminosity
       discs ($5\times10^{-4}-0.01$), with dust temperatures lower
       than 140~K and without age restrictions. The upper limits
       correspond to 99.7\% confidence levels.}
     \label{tab:FGK}
     \begin{tabular}{lccccccc} 
       \hline
       \hline
       Name   & Age [Myr] & $L_\star$ [$L_\odot$] & $f_\mathrm{IR}$  & $\rb$ [au] & $M_\mathrm{CO}$ [\Me]   & ref \\
       \hline
       HD~61005	   & 40     &  0.7             & $2.3\times10^{-3}$ & 66   &	$<6.\times10^{-6}$ 	& 1, 2, 3 \\
       HD~92945	   & 100-300 &  0.37            & $6.6\times10^{-4}$ & 87   & 	 $<3\times10^{-5}$	& 4 \\
       HD~107146   & 80-200 &   1.0             & $8.6\times10^{-4}$ & 89   &    $<5\times10^{-6}$	& 5 \\
       HD~111520   & 15     &  3.0             & $1.1\times10^{-3}$ & 96   &	$<3\times10^{-4}$	& 2, 3, 6 \\
       HD~145560   & 16     &   3.2            & $2.1\times10^{-3}$ & 88   &    $<2\times10^{-4}$	& 2, 3, 6 \\
       HD~146181   & 16     &    2.6           &  $2.2\times10^{-3}$ & 93  & 	$<3\times10^{-4}$	& 2, 3, 6 \\
       HD~146897   & 10. &      3.1             & $8.2\times10^{-3}$ & 81   &    $2.1\times10^{-4}$      & 6, 7 \\
       HD~170773   & 1500   &   3.6            &  $5.0\times10^{-4}$   & 193 &    $<1.4\times10^{-5}$	& 8\\
       HD~181327   & 23  &      2.9             & $2.1\times10^{-3}$ & 86    &	 $2.1\times10^{-6}$     & 9\\
       \hline
     \end{tabular}
     \begin{tablenotes}
     \item \textbf{References used in this table:} (1):
       \cite{Olofsson2016}; (2) \cite{Matra2019twa7}; (3): upper limit and
       mass derived in this work; (4) \cite{Marino2019}; (5):
       \cite{Marino2018hd107}; (6): \cite{Lieman-Sifry2016}; (7):
       \cite{Kral2017CO}; (8) \cite{Sepulveda2019}; (9) \cite{Marino2016}.
     \end{tablenotes}
   \end{threeparttable}  
  \end{adjustbox}
\end{table*}

%% If you want to present additional material which would interrupt the flow of the main paper,
%% it can be placed in an Appendix which appears after the list of references.

%%%%%%%%%%%%%%%%%%%%%%%%%%%%%%%%%%%%%%%%%%%%%%%%%%

% Don't change these lines
\bsp	% typesetting comment
\label{lastpage}
\end{document}